\documentclass[12pt]{article}
\usepackage{amssymb, bm, amsmath}
\usepackage{multirow}
\usepackage{epsfig}
\usepackage{float}

\renewcommand{\thefootnote}{\fnsymbol{footnote}}
\newcommand{\vev}[1]{{\langle{#1}\rangle}}

\usepackage{graphicx}
\usepackage{fancybox}
\usepackage{color}
\fancyput(-0.7in,-9.5in){
\color[rgb]{0.90,0.90,0.90}{
\rotatebox{52.5}{
}}}

\begin{document}
\title{
\begin{flushright}
\begin{minipage}{0.2\linewidth}
\normalsize
WU-HEP-14-02\\
EPHOU-14-007\\*[50pt]
\end{minipage}
\end{flushright}
{\Large \bf
Non-Abelian discrete flavor symmetries\\ of 10D SYM theory\\ with magnetized extra dimensions
\\*[20pt]}}

\author{
Hiroyuki~Abe$^{1,}$\footnote{
E-mail address: abe@waseda.jp}, \ 
Tatsuo~Kobayashi$^{2,3,}$\footnote{
E-mail address: kobayashi@particle.sci.hokudai.ac.jp}, \
Hiroshi~Ohki$^{4,}$\footnote{
E-mail address: ohki@kmi.nagoya-u.ac.jp}, \
Keigo~Sumita$^{1,}$\footnote{
E-mail address: k.sumita@moegi.waseda.jp}
\\ and \ 
Yoshiyuki~Tatsuta$^{1,}$\footnote{
E-mail address: y\_tatsuta@akane.waseda.jp}\\*[20pt]
$^1${\it \normalsize 
Department of Physics, Waseda University, 
Tokyo 169-8555, Japan} \\
$^2${\it \normalsize 
Department of Physics, Kyoto University, Kyoto 606-8502, Japan} \\
$^3${\it \normalsize 
Department of Physics, Hokkaido University, Sapporo 060-0810, Japan} \\
$^4${\it \normalsize 
Kobayashi-Maskawa Institute for the 
Origin of Particles and the Universe (KMI),} \\
{\it \normalsize 
Nagoya University, Nagoya 464-8602, Japan} \\*[50pt]}

\date{
\centerline{\small \bf Abstract}
\begin{minipage}{0.9\linewidth}
\medskip 
\medskip 
\small
We study discrete flavor symmetries of the models based on a ten-dimensional supersymmetric Yang-Mills (10D SYM) theory compactified on magnetized tori.
We assume non-vanishing non-factorizable fluxes as well as the orbifold projections.
These setups allow model-building with more various flavor structures.
Indeed, we show that there exist various classes of non-Abelian discrete flavor symmetries.
In particular, we find that $S_3$ flavor symmetries can be realized in the framework of the magnetized 10D SYM theory for the first time.
\end{minipage}
\vspace{-10pt}
}

\begin{titlepage}
\maketitle
\thispagestyle{empty}
\clearpage
\tableofcontents
\thispagestyle{empty}
\end{titlepage}

\renewcommand{\thefootnote}{\arabic{footnote}}
\setcounter{footnote}{0}

\section{Introduction}
The standard model (SM) of particle physics is a quite successful theory, which can explain experimental data so far.
However, there are still several mysteries and puzzles.
For example, the SM has many free parameters including the neutrino masses.
Most of such free parameters appear in Yukawa couplings of quarks and leptons, i.e., in the flavor sector.
Recent experiments of neutrino oscillations reported relatively large mixing angles in the lepton sector.
They are completely different from the quark mixing angles.
Therefore, it is quite important to study a realistic and natural model that can simultaneously explain such mixing patterns of quarks and leptons.
A certain symmetry could control Yukawa couplings among three generations.
Indeed, quark and lepton masses and mixing angles have been studied from the viewpoint of flavor symmetries, in particular non-Abelian discrete flavor symmetries \cite{Altarelli:2010gt, Ishimori:2010au, King:2013eh}.

Superstring theory is a promising candidate for the unified theory of all the interactions including gravity and all the matter fields and Higgs fields.
Superstring theory is defined in ten-dimensional (10D) spacetime and then predicts extra six dimensions compactified on some compact space in addition to the observed four-dimensional (4D) spacetime.
Furthermore, supersymmetric Yang-Mills (SYM) theory in higher dimensional spacetime appears as effective field theory of superstring theory.
That leads to quite interesting aspects from both theoretical and phenomenological points of view. (See Ref.~\cite{Ibanez:2012zz} for a review of superstring phenomenology.)
It is important to study the structure of such an internal compact space, especially, from the latter viewpoint. 
The detailed structure of the internal space determines important aspects of particle phenomenology in four-dimensional (4D) low-energy effective field theory (LEEFT), e.g., mass spectra including the generation number, coupling selection rules, coupling strength, symmetries in 4D LEEFT, etc.
For example, the toroidal compactification is one of the simplest compactifications, but 10D SYM theory on the 6D torus without any gauge background as well as superstring theory leads to $\mathcal N =4$ supersymmetry in 4D spacetime.
That is non-chiral theory and not realistic.
The orbifold compactification and the torus compactification with magnetic fluxes as well as the orbifold with magnetic fluxes can reduce the number of 4D supersymmetric currents and lead to 4D chiral theory.
Thus, these are quite interesting to study.

Recently, magnetic fluxes in extra dimensions have been receiving many attentions.\footnote{See for a review of phenomenological aspects in orbifold compactification \cite{Choi:2006qh}.}
The $\mathcal N=4$ supersymmetry is broken by magnetic fluxes down to $\mathcal{N}=0$, $1$ or $2$, which depends on the configuration of magnetic fluxes.
It is quite interesting that the simplest toroidal compactifications with magnetic fluxes in extra dimensions lead to 4D chiral spectra, starting from higher-dimensional SYM theories which might be obtained as LEEFT of superstring theories \cite{Bachas:1995ik, Cremades:2004wa}.
In addition, the structure of compact six dimensions determines generations of chiral matters, masses and couplings of the 4D LEEFT after dimensional reductions.
For example, the degeneracy of chiral zero-modes, i.e., the number of generation, is determined by the magnitude of magnetic fluxes, and the overlap integrals of localized zero-mode wavefunctions yield Yukawa couplings for chiral matter fields in the 4D LEEFT.
Indeed, many phenomenologically important properties of the SM, such as the 4D chirality, the number of generations and hierarchical Yukawa couplings \cite{Abe:2009vi,Abe:2012fj} could be originated from the magnetic fluxes.

Furthermore, it is known that magnetized D-brane models as well as intersecting D-brane models can derive certain non-Abelian discrete flavor symmetries such as $D_4$, $\Delta (27)$ and $\Delta (54)$ \cite{Abe:2009vi, Fujimoto:2013xha, Abe:2013bca, Abe:2008sx, BerasaluceGonzalez:2011wy, Marchesano:2013ega, Hamada:2014hpa, Abe:2009uz}.\footnote{See also Ref.~\cite{Higaki:2005ie}.}
Similar flavor symmetries can be obtained from heterotic orbifold models \cite{Kobayashi:2004ya, Kobayashi:2006wq,Ko:2007dz}.
Thus, non-Abelian discrete symmetries which play a role in particle physics can arise from the underlying theory, e.g., superstring theory.
In addition, non-Abelian discrete symmetries are interesting ideas for controlling flavor structures in model-building in the bottom-up approach as mentioned above.
These could provide a bridge between the low-energy phenomenology and the underlying theory, especially superstring theory.
Therefore, it is interesting and important to study the non-Abelian discrete flavor symmetry obtained from the magnetized brane models as the low-energy effective theory of superstring theory.

In our previous paper \cite{Abe:2013bba}, we studied the flavor structures realized by non-factorizable fluxes on toroidal extra dimensions.
That expanded the possibilities for new types of model building, and indeed we have obtained several new types of models with the SM particle content as massless modes.
Then, it turned out that non-factorizable fluxes can lead rich flavor structures in three-generation models of quarks and leptons.
Because of these facts, it is quite attractive to study the flavor symmetry realized in the magnetized models with the extension to non-factorizable fluxes.

This paper is organized as follows.
In Section \ref{sec:model}, we review the magnetized 10D SYM theory and the fields appearing in its action.
In addition, we explain the chiral zero-modes and Yukawa couplings in two cases with factorizable and non-factorizable fluxes, respectively.
Then we develop a way to label the zero-modes with non-factorizable fluxes in detail in Section \ref{sec:type}.
In Section \ref{sec:symmetry}, we show the non-Abelian discrete flavor symmetries realized in the 10D SYM theory with generic configurations of magnetic fluxes in extra dimensions.
In addition, we confirm that these flavor symmetries could be rederived from the perspective of the non-Abelian discrete gauge symmetry, in Section \ref{sec:gaugesymmetry}.
Section \ref{sec:conclusion} is devoted to discussions and conclusions.
In Appendix \ref{appendix}, we refer to the number of generation-types for the arbitrary degeneracy of zero-modes and give some interpretations for them.
In Appendix \ref{sec:egs} and \ref{sec:4gene}, we enumerate and discuss the examples of some configurations of magnetic fluxes in three- and four-generation models, respectively, which are not explained in Section \ref{sec:symmetry}.

\section{Magnetized brane models} \label{sec:model}
We start with 10D SYM theory.
We consider 4D flat Minkowski spacetime and factorizable three tori $T^2 \times T^2 \times T^2$, that is, $R^{3,1} \times (T^2)^3$.
The Lagrangian is given by
\begin{gather}
\mathcal{L} = -\frac{1}{4g^2} \textrm{Tr} \left( F^{MN}F_{MN} \right) + \frac{i}{2g^2}\textrm{Tr} \left( \bar{\lambda}\Gamma^M D_M \lambda \right),
\end{gather}
where $g$ is a 10D YM gauge coupling constant and $M, N=0,1, \ldots ,9$.
The field strength $F_{MN}$ and the covariant derivative $D_M$ are written by
\begin{align}
F_{MN} &= \partial_M A_N - \partial_N A_M - i [A_M, A_N],\\
D_M \lambda &= \partial_M \lambda - i[A_M, \lambda].
\end{align}
In the following, we use $x_i$ and $y_i$ as two real coordinates on the $i$-th $T^2$ for $i=1,2,3$. 
The SYM theory includes a 10D vector field $A_M$ and a 10D Majorana-Weyl spinor field $\lambda$.
The trace in the above Lagrangian acts the indices of YM gauge group.

For convenience we adopt complex coordinates $z_i$ and complex vector fields $A_i$ for $i=1,2,3$, which are defined as
\begin{gather}
z_i = \frac{1}{2}(x_i + \tau_i y_i ), \qquad A_i = \frac{1}{\textrm{Im} \, \tau_i} (A_{3+2i} - \bar{\tau}_i A_{2+2i}).
\end{gather}
The 10D SYM theory possesses $\mathcal{N}=4$ supersymmetry counted in terms of 4D supercharges.
The 10D vector field $A_M$ and Majorana-Weyl spinor field $\lambda$ are decomposed into 4D $\mathcal{N}=1$ single vector and triple chiral multiplets, i.e., $\bm{V}=\{A_\mu, \lambda_0 \}$ and $\bm{\phi}_i = \{A_i,\lambda_i \} \,\, (i=1,2,3)$.
For 4D positive chirality, these spinor fields $\lambda_0$, $\lambda_1$, $\lambda_2$ and $\lambda_3$ have the 6D chiralities, $(+,+,+)$, $(+,-,-)$, $(-,+,-)$, and $(-,-,+)$ on 6D spacetime $R^{3,1} \times T_i^2$ for $i=1,2,3$, respectively.
The 4D $\mathcal{N}=1$ single vector and triple chiral multiplets can be expressed in terms of vector superfield $V$ and chiral superfields $\phi_i \,\, (i=1,2,3)$.

\subsection{Magnetized torus model with factorizable fluxes}
We consider the 10D SYM theory with two types of magnetic fluxes, factorizable flux and non-factorizable flux.
In this subsection, we review the former factorizable case based on Ref.~\cite{Cremades:2004wa}, and assume the following magnetic background :
\begin{gather}
\langle A_i \rangle = \frac{\pi}{\mathrm{Im} \, \tau_i} \left(M^{(i)} \bar{z}_{i} + \bar{\zeta}_{i}\right), \qquad \langle A_\mu \rangle = \langle \lambda_0 \rangle =\langle \lambda_i \rangle =0,\label{vev_fact}
\end{gather}
where $M^{(i)}$ and $\zeta_i$ are $N \times N$ matrices of (Abelian) magnetic fluxes and Wilson-lines, respectively\footnote{For simplicity, we assume the following forms of magnetic fluxes and Wilson-lines, although those are not general forms. In general, we can choose the different forms of magnetic fluxes and Wilson-lines from each torus $T^2_i$. However, we do not require such general forms when we study only non-Abelian flavor symmetries.}, given as
\begin{gather}
M^{(i)} =
\begin{pmatrix}
M^{(i)}_1 \bm{1}_{N_1}&&&\\
&M^{(i)}_2 \bm{1}_{N_2}&&\\
&&\ddots&\\
&&&M^{(i)}_{n} \bm{1}_{N_n}
\end{pmatrix},\label{flux_fact}
\qquad
\zeta_i =
\begin{pmatrix}
\zeta^{(i)}_1 \bm{1}_{N_1}&&&\\
&\zeta^{(i)}_2 \bm{1}_{N_2}&&\\
&&\ddots&\\
&&&\zeta^{(i)}_{n} \bm{1}_{N_n}
\end{pmatrix},
\end{gather}
with a positive integer $N_a$ ($a=1,2,\ldots,n$) satisfying $\sum_{a=1}^{n} N_a=N$, and $\tau_i$ denotes the complex structure parameter that characterizes the shape of the $i$-th $T^2$.
When there are non-vanishing magnetic fluxes and Wilson-lines, the form of VEV (\ref{vev_fact}) leads to factorizable fluxes.
Here, the magnetic fluxes satisfy $M^{(i)}_1, M^{(i)}_2, \ldots, M^{(i)}_n \in \mathbb{Z}$ due to Dirac's quantization condition.
In the case that the magnetic fluxes $M^{(i)}_1, M^{(i)}_2, \ldots, M^{(i)}_n$ take different values from each other, $U(N)$ gauge group breaks into $U(N_1) \times U(N_2) \times \ldots \times U(N_n)$.
The same holds for Wilson-lines.
We use indices $a, b, \ldots$ for labeling the unbroken subgroups $U(N_a), U(N_b), \ldots$ of $U(N)$, respectively. 
The block off-diagonal part  $(\phi_i)_{ab}$ of chiral superfield $\phi_i$ is the bi-fundamental representation under $U(N_a) \times U(N_b)$ and the block diagonal part $(\phi_i)_{aa}$ is the adjoint representation under $U(N_a)$.

Next, we refer to the zero-mode equations for chiral matter superfields $\phi_j$.
The zero-mode equations are found as \cite{Abe:2012ya}
\begin{align}
\bar{\partial}_i f_j^{(i)} + \frac{1}{\sqrt{2}} \left[\vev{\bar{A}_i}, f_j^{(i)}\right] &=0 \quad (i=j),\label{eq_zero2}\\
\partial_i f_j^{(i)} - \frac{1}{\sqrt{2}} \left[\vev{A_i}, f_j^{(i)}\right] &=0 \quad (i\neq j),\label{eq_zero3}
\end{align}
where $f^{(i)}_j$ denotes the zero-mode wavefunction of the chiral superfield $\phi_j$ on the $i$-th $T^2$ and then $\partial_i \equiv \partial/\partial z_i$.
Note that a difference of the signs in Eqs.~(\ref{eq_zero2}) and (\ref{eq_zero3}) comes from the chirality structure.
The zero-mode wavefunctions $f_j^{(i)}$, in general, satisfy different equations on each $T_i^2$ for $i=1, 2, 3$.
Due to the existence of non-vanishing magnetic fluxes, chirality projection occurs.

For the zero-mode wavefunction $f_j^{(i=j)}$ in the $ab$-sector, if $M^{(i)}_{ab} \equiv M^{(i)}_a-M^{(i)}_b >0$, then there exist $|M^{(i)}_{ab}|$ solutions of zero-mode equation (\ref{eq_zero2}),
\begin{align}
(f_j^{(i)})_{ab} &= g \, \Theta_j^{I^{(i)}_{ab}, \, M^{(i)}_{ab}}(z'_i) \qquad (I^{(i)}_{ab} = 1,2, \ldots, |M^{(i)}_{ab}| ),\label{zeromode_factpositive}\\
\Theta_j^{I^{(i)}_{ab}, \, M^{(i)}_{ab}}(z_i) &= \mathcal{N} e^{i\pi M^{(i)}_{ab} z_i \textrm{Im} \, z_i/\textrm{Im}\, \tau_i} \cdot \vartheta \begin{bmatrix}I^{(i)}_{ab}/ M^{(i)}_{ab}\\[5pt] 0\end{bmatrix}(M^{(i)}_{ab}z_i, M^{(i)}_{ab}\tau_i),
\end{align}
where $z'_i \equiv z_i + \zeta^{(i)}_{ab}/ M^{(i)}_{ab}$, $\zeta^{(i)}_{ab} \equiv \zeta^{(i)}_a-\zeta^{(i)}_b$, and $\vartheta$ denotes Jacobi $\vartheta$-function,
\begin{gather}
\vartheta \begin{bmatrix}a\\[5pt] b\end{bmatrix}(\nu, \tau) = \sum_{l \in \mathbb{Z}} e^{\pi i(a+l)^2\tau} e^{2\pi i (a+l)(\nu +b)}.
\end{gather}
On the other hand, there is no  normalizable zero-mode wavefunction if $M^{(i)}_{ab} <0$ and, finally, the zero-mode wavefunction is constant if $M^{(i)}_{ab}=0$.

For the zero-mode wavefunction $f_j^{(i \ne j)}$ in the $ab$-sector, if $M^{(i)}_{ab} <0$, the zero-mode wavefunctions can be written as the complex conjugate of the wavefunction (\ref{zeromode_factpositive}).
There is no  normalizable zero-mode wavefunction if $M^{(i)}_{ab} >0$ and the zero-mode wavefunction is constant if $M^{(i)}_{ab}=0$.
Notice that the degeneracy of the zero-modes in the chiral superfield $\phi_j$ on the $i$-th torus is determined by the number of the magnetic fluxes $M^{(i)}_{ab}$ that the $\phi_j$ feels on the $i$-th torus.
With three toroidal compactifications, the total degeneracy  $N_{ab}$ of the chiral zero-modes in $(\phi_j)_{ab}$ can be written by $N_{ab} = \prod_{i=1}^3 \bigl| M^{(i)}_{ab} \bigr|$.

The Yukawa couplings between chiral zero-modes in the 4D effective theory are given by the overlap integrals,
\begin{gather}
\lambda_{\mathcal{I} \mathcal{J} \mathcal{K}} = \prod_{i=1}^{3} \int d^2z_i \sqrt{\det g^{(i)}} \, (f^{(i)}_1)_{ab} (f^{(i)}_3)_{bc} (f^{(i)}_2)_{ca} \label{yukawa_fact},
\end{gather}
where $g^{(i)}$ denotes the metric for the $i$-th torus  $T^2_i$ and $\mathcal{I} \equiv (I^{(1)}_{ab}, I^{(2)}_{ab}, I^{(3)}_{ab})$ labels the total generation of zero-modes in $ab$-sector.
The same holds for the other sectors.
We can calculate the overlap integral (\ref{yukawa_fact}) under $M^{(i)}_{ab} + M^{(i)}_{bc} + M^{(i)}_{ca} = 0$, which are evaluated as
\begin{align}
\lambda_{\mathcal{I} \mathcal{J} \mathcal{K}} &= \prod_{i=1}^{3} \lambda_{I^{(i)}_{ab} I^{(i)}_{ca} I^{(i)}_{bc}},\\
\lambda_{I^{(i)}_{ab} I^{(i)}_{ca} I^{(i)}_{bc}} &\propto \sum_{m=1}^{M^{(i)}_{ca}} \delta_{I^{(i)}_{ab} + I^{(i)}_{ca} +M^{(i)}_{ab}m, \, I^{(i)}_{bc}} \notag\\
&\hspace{30pt} \times \vartheta 
\begin{bmatrix}
\frac{M^{(i)}_{ca} I^{(i)}_{ab} - M^{(i)}_{ab}I^{(i)}_{ca} + M^{(i)}_{ab} M^{(i)}_{ca}m }{-M^{(i)}_{ab} M^{(i)}_{ca} M^{(i)}_{bc}} \\[10pt]
0
\end{bmatrix}
(M^{(i)}_{bc} \bar{\zeta}^{(i)}_{ca} - M^{(i)}_{ab} \bar{\zeta}^{(i)}_{bc}, -\bar{\tau}_i M^{(i)}_{ab} M^{(i)}_{ca} M^{(i)}_{bc}), \label{yukawa'_fact}
\end{align}
where we omit an overall factor, because the factor has no effect on the flavor symmetry in magnetized torus models.

\subsection{Magnetized torus model with non-factorizable fluxes}
Next, we review the generalization of the above results including non-factorizable fluxes, based on Refs. \cite{Cremades:2004wa, Antoniadis:2009bg}. 
We assume the following magnetic background,
\begin{gather}
\langle A_i \rangle = \frac{\pi}{\mathrm{Im} \, \tau_i} \left(M^{(i)} \bar{z}_{i} + M^{(ij)} \bar{z}_j + \bar{\zeta}_i\right),\label{vev_nonfact}\\
\langle A_\mu \rangle = \langle \lambda_0 \rangle =\langle \lambda_i \rangle =0,
\end{gather}
with $i \neq j$, where $M^{(ij)}$ is a $N \times N$ matrix of an additional (Abelian) magnetic fluxes,
\begin{gather}
M^{(ij)} =
\begin{pmatrix}
M^{(ij)}_1 \bm{1}_{N_1}&&&\\
&M^{(ij)}_2 \bm{1}_{N_2}&&\\
&&\ddots&\\
&&&M^{(ij)}_{n} \bm{1}_{N_{n'}}
\end{pmatrix},
\end{gather}
with a positive integer $N_{a'}$ $(a'=1,2, \ldots, n')$ satisfying $\sum_{a'}^{n'} N_{a'}=N$.\footnote{As mentioned in Eq.~(\ref{flux_fact}), these magnetic fluxes and Wilson-lines are not general forms.} 
It holds that $M^{(ij)}_1, M^{(ij)}_2, \ldots, M^{(ij)}_n$ $\in \mathbb{Z}$ due to Dirac's quantization condition.
The magnetic background (\ref{vev_nonfact}) is a straightforward extension of Eq.~(\ref{vev_fact}) and leads to non-factorizable magnetic fluxes.

We substitute the VEVs (\ref{vev_nonfact}) into zero-mode equations (\ref{eq_zero2}) and (\ref{eq_zero3})  and find that the zero-mode wavefunctions and the degeneracy of zero-modes are changed from the factorizable case.
Again, we focus on chiral superfields $\phi_i \,\, (i=1,2,3)$ and then explain their zero-mode wavefunctions in the following.
In this paper, we consider the case that only magnetic fluxes $M^{(12)}$ and $M^{(21)}$ in the first and the second tori $T^2_1 \times T^2_2$ are turned on.
The extensions to the other non-vanishing magnetic fluxes $M^{(ij)}$ are straightforward.
Now, we define the matrix
\begin{align}
\mathbb{N}_{ab} &\equiv
\begin{pmatrix}
M^{(1)}_{ab} & M^{(21)}_{ab}\\
M^{(12)}_{ab} & M^{(2)}_{ab}
\end{pmatrix},\\
M^{(i)}_{ab} &\equiv M^{(i)}_{a} - M^{(i)}_{b},\\
M^{(ij)}_{ab} &\equiv \frac{\textrm{Im} \, \tau_i}{\textrm{Im} \, \tau_j} (M^{(ij)}_a -M^{(ij)}_b) + (M^{(ji)}_a -M^{(ji)}_b),
\end{align}
which determines the degeneracy of zero-modes.
Note that diagonal elements of the matrix $\mathbb{N}_{ab}$ correspond to the magnetic fluxes defined in Eq.~(\ref{flux_fact}).

Next, in order to obtain the normalizable wavefunctions with the matrix $\mathbb{N}$ and complex structure parameters $\tau_i \,\, (i=1,2)$, we must impose the Riemann conditions
\begin{gather}
\mathbb{N}^{ij}_{ab} \in \mathbb{Z}, \qquad (\mathbb{N}_{ab} \cdot \textrm{Im}\,\Omega)^T = \mathbb{N}_{ab} \cdot \textrm{Im}\,\Omega, \qquad \mathbb{N}_{ab} \cdot \textrm{Im}\,\Omega>0, \qquad \forall a,b, \label{Riemanncondition}
\end{gather}
where $\Omega \equiv \textrm{diag} (\tau_1, \tau_2)$ is a $2 \times 2$ matrix constructed from complex structure parameters.
For a while, we consider the case with vanishing Wilson-lines, i.e., $\bar{\zeta}_1=\bar{\zeta}_2=0$.
Only if the matrix $\mathbb{N}_{ab}$ and the complex structure $\Omega$ satisfy the Riemann conditions (\ref{Riemanncondition}), there exist the normalizable zero-mode wavefunctions in the $ab$-sector on the first and second tori, which are expressed as
\begin{align}
(f^{(12)}_j)_{ab} &= g \, \Theta^{\vec{i}_{ab}, \, \mathbb{N}_{ab}}_j (\vec{z}),\label{zeromode_nonfact}\\
\Theta^{\vec{i}_{ab}, \, \mathbb{N}_{ab}}_j (\vec{z}) &= \mathcal{N} e^{\pi i (\mathbb{N}_{ab} \cdot \vec{z}) \cdot (\textrm{Im} \, \Omega)^{-1} \cdot (\textrm{Im} \, \vec{z})} \cdot \vartheta
\begin{bmatrix}
\vec{i}_{ab}\\[5pt] 0
\end{bmatrix}
(\mathbb{N}_{ab} \cdot \vec{z}, \mathbb{N}_{ab} \cdot \Omega),
\end{align}
where $\vec{z} \equiv (z_1, z_2)$ and $\vartheta$ denotes the Riemann $\vartheta$-function,
\begin{gather}
\vartheta
\begin{bmatrix}
\vec{a}\\[5pt] \vec{b}
\end{bmatrix}
(\vec{\nu}, \Omega) =
\sum_{\vec{l} \in \mathbb{Z}^2} e^{\pi i (\vec{l}+\vec{a}) \cdot \Omega \cdot (\vec{l} + \vec{a})} e^{2\pi i (\vec{l}+\vec{a}) \cdot (\vec{\nu} + \vec{b})}.
\end{gather}
The vector $\vec{i}_{ab}$ labels degenerated zero-modes (generations), and we will explain its meaning in detail in the next section.

Note that the expression of the wavefunction (\ref{zeromode_nonfact}) is for (totally) positive chirality matters, which namely have the chirality $(+, +)$ and $(-, -)$ on the first and second tori.
In 10D SYM theory with the superfield description \cite{Abe:2012ya} we adopt in this paper, the wavefunction (\ref{zeromode_nonfact}) is valid for a chiral superfield $\phi_3$ that has the chirality $(-, -)$ on the first and second tori.
For chiral superfields $\phi_1$ and $\phi_2$, they need to be mixed up to be the solution of the zero-mode equations.
As stated in Ref.~\cite{Antoniadis:2009bg}, we consider the following parameterizations,
\begin{gather}
(\phi_1)_{ab} = \alpha_{ab} \Phi_{ab}, \qquad (\phi_2)_{ab} = \beta_{ab} \Phi_{ab}.
\end{gather}
The Riemann conditions (\ref{Riemanncondition}) to obtain normalizable zero-mode wavefunctions and an explicit form of the zero-mode wavefunction (\ref{zeromode_nonfact}) can be also applied for $\Phi_{ab}$ by replacing the complex structure $\Omega$ with the effective complex structure $\tilde{\Omega} \equiv \hat{\Omega}_{ab} \cdot \Omega$, where
\begin{gather}
\hat{\Omega}_{ab} \equiv \frac{1}{1+q_{ab}^2}
\begin{pmatrix}
1-q_{ab}^2 & -2q_{ab}\\
-2q_{ab} & q_{ab}^2 -1
\end{pmatrix}
.
\end{gather}
Mixing parameters $q_{ab} \equiv \beta_{ab}/\alpha_{ab}$ are given for individual bi-fundamental representations labeled by $a$ and $b$ ($a \ne b$), and their values are determined by the second condition of the Riemann conditions.

On the third torus, the zero-mode wavefunction is the same as the expression (\ref{zeromode_factpositive}) or the complex conjugate to that.
Thus, the degeneracy of zero-modes $N_{ab}$ with non-factorizable fluxes is determined by the matrix $\mathbb{N}_{ab}$ and the flux $M^{(3)}_{ab}$, i.e., $N_{ab} = | \det \mathbb{N}_{ab} \times M^{(3)}_{ab} |$ for $M^{(3)}_{ab} \ne 0$ in the present situation.

Next, the Yukawa couplings in the 4D effective theory is also evaluated by the overlap integral
\begin{gather}
\lambda_{\mathcal{I} \mathcal{J} \mathcal{K}} = \lambda_{I^{(3)}_{ab}I^{(3)}_{ca}I^{(3)}_{bc}} \int d^2z_1d^2z_2 \sqrt{\det g^{(1)}g^{(2)}} \, (f^{(12)}_1)_{ab} (f^{(12)}_3)_{bc} (f^{(12)}_2)_{ca}, \label{yukawa_nonfact}
\end{gather}
where $\mathcal{I} \equiv (\vec{i}_{ab}, \, I^{(3)}_{ab})$ labels the total generation of zero-modes in $ab$-sector.
The same holds for the other sectors.
We consider the case that there are zero-modes with the total negative chirality on the first and second tori.
Then, we can calculate the overlap integrals (\ref{yukawa_nonfact}) under $\mathbb{N}_{ab} + \mathbb{N}_{bc} + \mathbb{N}_{ca}=0$ and $M^{(3)}_{ab} + M^{(3)}_{bc} + M^{(3)}_{ca}=0$ \cite{Cremades:2004wa, Antoniadis:2009bg}, which are evaluated as
\begin{align}
\lambda_{\mathcal{I} \mathcal{J} \mathcal{K}} &= \lambda_{\vec{i}_{ab} \vec{i}_{ca} \vec{i}_{bc}} \cdot \lambda_{I^{(3)}_{ab}I^{(3)}_{ca}I^{(3)}_{bc}},\\
\lambda_{\vec{i}_{ab} \vec{i}_{ca} \vec{i}_{bc}} &\propto \sum_{\vec{m}} \delta_{\vec{i}_{bc}, \, \mathbb{N}_{cb}^{-1} (\mathbb{N}_{ab} \vec{i}_{ab} + \mathbb{N}_{ca}\vec{i}_{ca}+ \mathbb{N}_{ab} \vec{m})} \notag\\
&\hspace{30pt} \times \int dy_1dy_2 \left[ e^{- \pi \vec{y} \cdot (\mathbb{N}_{ab} \tilde{\Omega}_{ab} + \mathbb{N}_{ca} \tilde{\Omega}_{ca} + \mathbb{N}_{bc} \Omega) \cdot \vec{y}}
\cdot \vartheta \begin{bmatrix}\vec{\bold{K}}\\[5pt] 0 \end{bmatrix}(i\vec{\bold{Y}}|i\vec{\bold{Q}}) \right], \label{yukawa'_nonfact}
\end{align}
where $\vec{y} \equiv (y_1, y_2)$ and $\vec{m}$ denote the integer points in the region spanned by
\begin{gather}
\vec{e'}_i \equiv \vec{e}_i \, (\det \mathbb{N}_{ab} \det \mathbb{N}_{ca}) \, \mathbb{N}_{ca}^{-1} (\mathbb{N}_{ab}+\mathbb{N}_{ca}) \mathbb{N}_{ab}^{-1},\\
\vec{e}_1 =
\begin{pmatrix}
1\\
0
\end{pmatrix}
, \quad \vec{e}_2 =
\begin{pmatrix}
0\\
1
\end{pmatrix}
, \label{units}
\end{gather}
and 
\begin{gather}
\vec{\bold{K}} \equiv \left(\begin{array}{c} \vec{i}_{bc}\\ (\vec{i}_{ab} -\vec{i}_{ca}+\vec{m}) \frac{\mathbb{N}_{ab}(\mathbb{N}_{ab}+\mathbb{N}_{ca})^{-1}\mathbb{N}_{ca}}{\det\mathbb{N}_{ab} \det\mathbb{N}_{ca}} \end{array} \right),\\
\vec{\bold{Y}} \equiv \left(\begin{array}{c} (\mathbb{N}_{ab} \tilde{\Omega}_{ab} +\mathbb{N}_{ca} \tilde{\Omega}_{ca} +\mathbb{N}_{bc} \Omega) \cdot \vec{y}\\ (\det\mathbb{N}_{ab} \det \mathbb{N}_{ca}) (\mathbb{N}_{ab} \tilde{\Omega}_{ab} (\mathbb{N}_{ab}^{-1})^T - \mathbb{N}_{ca} \tilde{\Omega}_{ca} (\mathbb{N}_{ca}^{-1})^T)\cdot \vec{y} \end{array} \right),\\
\vec{\bold{Q}} \equiv \left(\begin{array}{cc} \mathbb{N}_{ab} \tilde{\Omega}_{ab}+\mathbb{N}_{ca} \tilde{\Omega}_{ca} +\mathbb{N}_{bc}\Omega & (\det\mathbb{N}_{ab}\det\mathbb{N}_{ca}) (\mathbb{N}_{ab} \tilde{\Omega}_{ab} (\mathbb{N}_{ab}^{-1})^T - \mathbb{N}_{ca} \tilde{\Omega}_{ca} (\mathbb{N}_{ca}^{-1})^T)\\
(\det\mathbb{N}_{ab}\det\mathbb{N}_{ca})(\tilde{\Omega}_{ab} - \tilde{\Omega}_{ca})&(\det\mathbb{N}_{ab}\det\mathbb{N}_{ca})^2 (\tilde{\Omega}_{ab} \mathbb{N}_{ab}^{-1} +\tilde{\Omega}_{ca} \mathbb{N}_{ca}^{-1})  \end{array} \right).
\end{gather}
In Eq.~(\ref{yukawa'_nonfact}), again, we omit an overall factor, because of the same reason as the model with factorizable fluxes in the previous subsection.
Note that  the integrals over $z_1$ and $z_2$ are non-factorizable, while the one over $z_3$ is factorized in the Yukawa couplings (\ref{yukawa_nonfact}), as a consequence of the flux configuration assumed above. 
The overlap integral on the third torus yields the factor $\lambda_{I^{(3)}_{ab}I^{(3)}_{ca}I^{(3)}_{bc}}$ that is exactly the same as Eq.~(\ref{yukawa'_fact}) for $i=3$.
The property of non-factorizable fluxes appears in the overlap integral on the first and second tori.
Therefore it is interesting to investigate the factor $\lambda_{\vec{i}_{ab} \vec{i}_{ca} \vec{i}_{bc}}$ in Eq.~(\ref{yukawa'_nonfact}).

We have limited the above discussion to the case with vanishing Wilson-lines.
In this paragraph, we show the zero-mode wavefunction and the Yukawa coupling with non-vanishing Wilson-lines, i.e., $\bar{\zeta}_1, \bar{\zeta}_2 \neq 0$.
Indeed, by means of shifting the coordinates, such a zero-mode wavefunction can be obtained as
\begin{gather}
(f^{(12)}_j)_{ab} = g \, \Theta^{\vec{i}_{ab}, \, \mathbb{N}_{ab}}_j (\vec{z'}),
\end{gather}
where $\vec{z'} \equiv \vec{z} + \mathbb{N}_{ab}^{-1} \cdot \vec{\zeta}_{ab}$ and $\vec{\zeta}_{ab} \equiv (\zeta^{(1)}_{ab}, \zeta^{(2)}_{ab})$.
By calculating the overlap integral of the above zero-mode wavefunctions on the first and second tori, the relevant part of the Yukawa couplings in the 4D effective theory can be obtained as
\begin{align}
\lambda_{\vec{i}_{ab} \vec{i}_{ca} \vec{i}_{bc}} &\propto \sum_{\vec{m}} \delta_{\vec{i}_{bc}, \, \mathbb{N}_{bc}^{-1} (\mathbb{N}_{ab} \vec{i}_{ab} + \mathbb{N}_{ca}\vec{i}_{ca}+ \mathbb{N}_{ab} \vec{m})} \notag\\
&\hspace{30pt} \times \int dy_1dy_2 \left[ e^{- \pi (\vec{y'}_{ab} \cdot \mathbb{N}_{ab} \tilde{\Omega}_{ab} \cdot \vec{y'}_{ab} + \vec{y'}_{ca} \cdot \mathbb{N}_{ca} \tilde{\Omega}_{ca} \cdot \vec{y'}_{ca} + \vec{y'}_{bc} \cdot \mathbb{N}_{bc} \Omega \cdot \vec{y'}_{bc})}
\cdot \vartheta \begin{bmatrix}\vec{\bold{K}}\\[5pt] 0 \end{bmatrix}(i\vec{\bold{Y}}|i\vec{\bold{Q}}) \right],\label{yukawa_nonfactwl}
\end{align}
up to an overall factor.
Moreover, we should replace $\vec{\bold{Y}}$ in Eq.~(\ref{yukawa_nonfactwl}) with
\begin{gather}
\vec{\bold{Y}} \equiv \left(\begin{array}{c} \mathbb{N}_{ab} \tilde{\Omega}_{ab}\cdot \vec{y'}_{ab} +\mathbb{N}_{ca} \tilde{\Omega}_{ca}\cdot \vec{y'}_{ca} +\mathbb{N}_{bc} \Omega\cdot \vec{y'}_{bc}\\ (\det\mathbb{N}_{ab} \det \mathbb{N}_{ca}) (\mathbb{N}_{ab} \tilde{\Omega}_{ab} (\mathbb{N}_{ab}^{-1})^T\cdot \vec{y'}_{ab} - \mathbb{N}_{ca} \tilde{\Omega}_{ca} (\mathbb{N}_{ca}^{-1})^T\cdot \vec{y'}_{ca})\end{array} \right),
\end{gather}
where we define $\vec{y'}_{ab} \equiv \vec{y}_{ab} + (\mathbb{N}_{ab} \textrm{Im} \, \tilde{\Omega}_{ab})^{-1} \cdot \textrm{Im} \, \vec{\zeta}_{ab}$ for the zero-mode wavefunction $(f^{(12)}_1)_{ab}$.
The same holds for the zero-mode wavefunction $(f^{(12)}_2)_{ca}$.
For the $ca$-sector in chiral superfield $\phi_3$, we replace $\vec{y}_{bc}$ with $\vec{y'}_{bc} \equiv \vec{y}_{bc} + (\mathbb{N}_{bc} \textrm{Im} \, \Omega)^{-1} \cdot \textrm{Im} \, \vec{\zeta}_{bc}$.

\subsection{Magnetized orbifold model with non-factorizable fluxes}
Finaly in this section we review the orbifold projection with non-factorizable fluxes  based on Ref.~\cite{Abe:2013bba}.
In our previous paper \cite{Abe:2013bba}, we extend the model proposed in Ref.~\cite{Abe:2008fi} (see also Ref.~\cite{Abe:2008sx}) where the orbifold models with factorizable fluxes are constructed.
The number of the (degenerate) zero-modes is changed by the orbifold projection.
We consider the $T^6/Z_2$ orbifold where the $Z_2$ projection acts on the first and second tori.
It is constructed by dividing $T^6$ by the $Z_2$ projection $z_1 \rightarrow -z_1$ and $z_2 \rightarrow -z_2$, simultaneously.
Such an identification prohibits (continuous) non-vanishing Wilson-lines.
Here, we consider vanishing Wilson-lines.
On such an orbifold, we impose the following boundary conditions for 10D superfields $V$ and $\phi_i$,
\begin{align}
V(x_\mu, -z_1, -z_2, z_3) &= + P V(x_\mu, z_1, z_2, z_3)P^{-1},\\
\phi_1(x_\mu, -z_1, -z_2, z_3) &= - P \phi_1(x_\mu, z_1, z_2, z_3)P^{-1},\\
\phi_2(x_\mu, -z_1, -z_2, z_3) &= - P \phi_2(x_\mu, z_1, z_2, z_3)P^{-1},\\
\phi_3(x_\mu, -z_1, -z_2, z_3) &= + P \phi_3(x_\mu, z_1, z_2, z_3)P^{-1},
\end{align}
where a projection operator $P$ acts on the YM indices and satisfies $P^2 = \bm{1}_N$.
Then, either even- or odd-modes among the zero-modes can survive depending on $P$.
Instead of Eq.~(\ref{zeromode_nonfact}), we find the zero-mode wavefunctions in the following form,
\begin{align}
\Theta^{\vec{i}_{ab}}_\textrm{even} (\vec{z}) &= \Theta^{\vec{i}_{ab}, \, \mathbb{N}_{ab}}(\vec{z}) + \Theta^{\vec{e} - \vec{i}_{ab}, \, \mathbb{N}_{ab}}(\vec{z}),\label{evenmode_orbifold}\\
\Theta^{\vec{i}_{ab}}_\textrm{odd} (\vec{z}) &= \Theta^{\vec{i}_{ab}, \, \mathbb{N}_{ab}}(\vec{z}) - \Theta^{\vec{e} - \vec{i}_{ab}, \, \mathbb{N}_{ab}}(\vec{z}),\label{oddmode_orbifold}
\end{align}
up to a normalization factor, where we define $\vec{e} \equiv \vec{e}_1 + \vec{e}_2$ in terms of Eq.~(\ref{units}) and utilized the following formula,
\begin{gather}
\Theta^{\vec{i}_{ab}, \, \mathbb{N}_{ab}}(-\vec{z}) = \Theta^{\vec{e}-\vec{i}_{ab}, \, \mathbb{N}_{ab}}(\vec{z}).
\end{gather}
We will also explain the label of generation $\vec{i}_{ab}$ in the next section.
After the orbifold projection, the degeneracy of these zero-modes on the first and second tori is changed as shown in Table \ref{tab:degeneracy_nonfact}.
\begin{table}[H]
\centering
\begin{tabular}{cccccccccccc}\hline
$|\det\mathbb{N}_{ab}|$ & $0$ & $1$ & $2$ & $3$ & $4$ & $5$ & $6$ & $7$ & $8$ & $9$ & $10$ \\ \hline
even & $1$ & $1$ & $2$ & $2$ & $3$ & $3$ & $4$ & $4$ & $5$ & $5$ & $6$ \\
odd & $0$ & $0$ & $0$ & $1$ & $1$ & $2$ & $2$ & $3$ & $3$ & $4$ & $4$ \\ \hline
\end{tabular}
\caption{The degeneracy of zero-modes for even- and odd-modes.}
\label{tab:degeneracy_nonfact}
\end{table}
Note that Table \ref{tab:degeneracy_nonfact} is the same as the corresponding one given in Ref.~\cite{Abe:2008fi} by replacing $M$ with $\det\mathbb{N}$.
Because of this replacement, we can obtain more various flavor structures.
We remark that there are exceptions in the above Table \ref{tab:degeneracy_nonfact} that will be illustrated in the subsection \ref{sec:orb_nonfact} in detail after explaining the label $\vec{i}_{ab}$ in the next section.

\section{Degenerated structures of zero-modes} \label{sec:type}
In this section, we propose a way to investigate the properties of the degenerated zero-modes on the magnetized torus with non-factorizable fluxes and classify the degeneracy based on it.
\subsection{Generation-types} \label{subsec:generation-types}
The degeneracy of zero-modes generated by non-factorizable fluxes are labeled by $\vec{i}_{ab}$ appearing in Eq.~(\ref{zeromode_nonfact}).
Unlike the magnetized torus model with factorizable fluxes, the zero-mode label $\vec i_{ab}$ is more complicated.
In the magnetized model with non-factorizable flux, we can no longer naively count the degeneracy of zero-modes in terms of the label $I_{ab}$ shown in Eq.~(\ref{zeromode_factpositive}).
Since the degeneracy of zero-modes can be identified with the generation, it is quite important to be familiar with a suitable way for labeling them, when we discuss the flavor symmetry obtained from the models with non-factorizable fluxes.

For simplicity, hereafter in this subsection we omit the YM indices $a, b$ those becomes implicit.
First we consider the case where three zero-modes are induced by non-factorizable fluxes : $|\det\mathbb{N}| = 3$.
We can also extend the following analysis to the case that $\det\mathbb{N}$ equals to an arbitrary prime number.
We can generally parametrize the matrix $\mathbb{N}$ as
\begin{gather}
\mathbb{N} =
\begin{pmatrix}
3n_{11} + n_{11}' & 3n_{21} + n_{21}'\\
3n_{12} + n_{12}' & 3n_{22} + n_{22}'
\end{pmatrix}
,
\end{gather}
where $n_{11}, n_{12}, n_{21}, n_{22}$ are integers and each of $n_{11}', n_{12}', n_{21}', n_{22}'$ is either 0, 1, or 2.
For $\det\mathbb{N}=\pm 3$, we obtain the relation
\begin{gather}
n_{11}'n_{22}'-n_{12}'n_{21}'=0 \qquad (\textrm{mod} \,\, 3). \label{classrelation}
\end{gather}
We can easily find a trivial pattern $n_{11}'=n_{21}'=0$ or $n_{12}'=n_{22}'=0$ satisfying Eq.~(\ref{classrelation}).
In addition, we find four patterns of the non-trivial solution as shown in Table \ref{tab:generationtypes}.
\begin{table}[H]
\centering
\begin{tabular}{ccc}\hline
& $(n_{11}', n_{21}'), \, (n_{12}', n_{22}')$ \\ \hline
Type 1 & $(0,1)$ \, or \, $(0,2)$\\
Type 2 & $(1,2)$ \, or \, $(2,1)$\\
Type 3 & $(1,1)$ \, or \, $(2,2)$\\
Type 4 & $(2,0)$ \, or \, $(1,0)$\\ \hline
\end{tabular}
\caption{The integer sets satisfying Eq. (\ref{classrelation}).}
\label{tab:generationtypes}
\end{table}

From the condition 
\begin{gather}
\mathbb{N} \cdot \vec{i} \in \mathbb{Z},
\end{gather}
given in Ref.~\cite{Cremades:2004wa} in order to obtain the normalizable zero-mode wavefunctions, we find four types of the three-generation label $\vec{i} \equiv (i_1, i_2)$, which are given as

\vspace{10pt}
Type 1 :
\begin{gather}
\vec{i} = \begin{pmatrix}0\\ 0\end{pmatrix}, \begin{pmatrix}1/3\\ 0\end{pmatrix}, \begin{pmatrix}2/3\\ 0\end{pmatrix}, \hspace{70pt} \begin{minipage}{0.2\textwidth}
\setlength{\unitlength}{0.6mm}
\begin{picture}(35,37)(0,0)
\put(0,0){\vector(1,0){32}}
\put(0,0){\vector(0,1){32}}
\put(34,-1){$i_1$}
\put(-1,34){$i_2$}
\put(0,0){\circle*{3}}
\put(10,0){\circle*{3}}
\put(20,0){\circle*{3}} 
\end{picture}
\end{minipage}\label{type1}
\end{gather}

Type 2 :
\begin{gather}
\vec{i} = \begin{pmatrix}0\\ 0\end{pmatrix}, \begin{pmatrix}1/3\\ 1/3\end{pmatrix}, \begin{pmatrix}2/3\\ 2/3\end{pmatrix}, \hspace{70pt} \begin{minipage}{0.2\textwidth}
\setlength{\unitlength}{0.6mm}
\begin{picture}(35,37)(0,0)
\put(0,0){\vector(1,0){32}}
\put(0,0){\vector(0,1){32}}
\put(34,-1){$i_1$}
\put(-1,34){$i_2$}
\put(0,0){\circle*{3}}
\put(10,10){\circle*{3}}
\put(20,20){\circle*{3}} 
\end{picture}
\end{minipage}\label{type2}
\end{gather}

Type 3 :
\begin{gather}
\vec{i} = \begin{pmatrix}0\\ 0\end{pmatrix}, \begin{pmatrix}1/3\\ 2/3\end{pmatrix}, \begin{pmatrix}2/3\\ 1/3\end{pmatrix}, \hspace{70pt} \begin{minipage}{0.2\textwidth}
\setlength{\unitlength}{0.6mm}
\begin{picture}(35,37)(0,0)
\put(0,0){\vector(1,0){32}}
\put(0,0){\vector(0,1){32}}
\put(34,-1){$i_1$}
\put(-1,34){$i_2$}
\put(0,0){\circle*{3}}
\put(10,20){\circle*{3}}
\put(20,10){\circle*{3}} 
\end{picture}
\end{minipage}
\end{gather}

\begin{figure}[H]
\begin{minipage}{\textwidth}
\centering\vspace{5pt}
\begin{minipage}{0.31\textwidth}
\centering \includegraphics[width=0.9\textwidth]{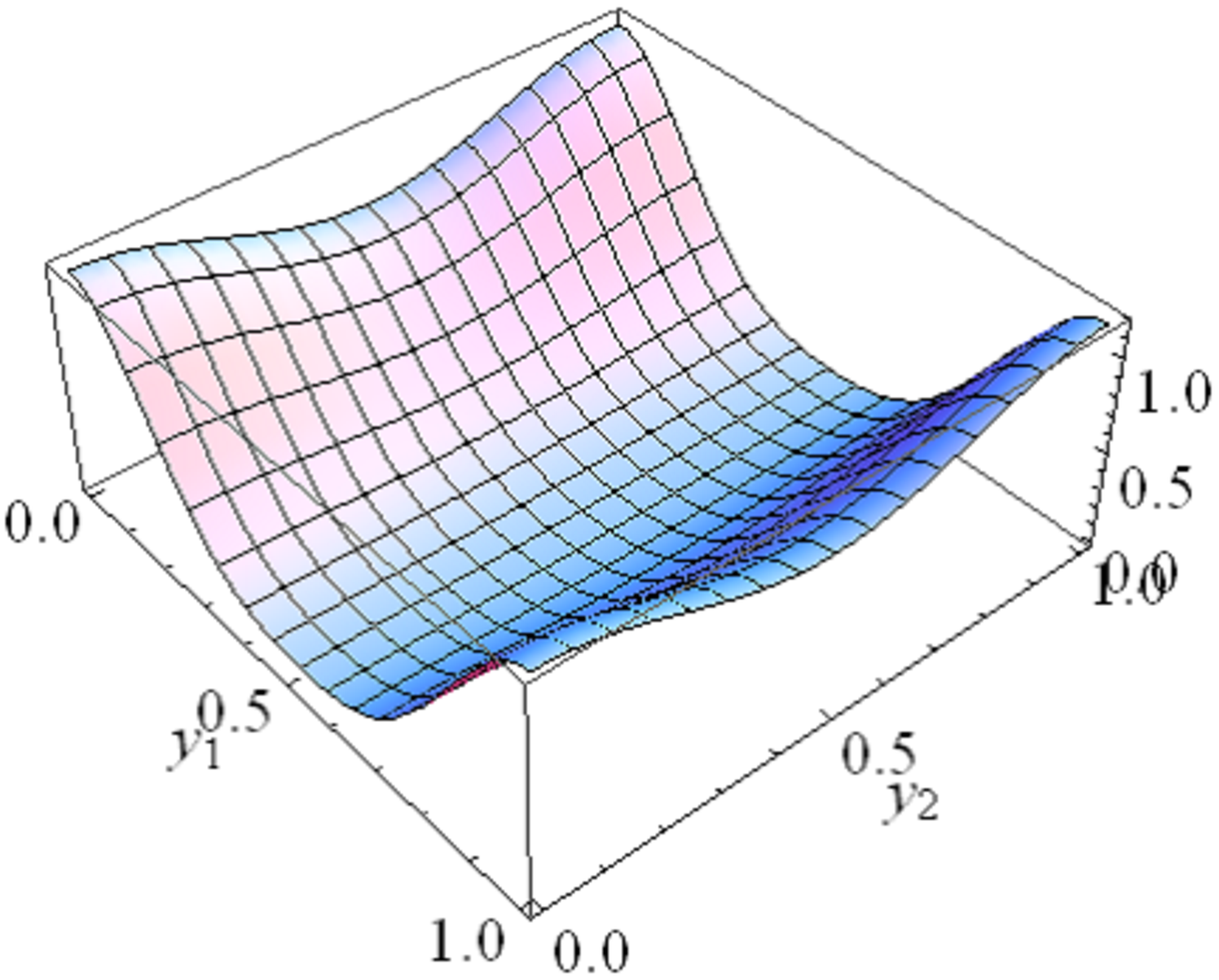}
\centering \bf \small (a) \,\, $\vec{i}=(0,0)$
\end{minipage}
\begin{minipage}{0.31\textwidth}
\centering \includegraphics[width=0.9\textwidth]{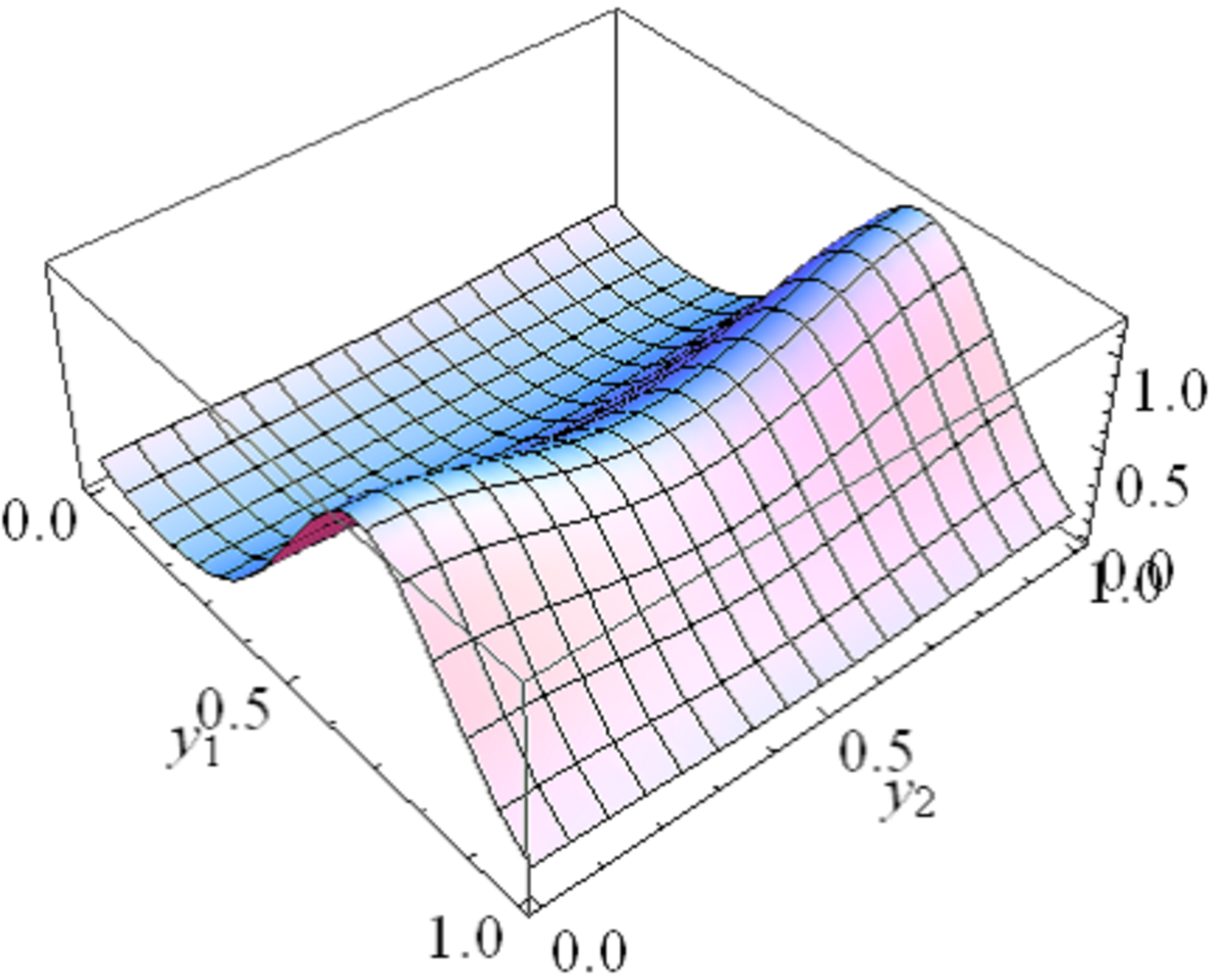}
\centering \bf \small (b) \,\, $\vec{i}=(1/3,0)$
\end{minipage}
\begin{minipage}{0.31\textwidth}
\centering \includegraphics[width=0.9\textwidth]{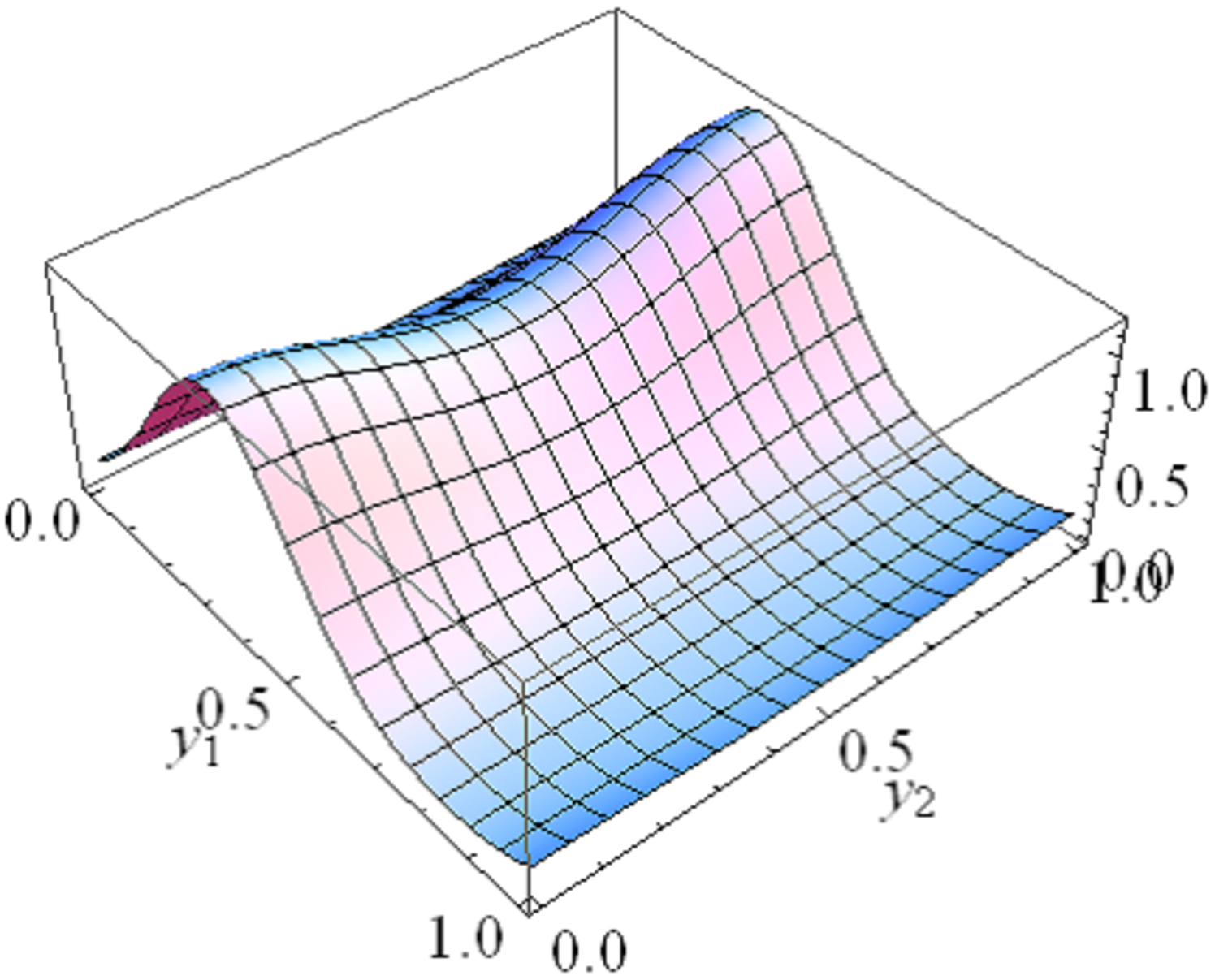}
\centering \bf \small (c) \,\, $\vec{i}=(2/3,0)$
\end{minipage}
\end{minipage}
\caption{The probability densities of zero-mode wavefunctions $\bigl| \Theta^{\vec{i}, \, \mathbb{N}}(\vec{z}) \bigr|^2$ on $(y_1, y_2)$-plane for Type 1, where we set $(x_1, x_2) =(0,0)$.
These figures show that the peaks of the probability densities are located at $\vec{y} = -\vec{i}$.}
\label{fig:zeromode_type1}
\end{figure}

\begin{figure}[H]
\begin{minipage}{\textwidth}
\centering\vspace{5pt}
\begin{minipage}{0.31\textwidth}
\centering \includegraphics[width=0.9\textwidth]{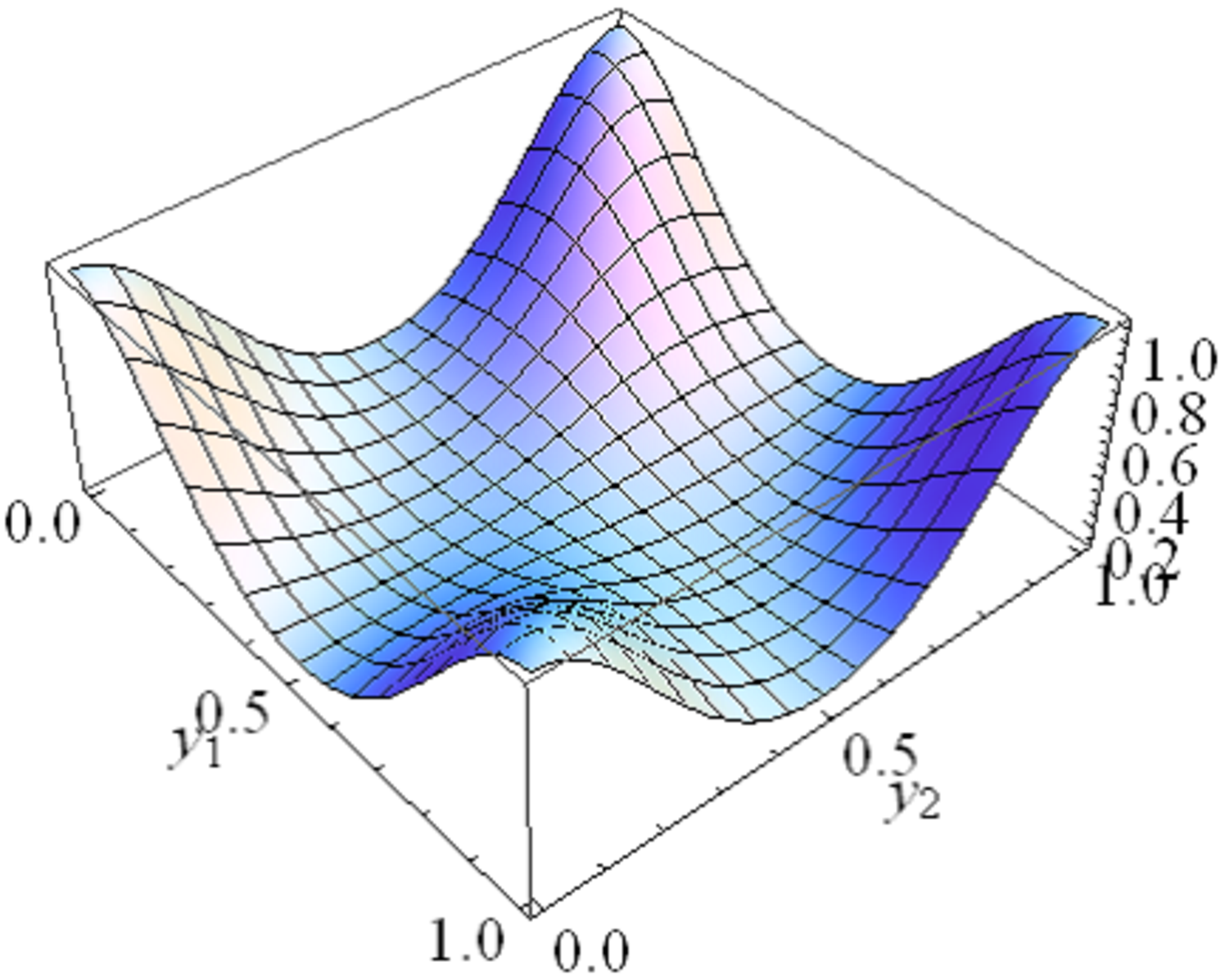}
\centering \bf \small (a) \,\, $\vec{i}=(0,0)$
\end{minipage}
\begin{minipage}{0.31\textwidth}
\centering \includegraphics[width=0.9\textwidth]{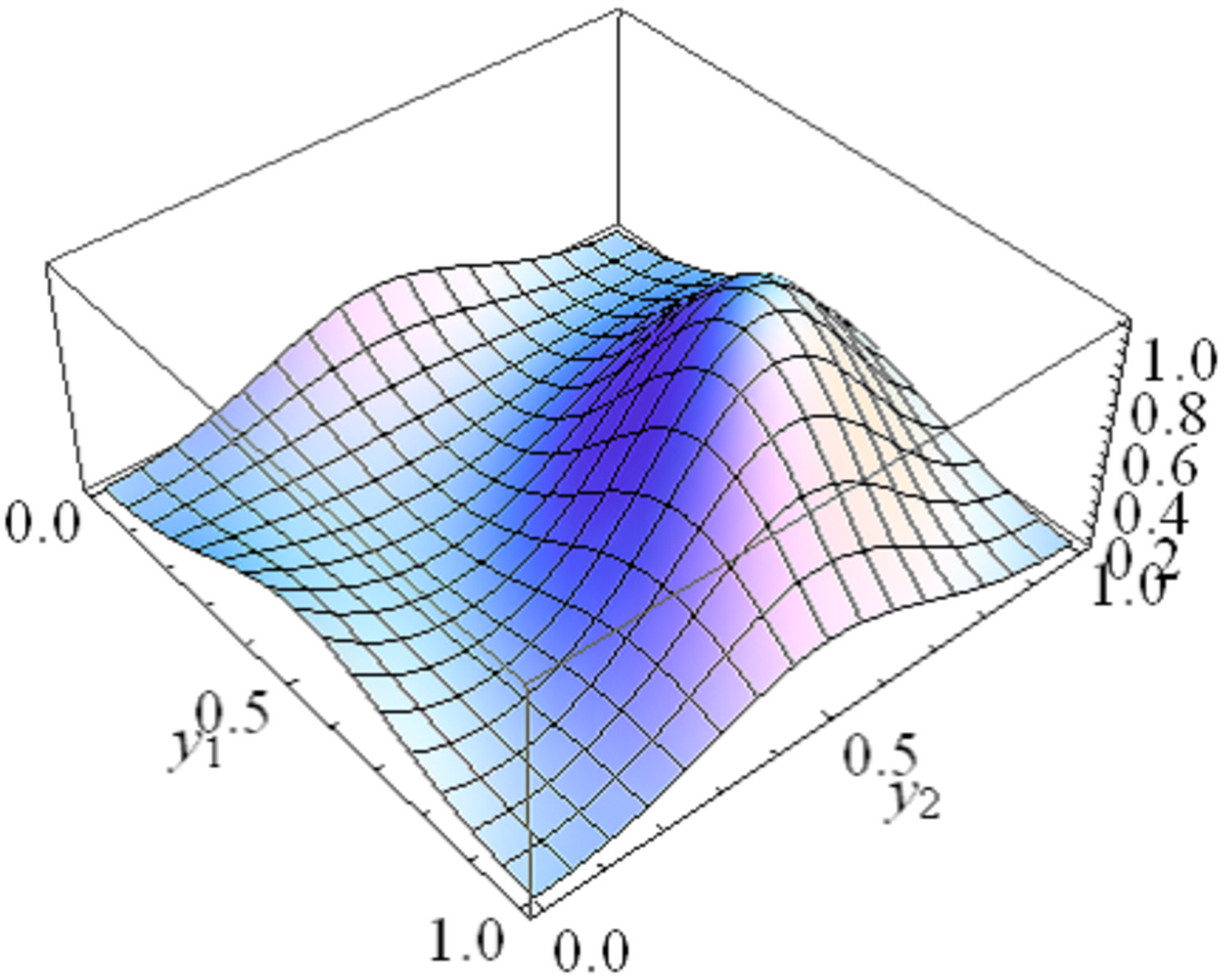}
\centering \bf \small (b) \,\, $\vec{i}=(1/3,1/3)$
\end{minipage}
\begin{minipage}{0.31\textwidth}
\centering \includegraphics[width=0.9\textwidth]{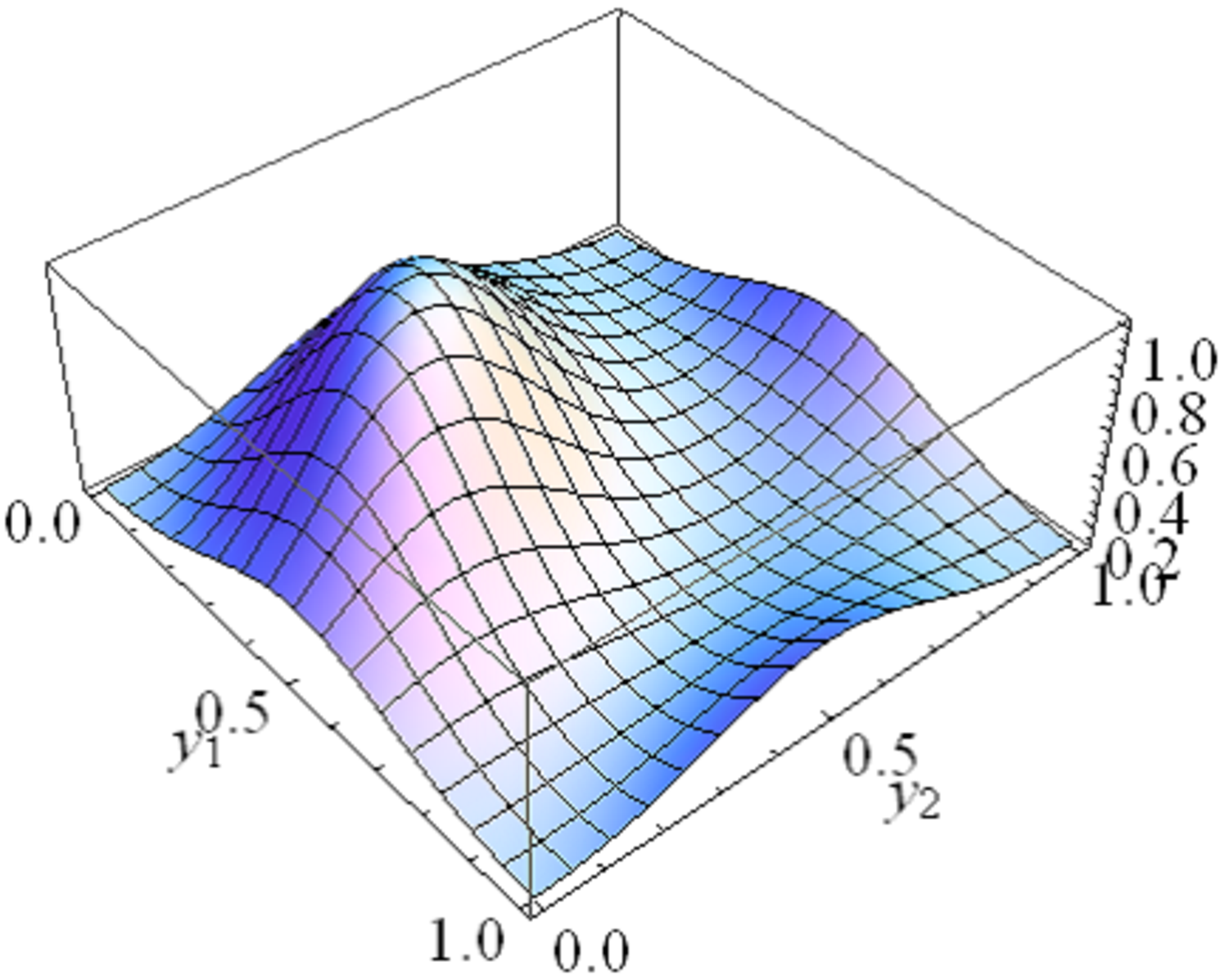}
\centering \bf \small (c) \,\, $\vec{i}=(2/3,2/3)$
\end{minipage}
\end{minipage}
\caption{The probability densities on $(y_1, y_2)$-plane for Type 2.
The peaks are located at $\vec{y} = -\vec{i}$.}
\label{fig:zeromode_type2}
\end{figure}

\begin{figure}[H]
\begin{minipage}{\textwidth}
\centering\vspace{5pt}
\begin{minipage}{0.31\textwidth}
\centering \includegraphics[width=0.9\textwidth]{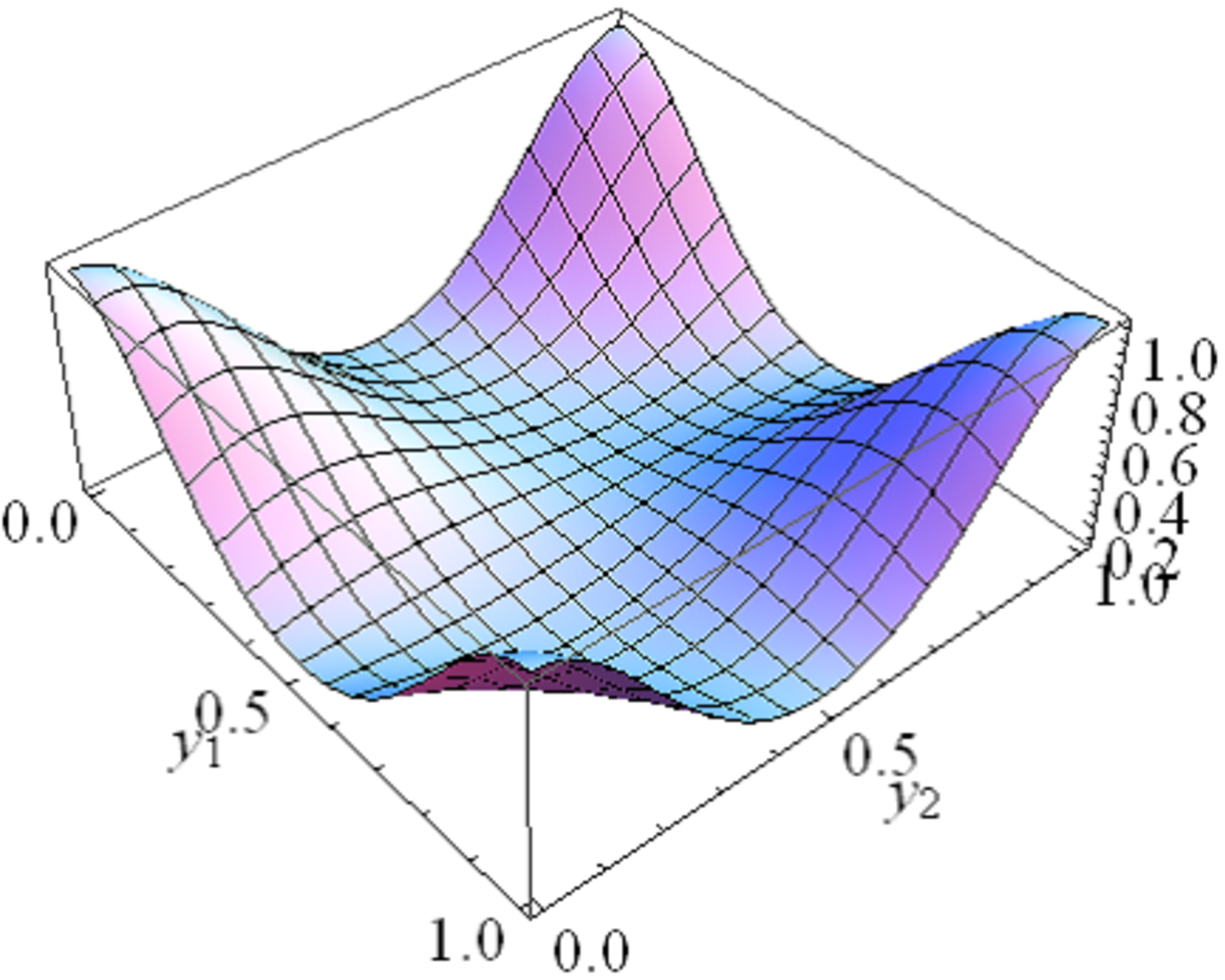}
\centering \bf (a) \,\, $\vec{i}=(0,0)$
\end{minipage}
\begin{minipage}{0.31\textwidth}
\centering \includegraphics[width=0.9\textwidth]{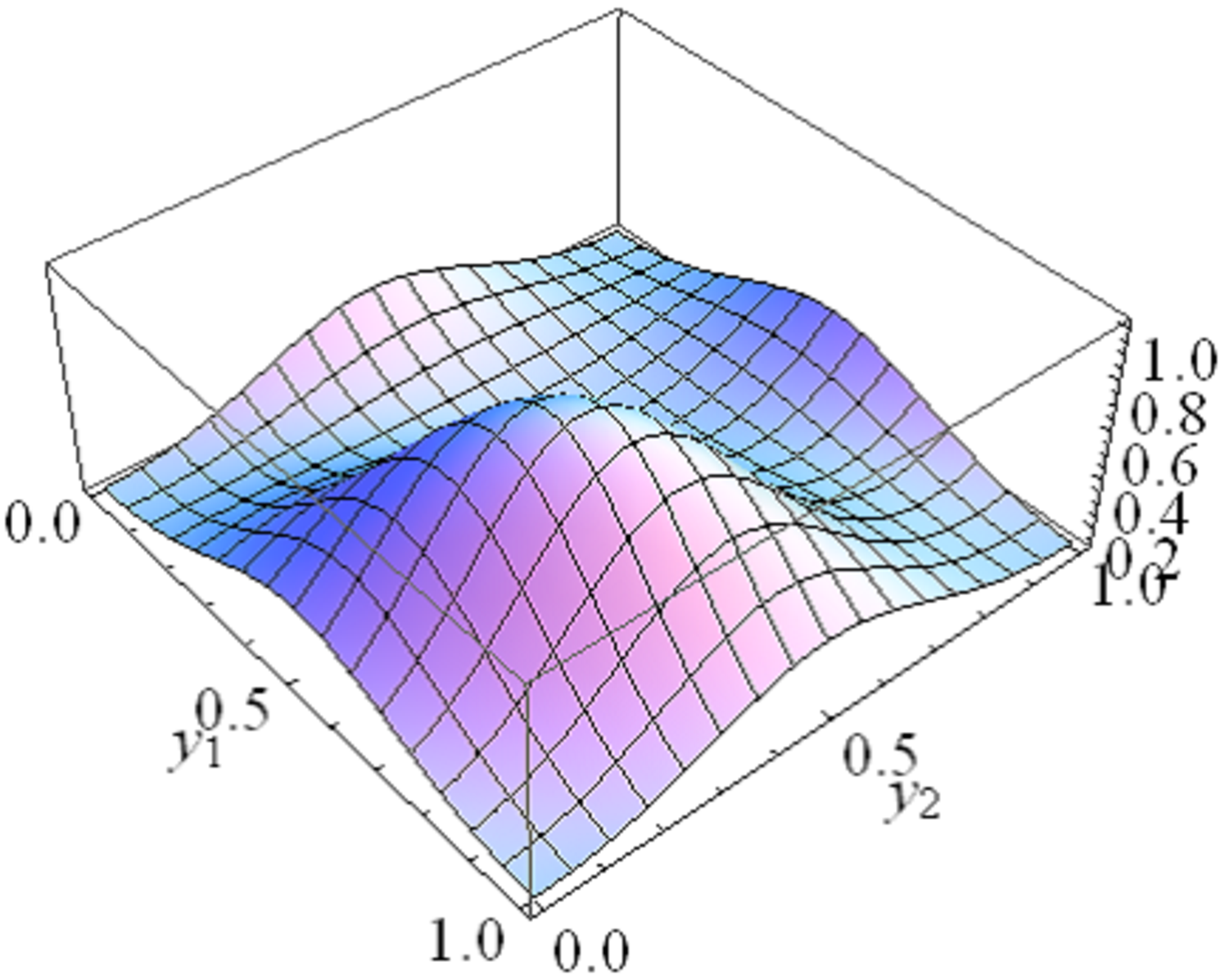}
\centering \bf (b) \,\, $\vec{i}=(1/3,2/3)$
\end{minipage}
\begin{minipage}{0.31\textwidth}
\centering \includegraphics[width=0.9\textwidth]{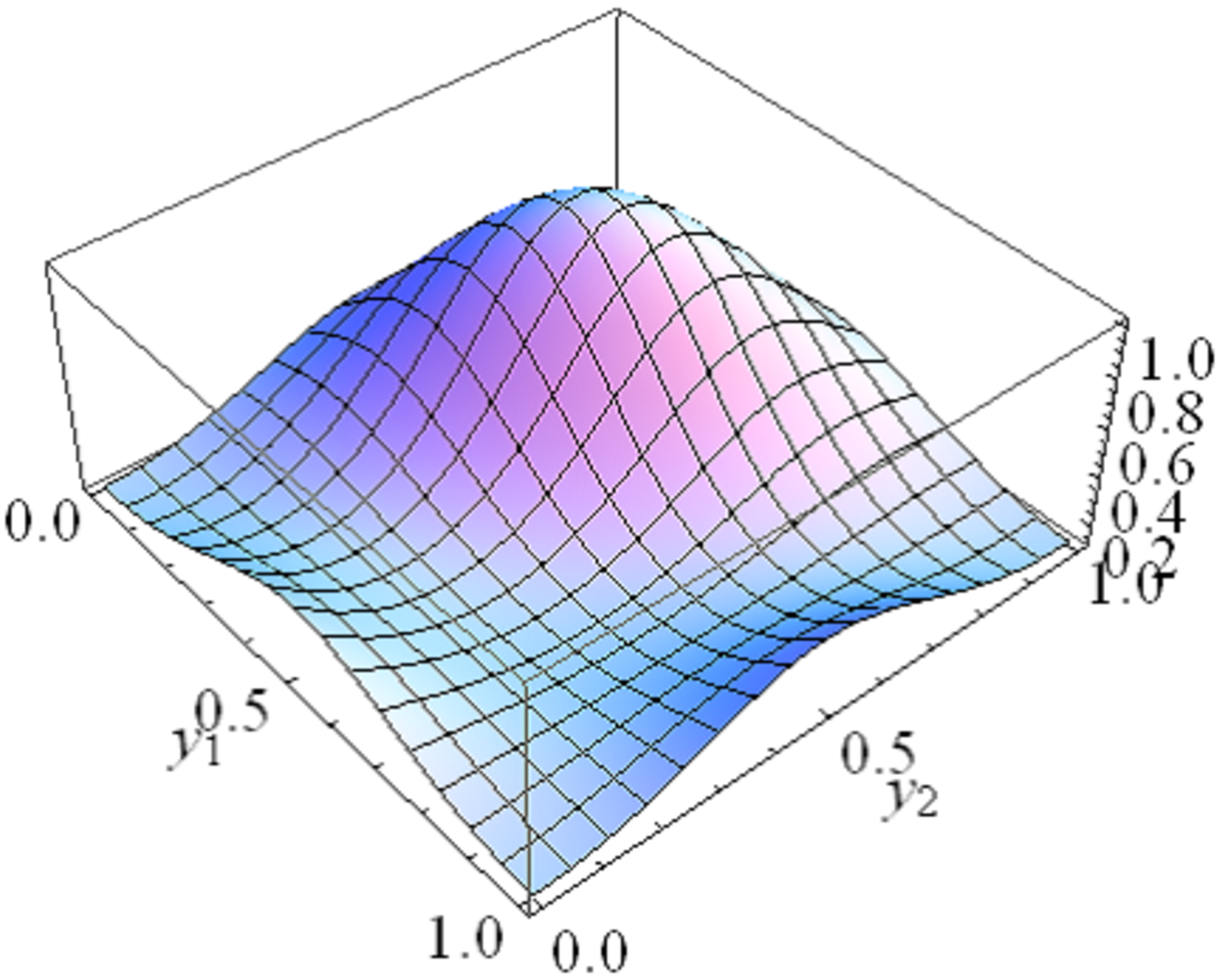}
\centering \bf (c) \,\, $\vec{i}=(2/3,1/3)$
\end{minipage}
\end{minipage}
\caption{The probability densities on $(y_1, y_2)$-plane for Type 3.
The peaks are located at $\vec{y} = -\vec{i}$.}
\label{fig:zeromode_type3}
\end{figure}

\begin{figure}[H]
\begin{minipage}{\textwidth}
\centering\vspace{5pt}
\begin{minipage}{0.31\textwidth}
\centering \includegraphics[width=0.9\textwidth]{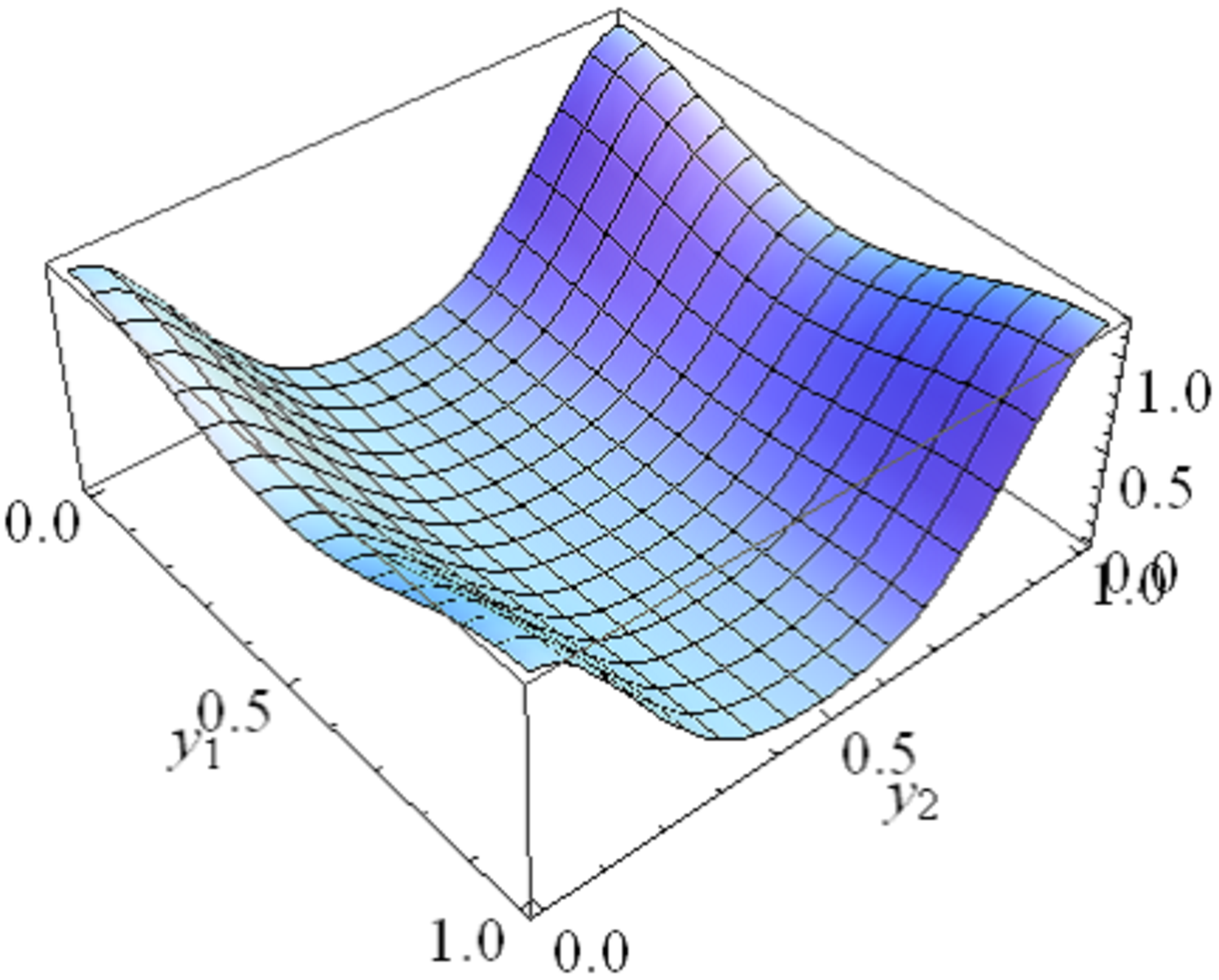}
\centering \bf (a) \,\, $\vec{i}=(0,0)$
\end{minipage}
\begin{minipage}{0.31\textwidth}
\centering \includegraphics[width=0.9\textwidth]{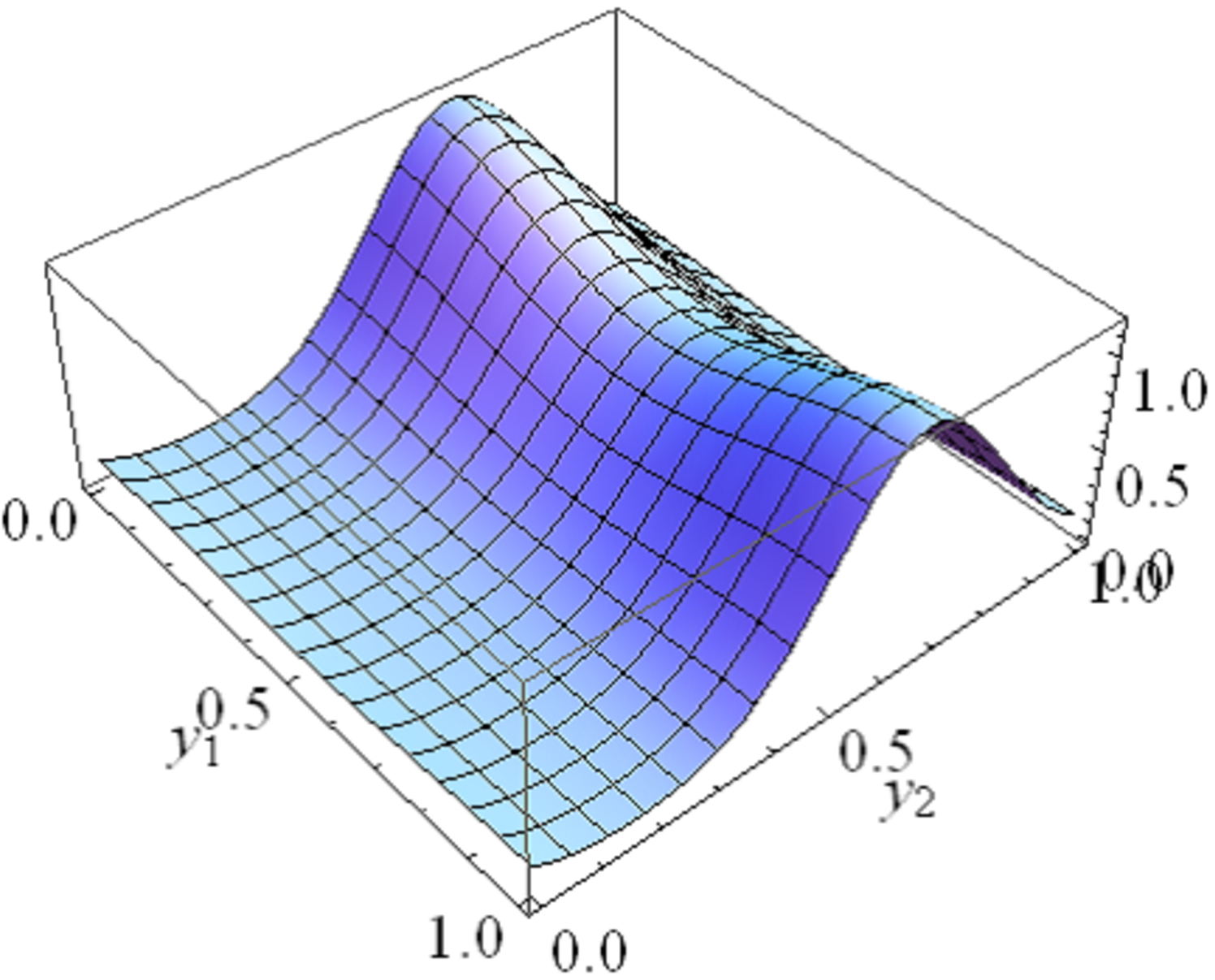}
\centering \bf (b) \,\, $\vec{i}=(0,1/3)$
\end{minipage}
\begin{minipage}{0.31\textwidth}
\centering \includegraphics[width=0.9\textwidth]{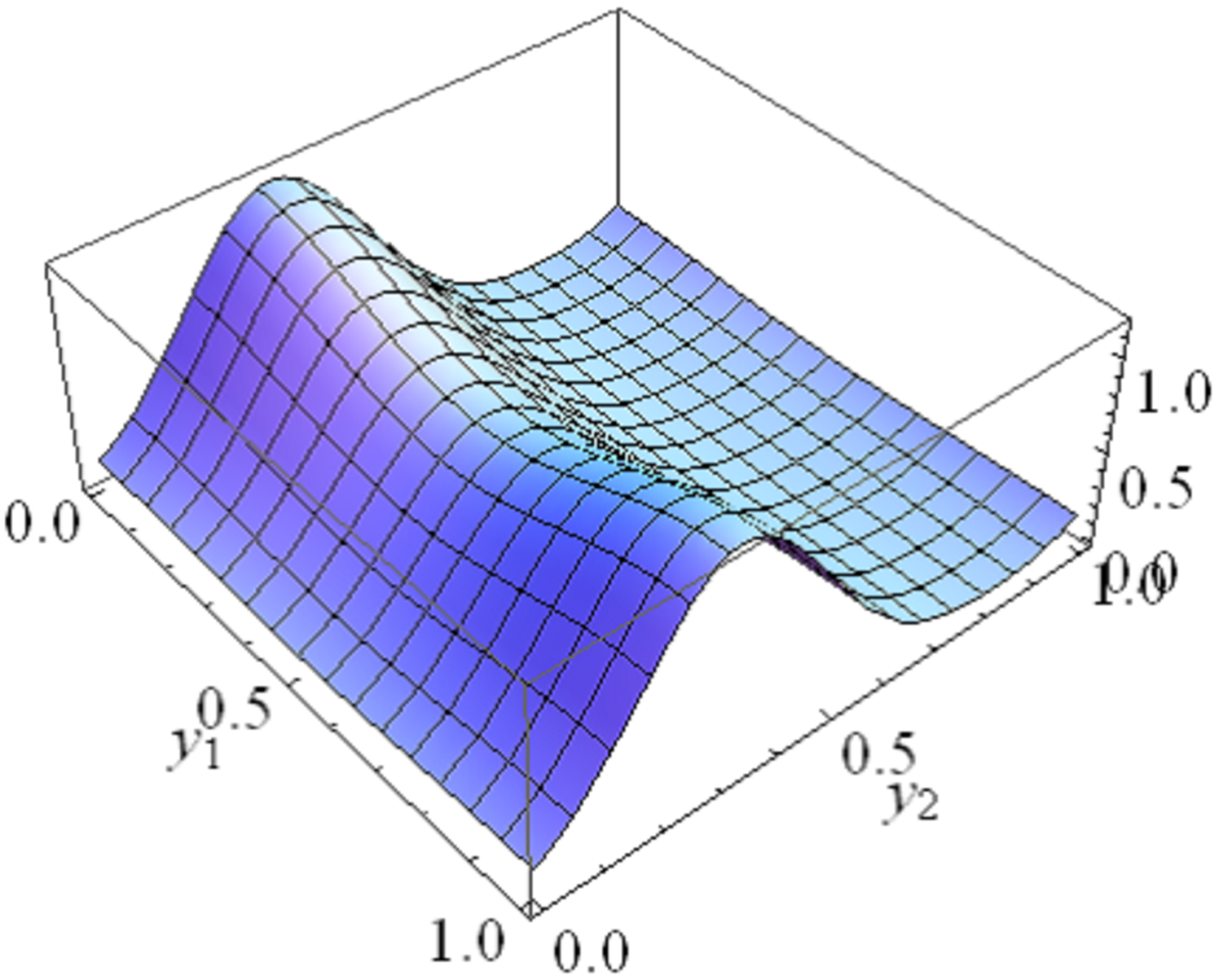}
\centering \bf (c) \,\, $\vec{i}=(0,2/3)$
\end{minipage}
\end{minipage}
\caption{The probability densities on $(y_1, y_2)$-plane for Type 4.
The peaks are located at $\vec{y} = -\vec{i}$.}
\label{fig:zeromode_type4}
\end{figure}

Type 4 :
\begin{gather}
\vec{i} = \begin{pmatrix}0\\ 0\end{pmatrix}, \begin{pmatrix}0\\ 1/3\end{pmatrix}, \begin{pmatrix}0\\ 2/3\end{pmatrix}. \hspace{70pt} \begin{minipage}{0.2\textwidth}
\setlength{\unitlength}{0.6mm}
\begin{picture}(35,37)(0,0)
\put(0,0){\vector(1,0){32}}
\put(0,0){\vector(0,1){32}}
\put(34,-1){$i_1$}
\put(-1,34){$i_2$}
\put(0,0){\circle*{3}}
\put(0,10){\circle*{3}}
\put(0,20){\circle*{3}} 
\end{picture}
\end{minipage}
\end{gather}
where the three sets of $(i_1,i_2)$ label three generations in every type.
We denominate these types of the label $\vec{i}$ {\em generation-types}.
Recall that the label $I$ has only single type, i.e., $I=1,2,3$ in magnetized torus model with factorizable fluxes.
In contrast to such a model, in general there are multiple generation-types in those with non-factorizable fluxes.
It is remarkable that the above labels represent the localization profiles of zero-mode wavefunctions on $(y_1, y_2)$-plane.
We show these profiles for each of generation-types in Figure \ref{fig:zeromode_type1}, \ref{fig:zeromode_type2}, \ref{fig:zeromode_type3} and \ref{fig:zeromode_type4}.

When we deal with the arbitrary degeneracy of zero-modes, $\det\mathbb{N} \in \mathbb{Z}$, the number of generation-types is given as the sum of divisors of $\det\mathbb{N} \in \mathbb{Z}$, that is, a divisor function in number theory.
We will mention it in Appendix \ref{appendix}.

\subsection{The relation between generation-types in each sectors}
We now identify $ab$-sector as left-handed matter sector, $ca$-sector as right-handed matter sector and $bc$-sector as Higgs sector, and study the degeneracy of Higgs fields in terms of the generation-types developed in the previous subsection.
In three generation models with $|\det\mathbb{N}_{ab}|=|\det\mathbb{N}_{ca}|=3$, we have 16 patterns of flavor structures, since each of left- and right-handed matter sectors has four generation-types.
Among these 16 patterns, if the generation-types for $ab$- and $ca$-sector are the same, the degeneracy of $bc$-sector is limited to $3n \,\, (n : \mbox{integer})$.
For example, we consider the case that the generation-types of $ab$- and $ca$-sector are both Type 2.
Then, we find 
\begin{gather}
(n'_{11}, n'_{21})_{bc} = (n'_{11}, n'_{21})_{ab} + (n'_{11}, n'_{21})_{ca} = (1,2) \quad \mbox{or} \quad (2,1) \quad \mbox{or}\quad  (0,0) \qquad (\mbox{mod} \,\, 3),\\
(n'_{12}, n'_{22})_{bc} = (n'_{12}, n'_{22})_{ab} + (n'_{12}, n'_{22})_{ca} = (1,2) \quad \mbox{or} \quad (2,1) \quad \mbox{or}\quad  (0,0) \qquad (\mbox{mod} \,\, 3).
\end{gather}
Thus, the generation-type of $bc$-sector is also Type 2 and we obtain
\begin{gather}
\det\mathbb{N}_{ca}=0 \qquad (\mbox{mod} \,\, 3).
\end{gather}
The same holds for the generation-types other than Type 2.
After all, we claim that the degeneracy of zero-modes in the $bc$-sector equals to $3n \,\, (n : \mbox{integer})$.
On the other hand, if the generation-types for $ab$- and $ca$-sector are different from each other, the degeneracy of $bc$-sector is not limited to $3n \,\, (n : \mbox{integer})$.

\subsection{Exceptional generation-types in magnetized orbifold model}\label{sec:orb_nonfact}
We refer to the exceptions for Table \ref{tab:degeneracy_nonfact} in the subsection \ref{subsec:generation-types}.
If $\det\mathbb{N}=4k \,\, (k : \mbox{integer})$, there are exceptional generation-types that is inconsistent with Table \ref{tab:degeneracy_nonfact}.
For example, we consider the case with $\det\mathbb{N}=4$ and then the degeneracy of zero-modes equals to four on magnetized torus model.
With the $Z_2$ projection, the degeneracy of zero-modes is expected to reduce to three for the periodic condition or one for the anti-periodic condition according to Table \ref{tab:degeneracy_nonfact}.
However, there exists the following generation-type for $\det\mathbb{N}=4$,
\vspace{10pt}
\begin{gather}
\vec{i} = \begin{pmatrix}0\\ 0 \end{pmatrix}, \quad \begin{pmatrix}0\\ 1/2 \end{pmatrix}, \quad\begin{pmatrix}1/2\\ 0 \end{pmatrix}, \quad \begin{pmatrix}1/2\\ 1/2 \end{pmatrix}. \qquad\quad
\begin{minipage}{0.25\textwidth}
\centering\setlength{\unitlength}{0.6mm}
\begin{picture}(35,37)(0,0)
\put(0,0){\vector(1,0){32}}
\put(0,0){\vector(0,1){32}}
\put(34,-1){$i_1$}
\put(-1,34){$i_2$}
\put(0,0){\circle*{3}}
\put(0,16){\circle*{3}}
\put(16,0){\circle*{3}} 
\put(16,16){\circle*{3}} 
\end{picture}
\end{minipage}
\label{exception}
\end{gather}

\vspace{10pt}
\noindent Then, four $Z_2$-even zero-mode wavefunctions
\begin{align}
\Theta^{[(0, \, 0)]}_\mathrm{even}(\vec{z}), \qquad \Theta^{[(0, \, 1/2)]}_\mathrm{even}(\vec{z}), \qquad \Theta^{[(1/2, \, 0)]}_\mathrm{even}(\vec{z}), \qquad \Theta^{[(1/2, \, 1/2)]}_\mathrm{even}(\vec{z}),
\end{align}
survive after the $Z_2$ projection.
On the other hand, no $Z_2$-odd zero-mode wavefunctions survive.
Because similar exceptions occur for $\det\mathbb{N}=4k$, in this case, we must count the number of zero-modes after the $Z_2$ projection in terms of the relations (\ref{evenmode_orbifold}) and (\ref{oddmode_orbifold}), instead of using Table \ref{tab:degeneracy_nonfact}.

\section{Non-Abelian discrete flavor symmetry on magnetized brane models} \label{sec:symmetry}
\subsection{Magnetized torus model with factorizable fluxes}
First of all, we refer to the non-Abelian discrete flavor symmetry from the magnetized torus model with only factorizable fluxes.
In this model, the flavor symmetry is investigated in detail in Ref.~\cite{Abe:2009vi}.
In this subsection, we briefly review the flavor symmetries revealed in Ref.~\cite{Abe:2009vi}.

First, we show the generic case with non-vanishing Wilson-lines.
For $\textrm{gcd}(M_{ab}, M_{ca}, M_{bc})=3$, there exists three $Z_3$ symmetries, which act $\sum_{J_{ab}=1}^3 X^{I_{ab}J_{ab}}\Theta^{J_{ab}, M_{ab}}$, where
\begin{gather}
X=Z, Z',C,\\
Z =
\begin{pmatrix}
1 & 0 & 0\\
0 & \omega & 0\\
0 & 0 & \omega^2
\end{pmatrix}
, \qquad
Z' =
\begin{pmatrix}
\omega & 0 & 0\\
0 & \omega & 0\\
0 & 0 & \omega
\end{pmatrix}
, \qquad C =
\begin{pmatrix}
0 & 1 & 0\\
0 & 0 & 1\\
1 & 0 & 0
\end{pmatrix}
,\label{fact_generator}
\end{gather}
and $\omega \equiv e^{2\pi i/3}$.
The generator $C$ acts on $\Theta^{I_{ab}, \, M_{ab}}$ as cyclic permutations
\begin{gather}
\sum_{J_{ab}=1}^3 (C^n)^{I_{ab}J_{ab}}\Theta^{J_{ab},M_{ab}}= \Theta^{I_{ab}+n, \, M_{ab}},\label{cycricperm}
\end{gather}
with an integer $n$.
The generators $Z$ and $C$ do not commute each other.
However, there exists the closed algebra consisting of $Z$, $Z'$ and $C$,
\begin{gather}
CZ = Z' ZC.\label{algebra_fact}
\end{gather}
In this closed algebra, diagonal matrices are denoted by $Z^nZ'^m$.
These generators generate the non-Abelian discrete flavor symmetry $(Z_3 \times Z_3') \rtimes Z_3^{(C)} \cong \Delta(27)$, which has 27 elements totally.
For gcd($M_{ab},M_{ca},M_{bc})=g$, there appears the flavor symmetry $(Z_g \times Z_g') \rtimes Z_g^{(C)}$.

In the remainder of this subsection, we consider the case with vanishing Wilson-lines.
In this case, we can define a $Z_2$ transformation which acts as
\begin{gather}
\Theta^{I_{ab}, \, M_{ab}} \rightarrow \Theta^{M_{ab}-I_{ab}, \, M_{ab}}.
\end{gather}
We denote the generator of this $Z_2$ transformation by $P$.
For simplicity again, we consider the case with $\textrm{gcd}(M_{ab}, M_{ca}, M_{bc})=3$ and the zero-modes of $ab$-sector with $|M_{ab}| = 3$.
Then the representations of four generators $Z$, $Z'$, $C$ and $P$ can be expressed as follows : 
\begin{gather}
Z =
\begin{pmatrix}
1 & 0 & 0\\
0 & \omega & 0\\
0 & 0 & \omega^2
\end{pmatrix}
, \qquad
Z' =
\begin{pmatrix}
\omega & 0 & 0\\
0 & \omega & 0\\
0 & 0 & \omega
\end{pmatrix}
, \qquad
C =
\begin{pmatrix}
0 & 1 & 0\\
0 & 0 & 1\\
1 & 0 & 0
\end{pmatrix}
, \qquad
P =
\begin{pmatrix}
1 & 0 & 0\\
0 & 0 & 1\\
0 & 1 & 0
\end{pmatrix}.
\end{gather}
The closed algebra for these generators is $\Delta(54) \cong (Z_3 \times Z_3') \rtimes S_3$.

For $\textrm{gcd}(M_{ab}, M_{ca}, M_{bc})=g$, notice that generators $Z$ and $P$ satisfy
\begin{gather}
PZ = Z^{-1} P,
\end{gather}
and the closed algebra of $C$ and $P$ is $D_g$.
Therefore the flavor symmetry, which is generated by $Z$, $Z'$, $C$ and $P$, is nothing but $(Z_g \times Z_g') \rtimes D_g$.
Note that in particular for $g=3$, $D_3 \cong S_3$ and then $(Z_3 \times Z_3') \rtimes S_3$ is isomorphic to $\Delta(54)$.

\subsection{Magnetized torus model with non-factorizable fluxes : aligned generation-types}
We study the magnetized torus model with non-factorizable fluxes.
In this model, Yukawa couplings are given by Eq.~(\ref{yukawa_nonfact}). 
Because the Yukawa couplings are written by Riemann $\vartheta$-function which is an extension of Jacobi $\vartheta$-function, the flavor symmetries possessed by these couplings would be different from those obtained in the factorizable case.
By focusing on and investigating the labels of generations, we study on the selection rule and the character in the Riemann $\vartheta$-function in order to analyze the flavor symmetry.
In the expression (\ref{yukawa_nonfact}) of Yukawa couplings, we look at the factors coming from the overlap integral on the first and second tori, that is $\lambda_{\vec{i}_{ab} \vec{i}_{ca} \vec{i}_{bc}}$ shown in Eq.~(\ref{yukawa'_nonfact}), because such factors reflect the property of non-factorizable fluxes.
As shown in Eq.~(\ref{yukawa'_nonfact}), these factors consist of the selection rule
\begin{gather}
\sum_{\vec{m}} \delta_{\vec{i}_{bc}, \, \mathbb{N}_{cb}^{-1} (\mathbb{N}_{ab} \vec{i}_{ab} + \mathbb{N}_{ca}\vec{i}_{ca}+ \mathbb{N}_{ab} \vec{m})}, \label{selectionrule}
\end{gather}
and the Riemann $\vartheta$-function
\begin{gather}
\vartheta \begin{bmatrix}\vec{\bold{K}}\\[5pt] 0 \end{bmatrix}(i\vec{\bold{Y}}|i\vec{\bold{Q}}).
\end{gather}
The value of the Riemann $\vartheta$-function is determined by the character of $\vartheta$-function
\begin{gather}
(\vec{i}_{ab} - \vec{i}_{ca} + \vec{m}) \frac{\mathbb{N}_{ab} (\mathbb{N}_{ab} + \mathbb{N}_{ca})^{-1} \mathbb{N}_{ca}}{\det\mathbb{N}_{ab} \det\mathbb{N}_{ca}}, \label{character}
\end{gather}
which appears in $\vec{\bold{K}}$.
Notice that the generation labels $\vec{i}_{ab}$, $\vec{i}_{ca}$ and $\vec{i}_{bc}$ appear only in the selection rule (\ref{selectionrule}) and the character (\ref{character}).
Accordingly, the flavor structure of the Yukawa coupling is completely determined by them.
Thus we focus on these parts for the purpose to identify the flavor symmetries.

We study the case with $\textrm{gcd}(\det\mathbb{N}_{ab}, \det\mathbb N_{ca}, \det\mathbb{N}_{bc})=3$.
In the following part of this subsection, we analyze the case that the generation-types of $ab$-, $ca$- and $bc$-sectors are aligned, and then study the case that the generation-types are not aligned in the next subsection.

First we consider a general case where Wilson-lines are turned on.
The selection rule (\ref{selectionrule}) reduces to the following relation,
\begin{gather}
\mathbb{N}_{ab} \vec{i}_{ab} + \mathbb{N}_{ca} \vec{i}_{ca} + \mathbb{N}_{bc} \vec{i}_{bc} = - \mathbb{N}_{ab} \vec{m} + \mathbb{N}_{bc} (l_1 \vec{e}_1 + l_2 \vec{e}_2),\label{selectionrule2}
\end{gather}
where $l_1$ and $l_2$ are integers and $\vec{e}_i \,\, (i=1,2)$ are defined in Eq.~(\ref{units}).
Therefore the selection rule (\ref{selectionrule}) yields a couple of constraints represented by two component equations in Eq.~(\ref{selectionrule2}) which restricts the flavor symmetry. 
On the other hand, we should notice that the selection rule (\ref{selectionrule}) remains intact under the following simultaneous translation,
\begin{gather}
\vec{i}_{ab} \rightarrow \vec{i}_{ab} + \vec{n}, \qquad \vec{i}_{ca} \rightarrow \vec{i}_{ca} + \vec{n}, \qquad \vec{i}_{bc} \rightarrow \vec{i}_{bc} + \vec{n}, \label{translation}
\end{gather}
with a 2-vector $\vec{n}$ determined by the generation-type in $ab$-sector.
By using the periodicity of the Riemann $\vartheta$-function, we can set the 2-vector $\vec{n}$ as the difference between the two of the set of the label $\{\vec{i}_{ab}\}$ without loss of generality, as shown in Table \ref{tab:nvector}.
Actually, we can confirm that the value of the character in the Riemann $\vartheta$-function would not change under such a translation, and then it preserves the value of the Yukawa coupling.
Such an invariance under the above translation is guaranteed by the relation $\mathbb{N}_{ab} + \mathbb{N}_{bc} +\mathbb{N}_{ca}=0$ for the zero-modes of the three matters that construct the Yukawa coupling.

\begin{table}[h]
\centering
\begin{tabular}{ccc} \hline
Generation-type & The set of the label $\{\vec{i}_{ab}\}$ & The 2-vector $\vec{n}$\\ \hline
Type 1 & $(0,0), \, (1/3, 0), \, (2/3, 0)$ & $(1/3, 0)$\\
Type 2 & $(0,0), \, (1/3, 1/3), \, (2/3, 2/3)$ & $(1/3, 1/3)$\\
Type 3 & $(0,0), \, (1/3, 2/3), \, (2/3, 1/3)$ & $(1/3, 2/3)$\\
Type 4 & $(0,0), \, (0, 1/3), \, (0, 2/3)$ & $(0, 1/3)$ \\ \hline
\end{tabular}
\caption{The 2-vector $\vec{n}$ determined by the matrix $\mathbb{N}_{ab}$.}
\label{tab:nvector}
\end{table}

\noindent Table \ref{tab:nvector} shows that the translation (\ref{translation}) is identified with the $Z_3^{(C)}$ transformation
\begin{gather}
\Theta^{\vec{i}_{ab}, \, \mathbb{N}_{ab}} \rightarrow \Theta^{\vec{i}_{ab}+\vec{n}, \, \mathbb{N}_{ab}}, \qquad
\Theta^{\vec{i}_{ca}, \, \mathbb{N}_{ca}} \rightarrow \Theta^{\vec{i}_{ca}+\vec{n}, \, \mathbb{N}_{ca}}, \qquad
\Theta^{\vec{i}_{bc}, \, \mathbb{N}_{bc}} \rightarrow \Theta^{\vec{i}_{bc}+\vec{n}, \, \mathbb{N}_{bc}}.
\end{gather}
The representation of the $Z_3^{(C)}$ generator is written as
\begin{gather}
C =
\begin{pmatrix}
0 & 1 & 0\\
0 & 0 & 1\\
1 & 0 & 0
\end{pmatrix}
,\label{C_nonfact}
\end{gather}
which acts on the basis
\begin{gather}
\begin{pmatrix}
\Theta^{\vec{i}_0, \, \mathbb{N}}\\
\Theta^{\vec{i}_1, \, \mathbb{N}}\\
\Theta^{\vec{i}_2, \, \mathbb{N}}
\end{pmatrix}
,\label{basis1}
\end{gather}
where we label the three-generation one by one, i.e., $\{\vec{i}\}\equiv\{\vec{i}_0, \vec{i}_1, \vec{i}_2 \}$.
There always exists a $Z_3$ invariance under $Z_3^{(C)}$ generator in the case with aligned generation-types.

In the following, we investigate additional $Z_3$ symmetries similar to those shown in as Eq.~(\ref{fact_generator}) in the factorizable case.
For concreteness, we consider the flux configuration with $|\det\mathbb N_{ab}|=|\det\mathbb N_{ca}|=3$ and $|\det\mathbb N_{bc}|=6$, e.g., 
\begin{gather}
\mathbb{N}_{ab}=
\begin{pmatrix}
-1 & -2\\
-1 & 1
\end{pmatrix}
, \qquad
\mathbb{N}_{ca} =
\begin{pmatrix}
5 & 4\\
2 & 1 
\end{pmatrix}
, \qquad
\mathbb{N}_{bc} =
\begin{pmatrix}
-4 & -2\\
-1 & -2
\end{pmatrix}.\label{eg:delta_336}
\end{gather}
In this case, the generation-types in both the $ab$- and the $ca$-sectors are of the Type 2 and the generation labels in $bc$-sector are given as
\begin{align}
\vec{i}_{bc,0} =
\begin{pmatrix}
0\\ 0
\end{pmatrix}, \qquad
\vec{i}_{bc,1}&=
\begin{pmatrix}
0 \\ 1/2
\end{pmatrix}, \qquad
\vec{i}_{bc,2}=
\begin{pmatrix}
1/3 \\ 1/3
\end{pmatrix},\\
\vec{i}_{bc,3} =
\begin{pmatrix}
1/3\\ 5/6
\end{pmatrix}, \qquad
\vec{i}_{bc,4} &=
\begin{pmatrix}
2/3 \\ 2/3
\end{pmatrix}, \qquad
\vec{i}_{bc,5}=
\begin{pmatrix}
2/3 \\ 1/6
\end{pmatrix}.
\end{align}
Then, the relevant factors in the Yukawa couplings are written as
\begin{gather}
\lambda_{\vec{i}_{ab}\vec{i}_{ca}\vec{i}_{bc, 0}}=\lambda_{\vec{i}_{ab}\vec{i}_{ca}\vec{i}_{bc, 1}}=
\begin{pmatrix}
\lambda_0 & 0 & 0\\
0& 0 & \lambda_1\\
0 & \lambda_2 & 0
\end{pmatrix}, \quad
\lambda_{\vec{i}_{ab}\vec{i}_{ca}\vec{i}_{bc,2}}=\lambda_{\vec{i}_{ab}\vec{i}_{ca}\vec{i}_{bc,3}}=
\begin{pmatrix}
0 & \lambda_1 & 0\\
\lambda_2& 0 & 0\\
0 & 0 & \lambda_0
\end{pmatrix},\\
\lambda_{\vec{i}_{ab}\vec{i}_{ca}\vec{i}_{ca,4}}=\lambda_{\vec{i}_{ab}\vec{i}_{ca}\vec{i}_{bc,5}}=
\begin{pmatrix}
0 & 0 & \lambda_2\\
0& \lambda_0 & 0\\
\lambda_1 & 0 & 0
\end{pmatrix},
\end{gather}
where values of $\lambda_0, \lambda_1$ and $\lambda_2$ are different from each other.
The numerical values of $\lambda_0, \lambda_1$ and $\lambda_2$ can be calculated in terms of the fluxes $\mathbb{N}_{ab}, \mathbb{N}_{ca}$ and $\mathbb{N}_{bc}$, however, they are irrelevant to the flavor symmetry itself which the above Yukawa couplings possess.

\begin{figure}[H]
\begin{minipage}{\textwidth}
\centering\vspace{100pt}
\begin{minipage}{0.31\textwidth}
\includegraphics[width=0.99\textwidth]{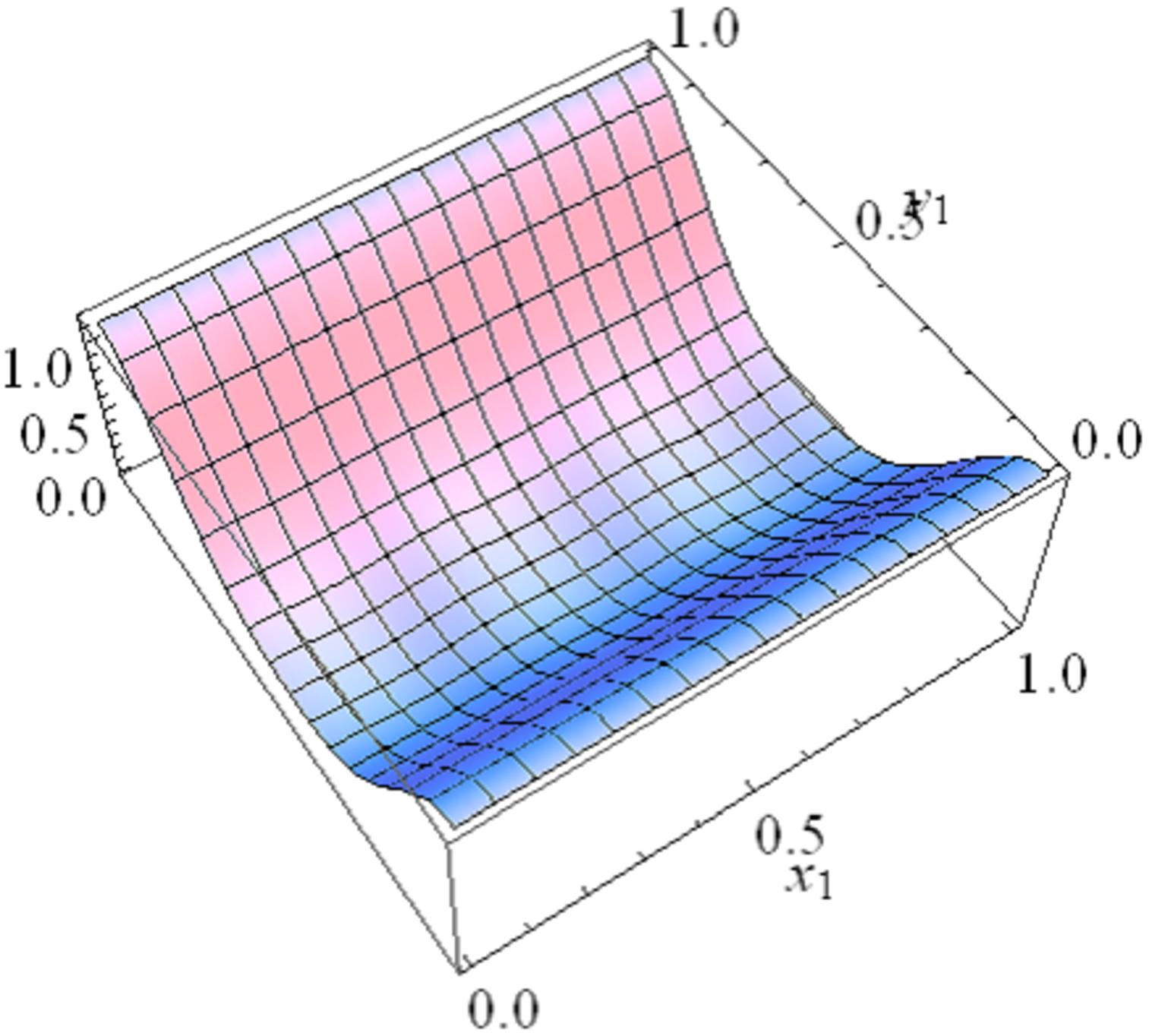}
\centering \bf \small (a) \,\, $\vec{i}=(0,0)$
\end{minipage}
\begin{minipage}{0.31\textwidth}
\includegraphics[width=0.99\textwidth]{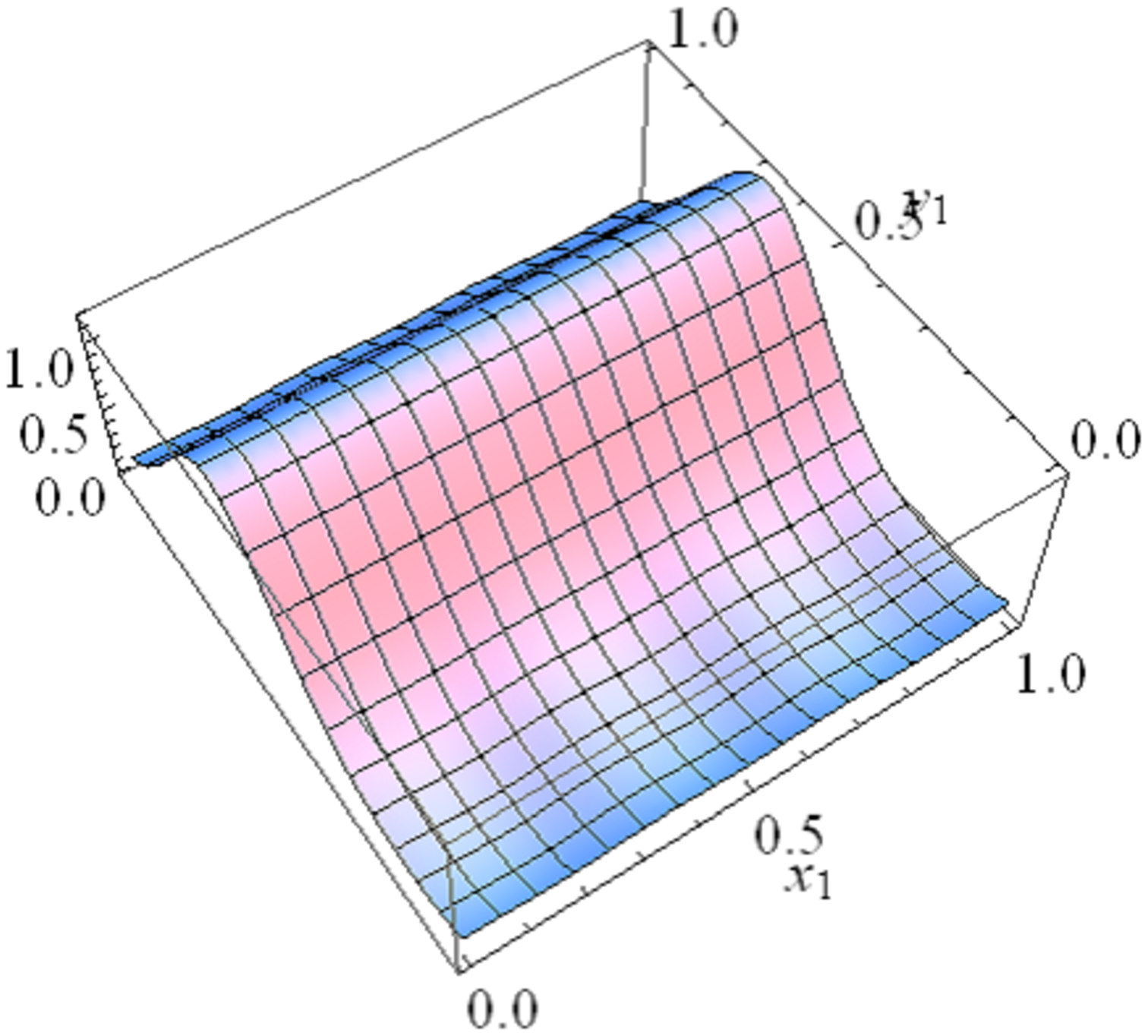}
\centering \bf \small (b) \,\, $\vec{i}=(1/3,0)$
\end{minipage}
\begin{minipage}{0.31\textwidth}
\includegraphics[width=0.99\textwidth]{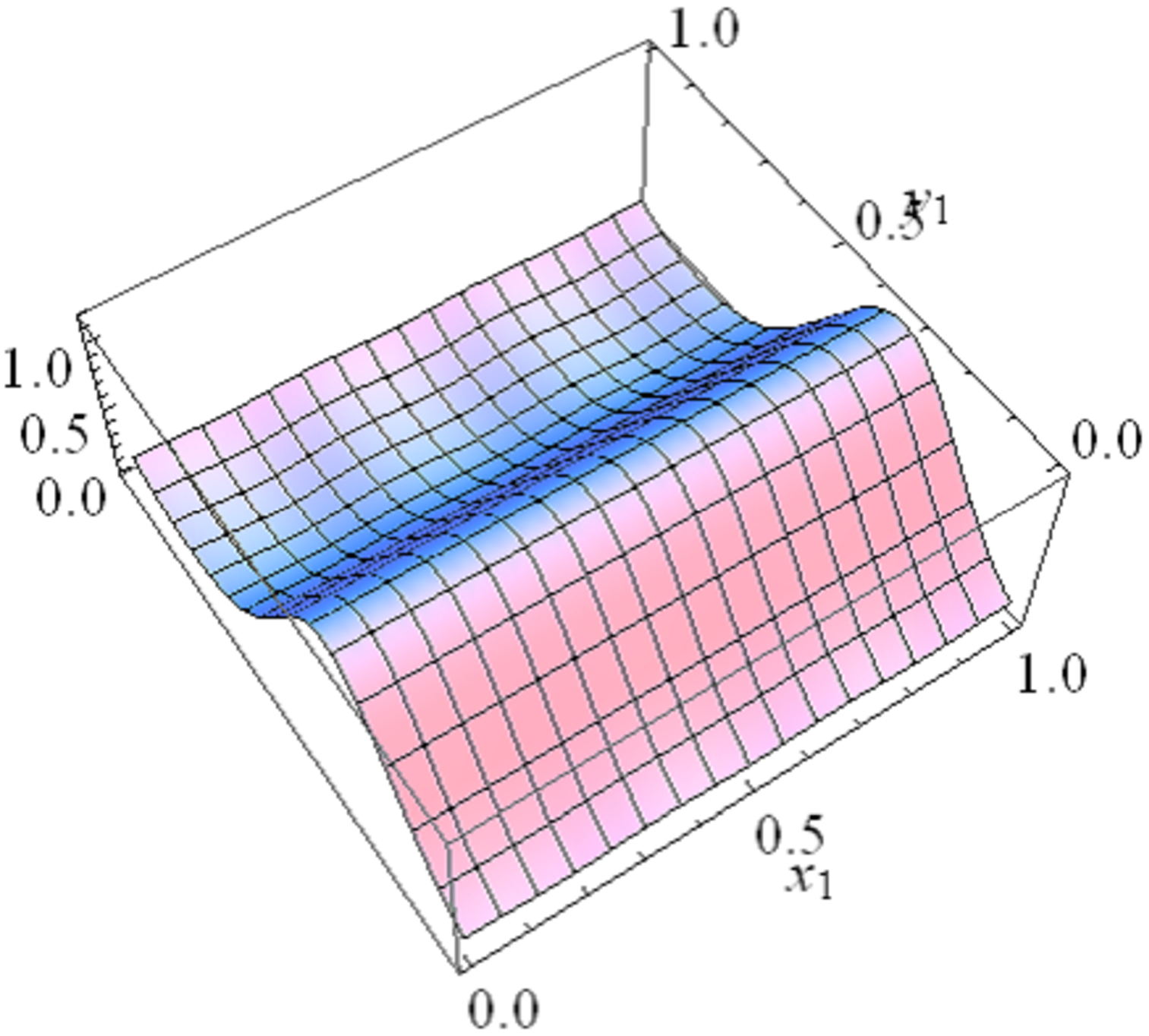}
\centering \bf \small (c) \,\, $\vec{i}=(2/3,0)$
\end{minipage}
\vspace{20pt}
\end{minipage}
\begin{minipage}{\textwidth}
\centering
\begin{minipage}{0.31\textwidth}
\includegraphics[width=0.99\textwidth]{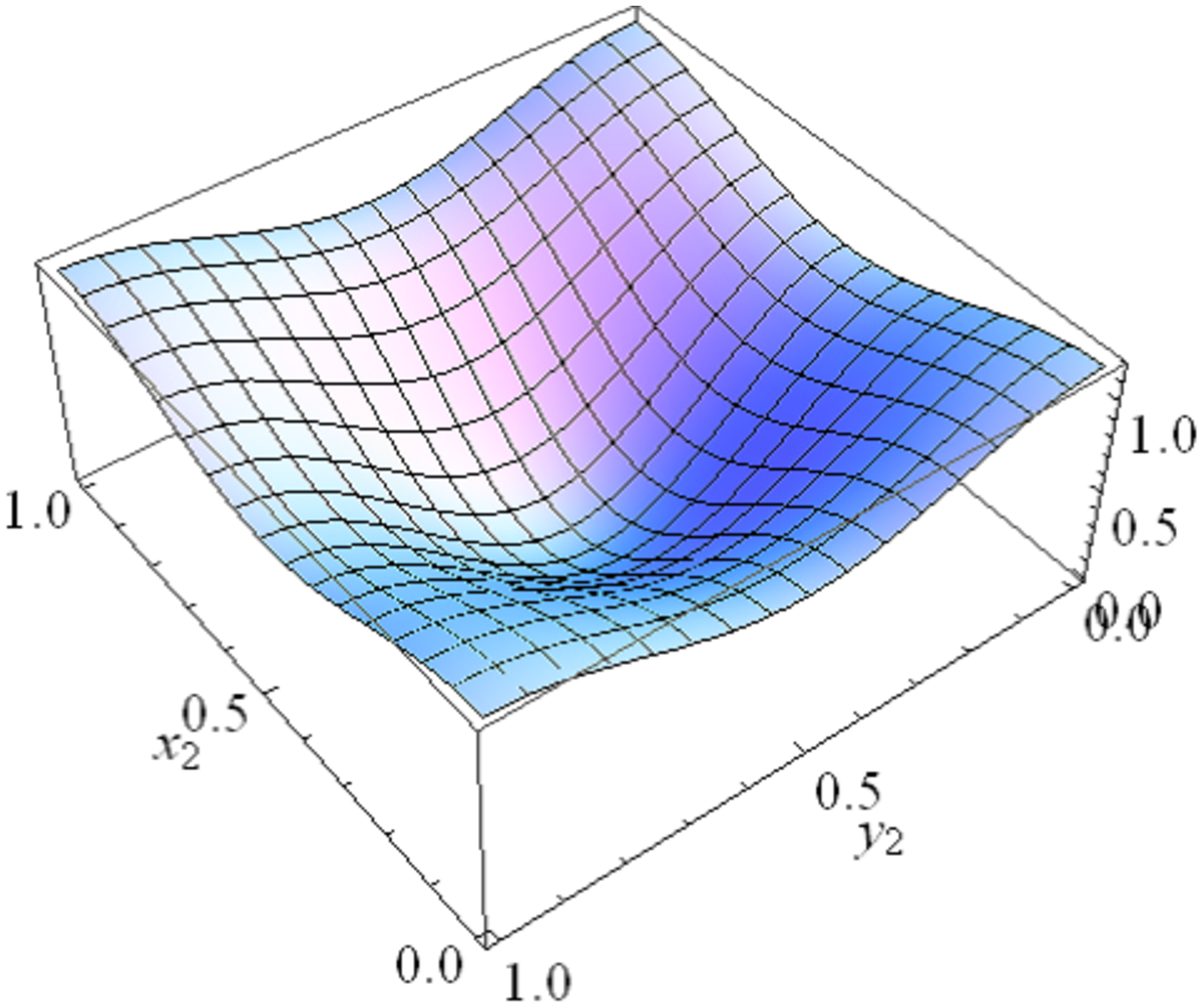}
\centering \bf \small (d) \,\, $\vec{i}=(0,0)$
\end{minipage}
\begin{minipage}{0.31\textwidth}
\includegraphics[width=0.99\textwidth]{2nd_type1_0.eps}
\centering \bf \small (e) \,\, $\vec{i}=(1/3,0)$
\end{minipage}
\begin{minipage}{0.31\textwidth}
\includegraphics[width=0.99\textwidth]{2nd_type1_0.eps}
\centering \bf \small (f) \,\, $\vec{i}=(2/3,0)$
\end{minipage}
\end{minipage}
\caption{The probability densities of zero-mode wavefunctions $\bigl|\Theta^{\vec{i}, \, \mathbb{N}}(\vec{z})\bigr|^2$ on $(x_1, y_1)$-plane for Type 1 are shown in (a), (b) and (c), while those on $(x_2, y_2)$-plane are depicted in (d), (e) and (f).}
\label{fig:profile_type1}
\end{figure}

\begin{figure}[H]
\begin{minipage}{\textwidth}
\centering\vspace{100pt}
\begin{minipage}{0.31\textwidth}
\includegraphics[width=0.99\textwidth]{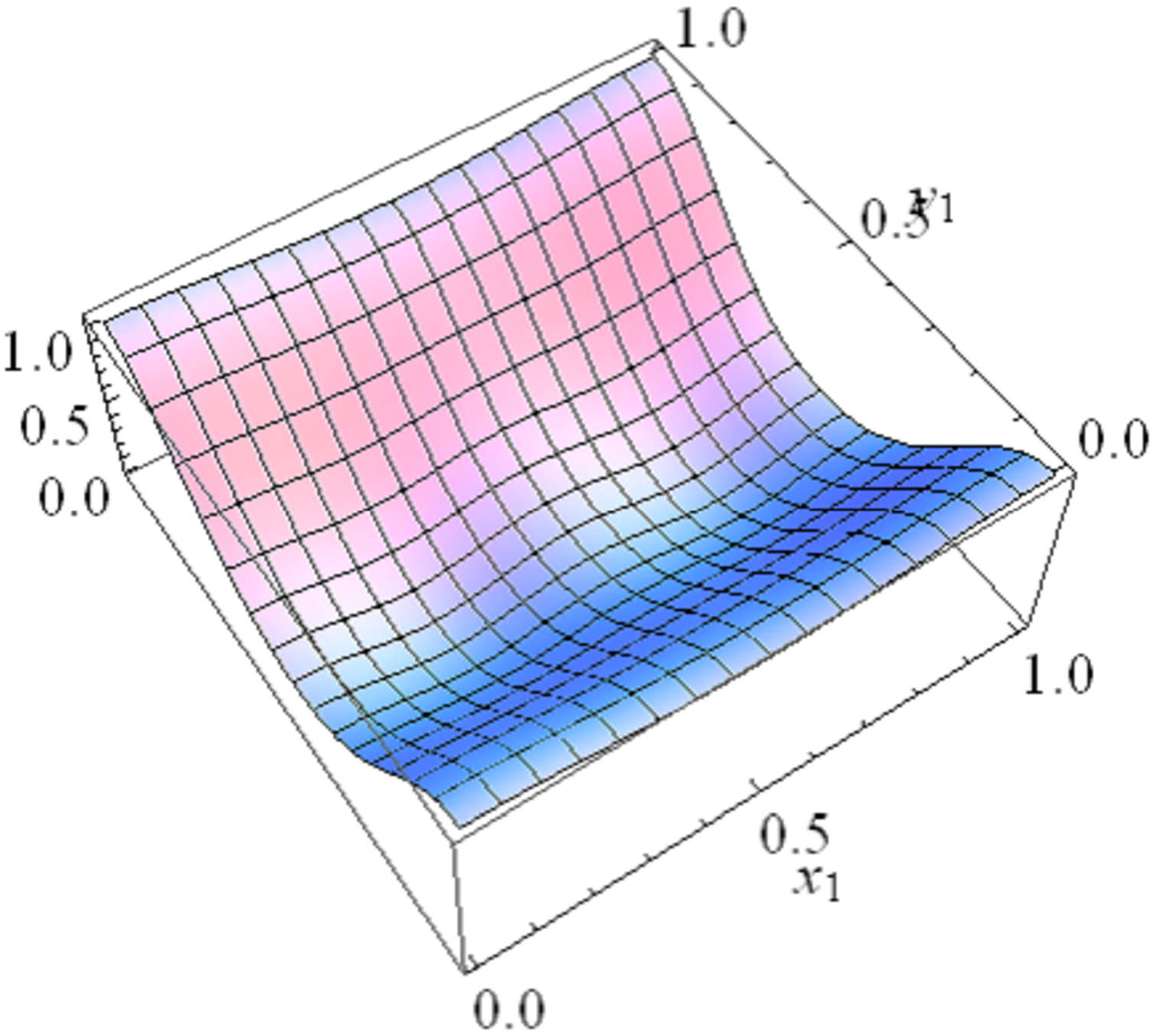}
\centering \bf \small (a) \,\, $\vec{i}=(0,0)$
\end{minipage}
\begin{minipage}{0.31\textwidth}
\includegraphics[width=0.99\textwidth]{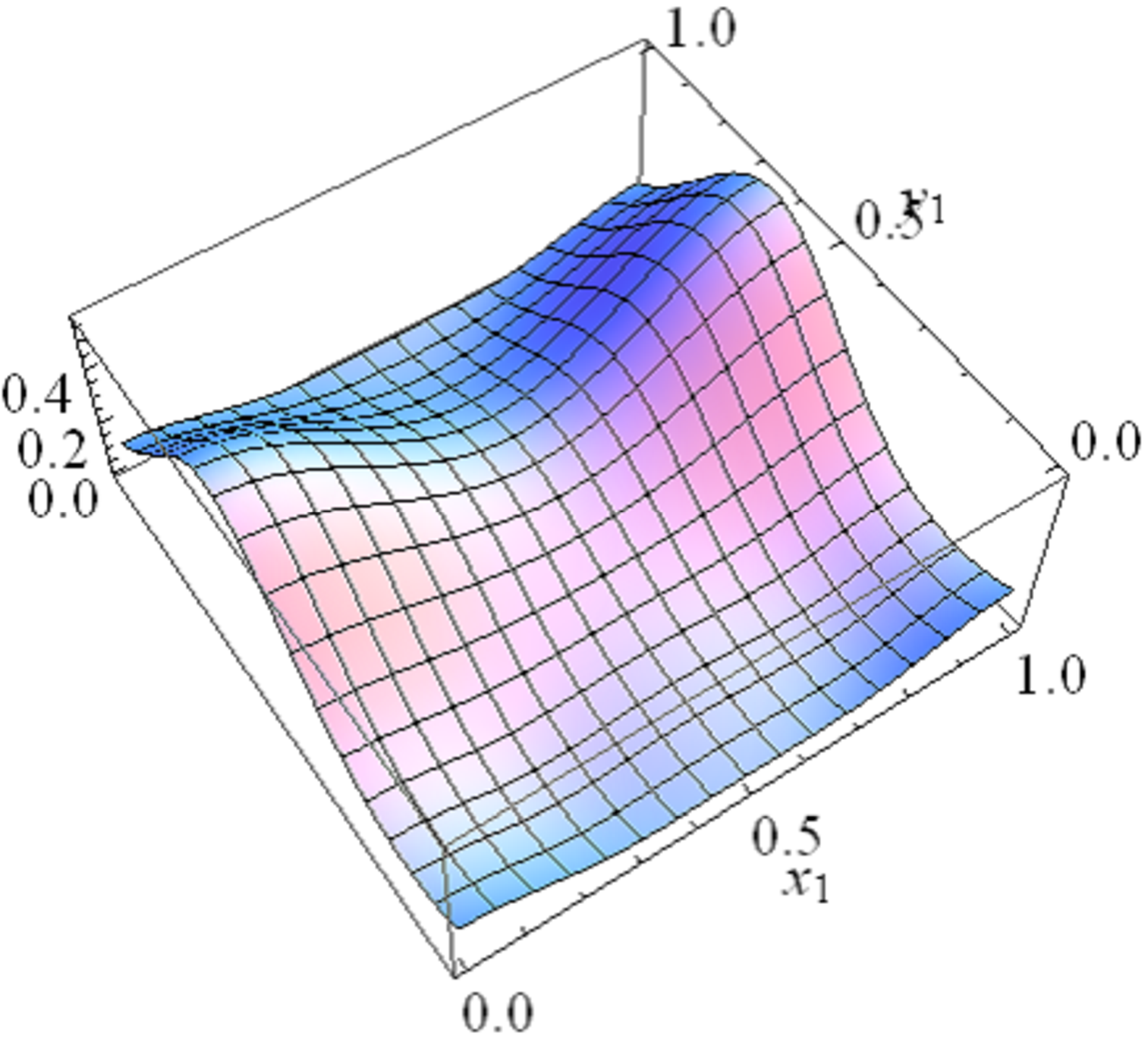}
\centering \bf \small (b) \,\, $\vec{i}=(1/3,1/3)$
\end{minipage}
\begin{minipage}{0.31\textwidth}
\includegraphics[width=0.99\textwidth]{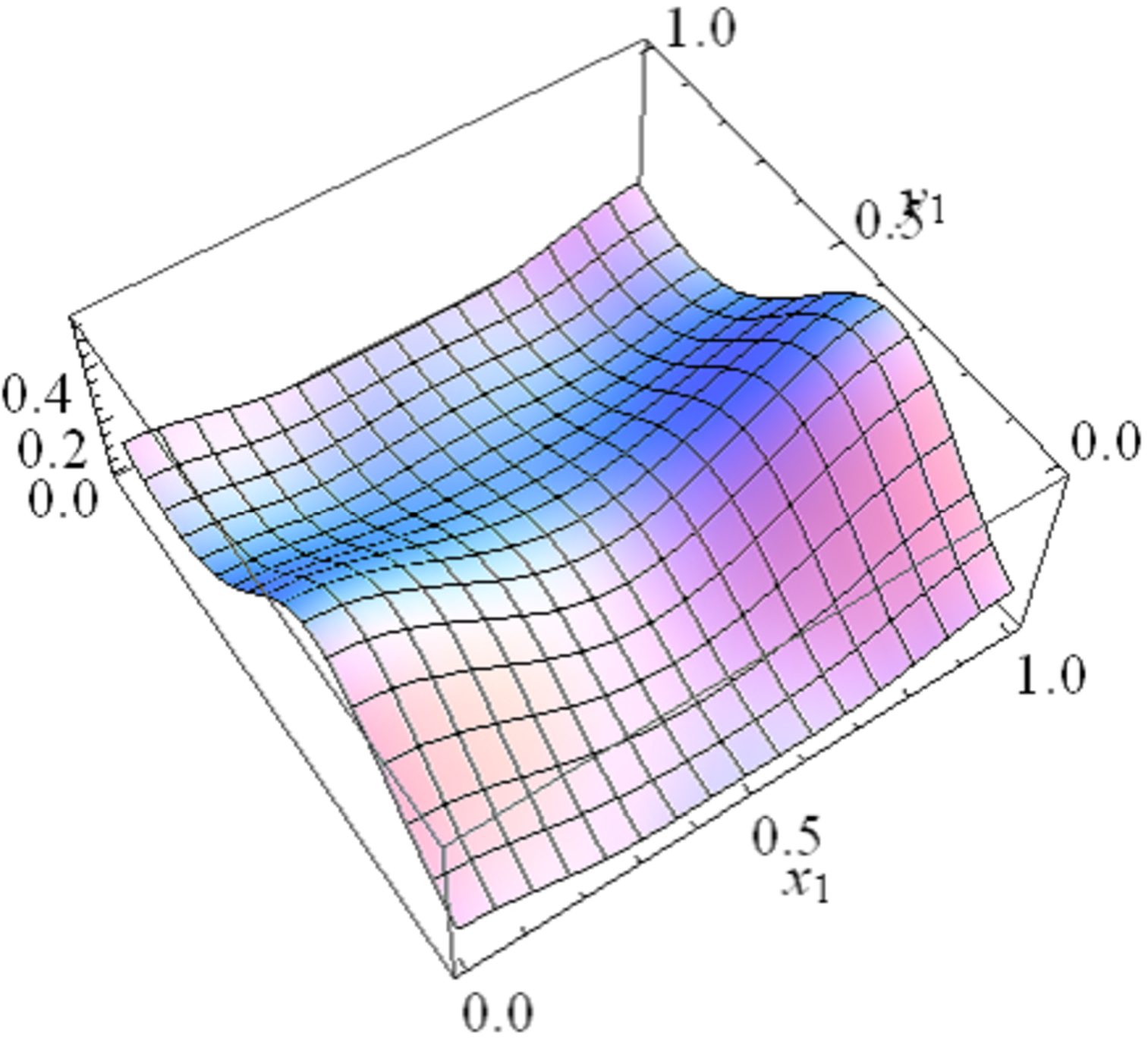}
\centering \bf \small (c) \,\, $\vec{i}=(2/3,2/3)$
\end{minipage}
\vspace{20pt}
\end{minipage}
\begin{minipage}{\textwidth}
\centering
\begin{minipage}{0.31\textwidth}
\includegraphics[width=0.99\textwidth]{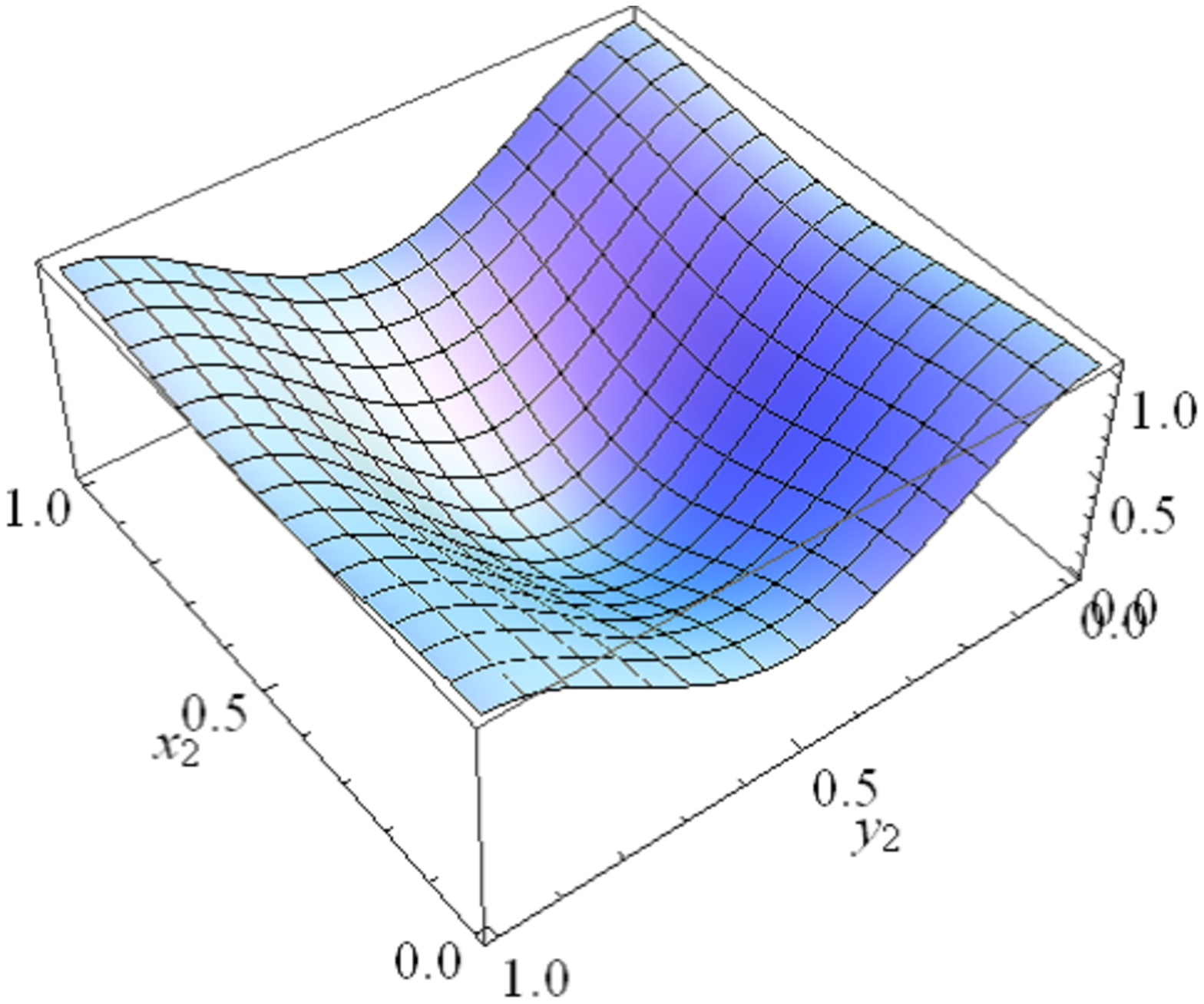}
\centering \bf \small (d) \,\, $\vec{i}=(0,0)$
\end{minipage}
\begin{minipage}{0.31\textwidth}
\includegraphics[width=0.99\textwidth]{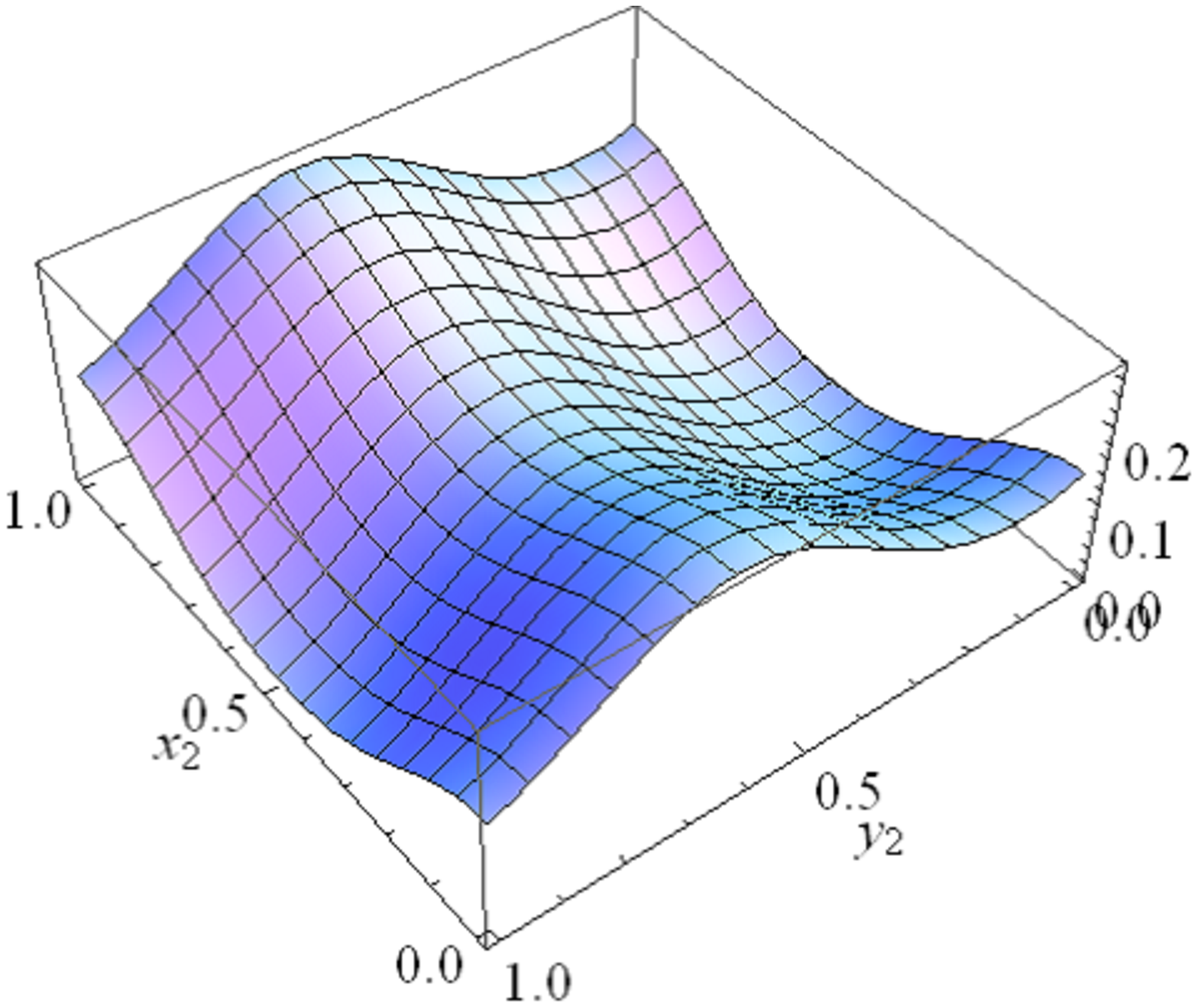}
\centering \bf \small (e) \,\, $\vec{i}=(1/3,1/3)$
\end{minipage}
\begin{minipage}{0.31\textwidth}
\includegraphics[width=0.99\textwidth]{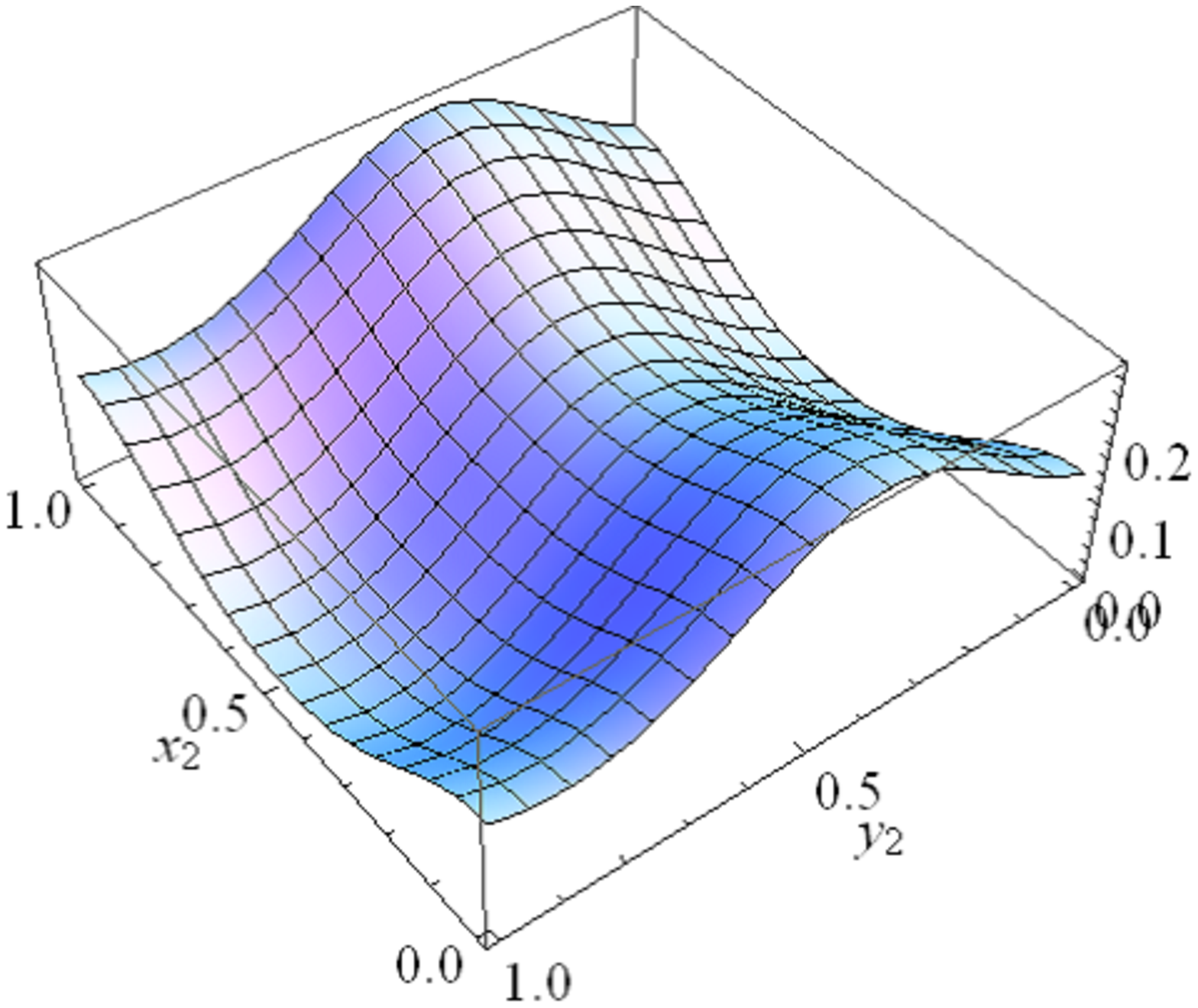}
\centering \bf \small (f) \,\, $\vec{i}=(2/3,2/3)$
\end{minipage}
\end{minipage}
\caption{The probability densities of zero-mode wavefunctions $\bigl|\Theta^{\vec{i}, \, \mathbb{N}}(\vec{z})\bigr|^2$ on $(x_1, y_1)$-plane for Type 2 are shown in (a), (b) and (c), while those on $(x_2, y_2)$-plane are depicted in (d), (e) and (f).}
\label{fig:profile_type2}
\end{figure}

In this example, in addition to the above $Z_3$ generator $C$ shown in Eq.~(\ref{C_nonfact}), there exists the $Z_3$ symmetry under the generator $Z$ defined by
\begin{gather}
Z=
\begin{pmatrix}
1 & 0 & 0\\
0 & \omega & 0\\
0 & 0 & \omega^2
\end{pmatrix}
\label{Z_nonfact},
\end{gather}
with $\omega=e^{2 \pi i /3}$.
Thus we can obtain non-Abelian discrete flavor symmetry, because these generators $C$ and $Z$ do not commute each other,
\begin{gather}
CZ=\omega ZC.
\end{gather}
Similarly to the argument in the previous subsection, the closed algebra of $Z$ and $C$ is $\Delta(27)$ with the generator of another $Z_3$ transformation,
\begin{gather}
Z'=
\begin{pmatrix}
\omega & 0 & 0\\
0 & \omega & 0\\
0 & 0 & \omega
\end{pmatrix}
.\label{Z'_nonfact}
\end{gather}
Thus, in the aligned case with $|\det\mathbb{N}|=3$ for all sectors, we have the possibility to obtain $\Delta(27)$ flavor symmetry in 4D effective theory from the magnetized model with non-factorizable fluxes.
Notice that, because the flux configuration yielding the same value of the determinant of flux matrices is not unique, various flavor structures are possible with such $\Delta(27)$ symmetry, due to the variety of the generation-types.
Yukawa couplings are written by the overlap integral of zero-modes on toroidal extra dimensions, as stated before.
The localization profiles of the zero-modes which govern the generation-types are determined by the configuration of magnetic fluxes.
We show the probability densities of zero-mode wavefunctions $\bigl|\Theta^{\vec{j}, \, \mathbb{N}}(\vec{z})\bigr|^2$ on each torus in Figure \ref{fig:profile_type1} and \ref{fig:profile_type2}, for two different configurations of magnetic fluxes.
Those figures imply that if the generation-type is different, the overlap integral of zero-modes on tori would be different, because the profiles of zero-modes in Type 1 are universal among three generations on the second torus $(x_2, y_2)$, while those in Type 2 are dependent on generations on the same torus $(x_2, y_2)$.

Furthermore, when the Wilson-lines are all vanishing in the present situation, we can define a $Z_2$ generator which acts as $\Theta^{\vec{i}_{ab}, \, \mathbb{N}_{ab}} \rightarrow \Theta^{\vec{e}-\vec{i}_{ab}, \, \mathbb{N}_{ab}}$.
The $Z_2$ generator is given by
\begin{gather}
P=
\begin{pmatrix}
1&0&0\\
0&0&1\\
0&1&0
\end{pmatrix}.\label{P_nonfact}
\end{gather}
If the intersection of three sets of labels $\{\vec{i}\}$ in each sector corresponds to the labels of any one from Type 1 to Type 4, the flavor symmetry is enhanced to $\Delta(54) \cong (Z_3 \times Z'_3) \rtimes S_3$ in 4D effective theory, as in the magnetized torus model with factorizable fluxes.

However, we remark that there does not always exist an invariance under the $Z_3$ transformation generated by $Z$.
Here and hereafter, we assume vanishing Wilson-lines in the expressions of Yukawa couplings.
We consider the flux configuration with $|\det \mathbb{N}_{ab}|=|\det\mathbb{N}_{ca}|=|\det\mathbb{N}_{bc}|=3$, e.g.,
\begin{gather}
\mathbb{N}_{ab}=
\begin{pmatrix}
-1 & -1\\
-3 & 0
\end{pmatrix}
, \qquad
\mathbb{N}_{ca} =
\begin{pmatrix}
5 & 2\\
4 & 1 
\end{pmatrix}
, \qquad
\mathbb{N}_{bc} =
\begin{pmatrix}
-4 & -1\\
-1 & -1
\end{pmatrix}.
\end{gather}
Then, generation-types are all of the Type 3, i.e.,
\begin{align}
\vec{i}_{ab,0} =
\begin{pmatrix}
0\\ 0
\end{pmatrix}, \qquad
\vec{i}_{ab,1}=
\begin{pmatrix}
1/3 \\ 2/3
\end{pmatrix}, \qquad
\vec{i}_{ab,2}=
\begin{pmatrix}
2/3 \\ 1/3
\end{pmatrix},
\end{align}
and the same holds for $ca$- and $bc$-sector.
The Yukawa couplings are written as
\begin{gather}
\lambda_{\vec{i}_{ab}\vec{i}_{ca}\vec{i}_{bc,0}}=
\begin{pmatrix}
\lambda_0 & \lambda_1 & \lambda_1\\
\lambda_2& \lambda_2 & \lambda_3\\
\lambda_2 & \lambda_3 & \lambda_2
\end{pmatrix}, \quad
\lambda_{\vec{i}_{ab}\vec{i}_{ca}\vec{i}_{bc,1}}=
\begin{pmatrix}
\lambda_2 & \lambda_3 & \lambda_2\\
\lambda_3& \lambda_2 & \lambda_2\\
\lambda_1 & \lambda_1 & \lambda_0
\end{pmatrix}, \quad
\lambda_{\vec{i}_{ab}\vec{i}_{ca}\vec{i}_{bc,2}}=
\begin{pmatrix}
\lambda_2 & \lambda_2 & \lambda_3\\
\lambda_1& \lambda_0 & \lambda_1\\
\lambda_3 & \lambda_2 & \lambda_2
\end{pmatrix},
\end{gather}
where values of $\lambda_0$, $\lambda_1$, $\lambda_2$ and $\lambda_3$ are different from each other.
In the above example, Yukawa couplings are not symmetric under the generator $Z$ and the flavor symmetry is $(Z_3^{(C)} \rtimes Z_2) \times Z_3' \cong S_3 \times Z_3'$ ($Z_3^{(C)} \times Z_3'$ in the case with non-vanishing Wilson-lines).
We give another example, which corresponds to $|\det\mathbb N_{ab}|=|\det\mathbb N_{ca}|=3$ and $|\det\mathbb N_{bc}|=6$, but does not have the $Z_3$ invariance under $Z$ generator (\ref{Z_nonfact}).
The magnetic fluxes
\begin{gather}
\mathbb{N}_{ab}=
\begin{pmatrix}
3 &0\\
1 & -1
\end{pmatrix}
, \qquad
\mathbb{N}_{ca} =
\begin{pmatrix}
1 & 2\\
4 & 5 
\end{pmatrix}
, \qquad
\mathbb{N}_{bc} =
\begin{pmatrix}
-4 & -2\\
-5 & -4
\end{pmatrix},\label{eg:s3_336}
\end{gather}
yield the same generation-types as those in the previous example (\ref{eg:delta_336}).
Then, the Yukawa couplings are written by
\begin{gather}
\lambda_{\vec{i}_{ab}\vec{i}_{ca}\vec{i}_{bc,0}}=\lambda_{\vec{i}_{ab}\vec{i}_{ca}\vec{i}_{bc,1}}=
\begin{pmatrix}
\lambda_0 & \lambda_1 & \lambda_1\\
\lambda_2& \lambda_1 & \lambda_3\\
\lambda_2 & \lambda_3 & \lambda_3
\end{pmatrix}, \quad
\lambda_{\vec{i}_{ab}\vec{i}_{ca}\vec{i}_{bc,2}}=\lambda_{\vec{i}_{ab}\vec{i}_{ca}\vec{i}_{bc,3}}=
\begin{pmatrix}
\lambda_1 & \lambda_3 & \lambda_2\\
\lambda_3& \lambda_1 & \lambda_2\\
\lambda_3 & \lambda_1 & \lambda_0
\end{pmatrix},\\
\lambda_{\vec{i}_{ab}\vec{i}_{ca}\vec{i}_{ca,4}}=\lambda_{\vec{i}_{ab}\vec{i}_{ca}\vec{i}_{bc,5}}=
\begin{pmatrix}
\lambda_1 & \lambda_2 & \lambda_3\\
\lambda_1& \lambda_0 & \lambda_1\\
\lambda_3 & \lambda_2 & \lambda_1
\end{pmatrix}.
\end{gather}
Such Yukawa couplings have $(Z_3^{(C)} \rtimes Z_2) \times Z_3' \cong S_3 \times Z_3'$ symmetry.
Therefore, the same degeneracies and generation-types of zero-modes do not always yield the same flavor symmetry.
Accordingly, we investigate the flavor symmetries for the other flux configurations as systematically as possible, which is shown in Appendix \ref{sec:egs}.

Next, we extend the above argument and consider the case with $\textrm{gcd}(\det\mathbb{N}_{ab}, \det\mathbb{N}_{ca}, \det\mathbb{N}_{bc})=g >3$.
We substitute the following generators in the representation of a $g \times g$ matrix, 
\begin{gather}
Z=
\begin{pmatrix}
1 &&&&\\
&\rho&&&\\
&&\rho^2&&\\
&&&\ddots&\\
&&&&\rho^{g-1}
\end{pmatrix}
,\label{Zg_nonfact}
\end{gather}
with $\rho \equiv e^{2\pi i /g}$, and
\begin{gather}
C=
\begin{pmatrix}
0 & 1 & 0 & \cdots & 0\\
0 & 0 & 1 & \cdots & 0\\
&&& \ddots &\\
1 & 0 & 0 & \cdots & 0
\end{pmatrix}
,
\end{gather}
those are the generalizations of Eqs.~(\ref{Z_nonfact}) and (\ref{C_nonfact}).
Then, $Z_g^{(C)}$ or $(Z_g \times Z_g') \rtimes Z_g^{(C)}$ can be realized as the flavor symmetry in 4D effective theory.\footnote{There is an additional $Z_3$ symmetry ($Z_3'$) which is allowed due to the fact that Yukawa couplings are three-point couplings.}
Similarly to the case with $g=3$, when the Wilson-lines are all vanishing, we can obtain $(Z_g^{(C)} \rtimes Z_2) \cong D_g$ or $(Z_g \times Z_g') \rtimes D_g$ flavor symmetry enhanced by the $Z_2$ generator which is the generalization of $P$.
For $g=4$, we can obtain $D_4$ flavor symmetry for which some examples are shown in Appendix \ref{sec:4gene}.

So far we have shown the flavor symmetry obtained from the magnetic fluxes yielding aligned generation-types, by identifying the explicit forms of its generators.
Now we discuss about the representations realized under this symmetry.
The most typical example is the representation under the $\Delta(27)$ flavor symmetry for $g=3$.\footnote{The irreducible representation under the $\Delta(54)$ symmetry is equivalent to Table 4 in Ref.~\cite{Abe:2009vi}.}
We focus on a single sector and then omit the YM indices like $ab$.
We first consider a sector where three zero-modes are generated by $|\det\mathbb{N}|=3$ and label them with $\{\vec{i}\} \equiv \{\vec{i}_0, \vec{i}_1, \vec{i}_2\}$.
These three generations of zero-modes are represented as
\begin{gather}
| \Theta^3 \rangle_1 =
\begin{pmatrix}
\Theta^{\vec{i}_0, \, \mathbb{N}}\\
\Theta^{\vec{i}_1, \, \mathbb{N}}\\
\Theta^{\vec{i}_2, \, \mathbb{N}}
\end{pmatrix}
,\label{nonfact_3}
\end{gather}
that is identified with the triplet representation $\bm{3}$ under $\Delta(27)$.

Next, we consider a sector where six zero-modes are generated by $|\det\mathbb{N}|=6$ and label them with $\{\vec{i}\} \equiv \{\vec{i}_0, \vec{i}_1, \ldots, \vec{i}_5\}$.
We can decompose these six zero-modes into two triplet representations,
\begin{gather}
| \Theta^6 \rangle_1 =
\begin{pmatrix}
\Theta^{\vec{i}_0, \, \mathbb{N}}\\
\Theta^{\vec{i}_2, \, \mathbb{N}}\\
\Theta^{\vec{i}_4, \, \mathbb{N}}
\end{pmatrix}
, \qquad | \Theta^6 \rangle_2 =
\begin{pmatrix}
\Theta^{\vec{i}_3, \, \mathbb{N}}\\
\Theta^{\vec{i}_5, \, \mathbb{N}}\\
\Theta^{\vec{i}_1, \, \mathbb{N}}
\end{pmatrix}
.
\end{gather}
The generator $C$ of the $Z_3$ transformation $Z_3^{(C)}$ acts as $C| \Theta^3 \rangle_1$ for $|\det\mathbb{N}|=3$ and $C| \Theta^6 \rangle_i \,\, (i=1,2)$ for $|\det\mathbb{N}|=6$.
On the other hand, the representations $| \Theta^6 \rangle_i \,\, (i=1,2)$ behave as the complex conjugate to the triplet representation $| \Theta^3 \rangle_1$.
Accordingly, both $| \Theta^6 \rangle_i \,\, (i=1,2)$ are $\bar{\bm{3}}$ representation under $\Delta(27)$.

We further mention about a sector where nine zero-modes are generated by $|\det\mathbb{N}|=9$ and label them with $\{\vec{i}\} \equiv \{\vec{i}_0, \vec{i}_1, \ldots, \vec{i}_8\}$.
Also in this case, we decompose these nine zero-modes into three triplet representations,
\begin{gather}
| \Theta^9 \rangle_1 =
\begin{pmatrix}
\Theta^{\vec{i}_0, \, \mathbb{N}}\\
\Theta^{\vec{i}_3, \, \mathbb{N}}\\
\Theta^{\vec{i}_6, \, \mathbb{N}}
\end{pmatrix}
, \qquad | \Theta^9 \rangle_{\omega} =
\begin{pmatrix}
\Theta^{\vec{i}_1, \, \mathbb{N}}\\
\Theta^{\vec{i}_4, \, \mathbb{N}}\\
\Theta^{\vec{i}_7, \, \mathbb{N}}
\end{pmatrix}
, \qquad | \Theta^9 \rangle_{\omega^2} =
\begin{pmatrix}
\Theta^{\vec{i}_2, \, \mathbb{N}}\\
\Theta^{\vec{i}_5, \, \mathbb{N}}\\
\Theta^{\vec{i}_8, \, \mathbb{N}}
\end{pmatrix}
,
\end{gather}
where $\omega \equiv e^{2\pi i /3}$.
Note that these triplet representations are reducible.
The triplets $| \Theta^9 \rangle_{\omega^n}$ have $Z_3$ charges $n$, and they are decomposed into nine singlets, which are expressed as
\begin{gather}
\bm{1}_{\omega^n, \, \omega^m} : \quad \Theta^{\vec{i}_n, \, \mathbb{N}}+\omega^m \Theta^{\vec{i}_{n+3m}, \, \mathbb{N}} + \omega^{2m} \Theta^{\vec{i}_{n+6m}, \, \mathbb{N}},
\end{gather}
up to normalization factors.
We find that no new representation other than the above three appears for $|\det\mathbb{N}|>9$, because these three appear repeatedly, as shown in Table \ref{tab:representation}.
Table \ref{tab:representation} is exactly the same as that in Ref.~\cite{Abe:2009vi} if we replace $M$ with $\det\mathbb{N}$.

\begin{table}[h]
\centering
\begin{tabular}{cc} \hline
$|\det\mathbb{N}|$ & Representation under $\Delta(27)$\\ \hline
$3$ & $\bm{3}$\\
$6$ & $2 \times \bar{\bm{3}}$\\
$9$ & $\bm{1}_1, \bm{1}_2, \bm{1}_3, \bm{1}_4, \bm{1}_5, \bm{1}_6, \bm{1}_7, \bm{1}_8, \bm{1}_9$\\
$12$ & $4 \times \bm{3}$\\
$15$ & $5 \times \bar{\bm{3}}$\\
$18$ & $2 \times \{\bm{1}_1, \bm{1}_2, \bm{1}_3, \bm{1}_4, \bm{1}_5, \bm{1}_6, \bm{1}_7, \bm{1}_8, \bm{1}_9 \}$\\ \hline
\end{tabular}
\caption{Examples of $\Delta(27)$ representations consisting of the zero-modes for $g=3$.}
\label{tab:representation}
\end{table}

In the remainder of this subsection, we explain irreducible representations under $Z_3^{(C)} \rtimes Z_2 \cong S_3$ constructed by zero-mode wavefunctions.
It is known that the irreducible representations under $S_3$ are two singlets $\bm{1}, \bm{1}'$ and single doublet $\bm{2}$.
Since the triplet (\ref{nonfact_3}) for $|\det\mathbb{N}|=3$ is reducible representations under $S_3$, it is decomposed a singlet
\begin{gather}
\bm{1} : \quad \Theta^{\vec{i}_0, \, \mathbb{N}} + \Theta^{\vec{i}_1, \, \mathbb{N}} + \Theta^{\vec{i}_2, \, \mathbb{N}},
\end{gather}
and a doublet
\begin{gather}
\bm{2} : \quad
\begin{pmatrix}
\Theta^{\vec{i}_2, \, \mathbb{N}}-\Theta^{\vec{i}_0, \, \mathbb{N}}\\
\Theta^{\vec{i}_1, \, \mathbb{N}}-\Theta^{\vec{i}_0, \, \mathbb{N}}
\end{pmatrix}.
\end{gather}
On the other hand, the irreducible representations extracted from six zero-modes for $|\det \mathbb{N}|=6$ are found as follows.
With the flux matrix (\ref{eg:delta_336}), the $Z_2$ generator $P$ is written as
\begin{gather}
P=
\begin{pmatrix}
1&0&0&0&0&0\\
0&1&0&0&0&0\\
0&0&0&0&1&0\\
0&0&0&0&0&1\\
0&0&1&0&0&0\\
0&0&0&1&0&0
\end{pmatrix},
\end{gather}
on the basis
\begin{gather}
|\Theta^6\rangle =
\begin{pmatrix}
\Theta^{\vec{i}_0, \, \mathbb{N}}\\
\Theta^{\vec{i}_1, \, \mathbb{N}}\\
\Theta^{\vec{i}_2, \, \mathbb{N}}\\
\Theta^{\vec{i}_3, \, \mathbb{N}}\\
\Theta^{\vec{i}_4, \, \mathbb{N}}\\
\Theta^{\vec{i}_5, \, \mathbb{N}}
\end{pmatrix}.\label{sextet}
\end{gather}
This sextet (\ref{sextet}) is a reducible representation and it is decomposed into two singlets
\begin{gather}
\bm{1} : \quad \Theta^{\vec{i}_0, \, \mathbb{N}}+\Theta^{\vec{i}_2, \, \mathbb{N}}+\Theta^{\vec{i}_4, \, \mathbb{N}}, \,\, \Theta^{\vec{i}_1, \, \mathbb{N}}+\Theta^{\vec{i}_3, \, \mathbb{N}}+\Theta^{\vec{i}_5, \, \mathbb{N}},
\end{gather}
and two doublets
\begin{gather}
\bm{2} : \quad
\begin{pmatrix}
\Theta^{\vec{i}_4, \, \mathbb{N}}-\Theta^{\vec{i}_0, \, \mathbb{N}}\\
\Theta^{\vec{i}_2, \, \mathbb{N}}-\Theta^{\vec{i}_0, \, \mathbb{N}}
\end{pmatrix}, \,\,
\begin{pmatrix}
\Theta^{\vec{i}_5, \, \mathbb{N}}-\Theta^{\vec{i}_1, \, \mathbb{N}}\\
\Theta^{\vec{i}_3, \, \mathbb{N}}-\Theta^{\vec{i}_1, \, \mathbb{N}}
\end{pmatrix}.
\end{gather}
Even for $|\det\mathbb{N}|>6$, we can not obtain the remaining irreducible representation, i.e., singlet $\bm{1}'$.
Singlet $\bm{1}$ and doublet $\bm{2}$ appear repeatedly for $|\det\mathbb{N}|>6$, as shown in Table \ref{tab:representation2'}.

\begin{table}[h]
\centering
\begin{tabular}{cc} \hline
$|\det\mathbb{N}|$ & Representation under $S_3$\\ \hline
$3$ & $\bm{1}$, $\bm{2}$\\
$6$ & $2 \times \{\bm{1}, \bm{2}\}$\\
$9$ & $3 \times \{\bm{1}, \bm{2}\}$\\
$12$ & $4 \times \{\bm{1}, \bm{2}\}$\\ \hline
\end{tabular}
\caption{Examples of irreducible representations under $S_3 \cong Z_3^{(C)} \rtimes Z_2$.}
\label{tab:representation2'}
\end{table}

\subsection{Magnetized torus model with non-factorizable fluxes : not-aligned generation-types}
In this subsection, we study the case that generation-types in three sectors are not aligned.
Indeed, since there is no systematic way in general to identify the charges under the $Z_3$ transformations, we explain the flavor symmetry by means of concrete examples.
It is still interesting to consider the three-generation model of quarks and leptons, and we focus on the case with $\textrm{gcd}(\det\mathbb{N}_{ab}, \det\mathbb{N}_{ca}, \det\mathbb{N}_{bc})=3$ in this paper.

First, we focus on the case with $|\det\mathbb{N}|=3$ for each sector and non-vanishing Wilson-lines, e.g.,
\begin{gather}
\mathbb{N}_{ab} =
\begin{pmatrix}
1&-1\\
-3&0
\end{pmatrix}
, \qquad
\mathbb{N}_{ca} =
\begin{pmatrix}
0&1\\
3&3
\end{pmatrix}
, \qquad
\mathbb{N}_{bc} =
\begin{pmatrix}
-1&0\\
0&-3
\end{pmatrix}
.
\end{gather}
Then, a generation-type for $ab$-sector are identified as Type 2, for $ca$-sector Type 1 and for $bc$-sector Type 4.
We label the localization points of zero-modes as
\begin{align}
\vec{i}_{ab,0}=
\begin{pmatrix}
0\\ 0
\end{pmatrix}, \qquad
\vec{i}_{ab,1}&=
\begin{pmatrix}
1/3\\ 1/3
\end{pmatrix}, \qquad
\vec{i}_{ab,2}=
\begin{pmatrix}
2/3\\ 2/3
\end{pmatrix}, \qquad\\
\vec{i}_{ca,0}=
\begin{pmatrix}
0\\ 0
\end{pmatrix}, \qquad
\vec{i}_{ca,1}&=
\begin{pmatrix}
1/3\\ 0
\end{pmatrix}, \qquad
\vec{i}_{ca,2}=
\begin{pmatrix}
2/3\\ 0
\end{pmatrix}, \qquad\\
\vec{i}_{bc,0}=
\begin{pmatrix}
0\\ 0
\end{pmatrix}, \qquad
\vec{i}_{bc,1}&=
\begin{pmatrix}
0\\ 1/3
\end{pmatrix}, \qquad
\vec{i}_{bc,2}=
\begin{pmatrix}
0\\ 2/3
\end{pmatrix}.
\end{align}
Yukawa couplings $\lambda_{\vec{i}_{ab}\vec{i}_{ca}\vec{i}_{bc}}$ are written as
\begin{gather}
\lambda_{\vec{i}_{ab}\vec{i}_{ca}\vec{i}_{bc,0}} =
\begin{pmatrix}
\lambda_0&0&0\\
0&\lambda_1&0\\
0&0&\lambda_1
\end{pmatrix}, \quad
\lambda_{\vec{i}_{ab}\vec{i}_{ca}\vec{i}_{bc,1}} =
\begin{pmatrix}
0&0&\lambda_2\\
\lambda_3&0&0\\
0&\lambda_3&0
\end{pmatrix}, \quad
\lambda_{\vec{i}_{ab}\vec{i}_{ca}\vec{i}_{bc,2}} =
\begin{pmatrix}
0&\lambda_2&0\\
0&0&\lambda_3\\
\lambda_3&0&0
\end{pmatrix},
\end{gather}
where values of $\lambda_0, \lambda_1, \lambda_2$ and $\lambda_3 $ are different from each other.
We find a $Z_3$ symmetry, the generator of which is $Z$ defined in Eq.~(\ref{Z_nonfact}).
Thus, $Z_3 \times Z_3'$ symmetry is realized in 4D effective theory.
Note that there does not exist an invariance under the $Z_3$ generator $C$ in the case that generation-types for each gauge sector are not uniformly aligned.
If Wilson-lines are turned off, we can find an invariance under the $Z_2$ generator (\ref{P_nonfact}).
The discrete group generated by $Z, Z'$ and $P$ is $(Z_3 \rtimes Z_2) \times Z_3' \cong S_3 \times Z_3'$.

In the remainder of this subsection, we consider irreducible representations under $Z_3 \rtimes Z_2 \cong S_3$ constructed by zero-mode wavefunctions, those are two singlets $\bm{1}, \bm{1}'$ and single doublet $\bm{2}$.
Since the triplet (\ref{nonfact_3}) for $|\det\mathbb{N}|=3$ is a reducible representation, we decompose it into a singlet
\begin{gather}
\bm{1} : \quad \Theta^{\vec{i}_0, \, \mathbb{N}},
\end{gather}
and a doublet
\begin{gather}
\bm{2} : \quad
\begin{pmatrix}
\Theta^{\vec{i}_1, \, \mathbb{N}}\\
\Theta^{\vec{i}_2, \, \mathbb{N}}
\end{pmatrix}.
\end{gather}
In order to find the representations for $|\det\mathbb{N}|=6$, we consider the flux configuration as
\begin{gather}
\mathbb{N}_{ab}=
\begin{pmatrix}
2&1\\
3&0
\end{pmatrix}, \qquad
\mathbb{N}_{ca}=
\begin{pmatrix}
0&-1\\
-3&3
\end{pmatrix}, \qquad
\mathbb{N}_{bc}=
\begin{pmatrix}
-2&0\\
0&-3
\end{pmatrix},
\end{gather}
where $\det\mathbb{N}_{bc}=6$ and
\begin{align}
\vec{i}_0=
\begin{pmatrix}
0\\ 0
\end{pmatrix}, \qquad
\vec{i}_1&=
\begin{pmatrix}
1/2\\ 0
\end{pmatrix}, \qquad
\vec{i}_2=
\begin{pmatrix}
0\\ 1/3
\end{pmatrix}, \qquad\\
\vec{i}_3=
\begin{pmatrix}
1/2\\ 1/3
\end{pmatrix}, \qquad
\vec{i}_4&=
\begin{pmatrix}
0\\ 2/3
\end{pmatrix}, \qquad
\vec{i}_5=
\begin{pmatrix}
1/2\\ 2/3
\end{pmatrix}
\end{align}
for $bc$-sector.
Then zero-modes in $ab$- and $ca$-sector with $\det\mathbb{N}_{ab} = \det\mathbb{N}_{ca}=-3$, can be decomposed into singlets $\bm{1}$ and doublets $\bm{2}$.
The irreducible representations constructed by six zero-modes in $bc$-sector are found as follows.
In the present case, the $Z_2$ generator $P$ is expressed as
\begin{gather}
P=
\begin{pmatrix}
1&0&0&0&0&0\\
0&1&0&0&0&0\\
0&0&0&0&1&0\\
0&0&0&0&0&1\\
0&0&1&0&0&0\\
0&0&0&1&0&0
\end{pmatrix},
\end{gather}
on the basis
\begin{gather}
|\Theta^6\rangle =
\begin{pmatrix}
\Theta^{\vec{i}_0, \, \mathbb{N}}\\
\Theta^{\vec{i}_1, \, \mathbb{N}}\\
\Theta^{\vec{i}_2, \, \mathbb{N}}\\
\Theta^{\vec{i}_3, \, \mathbb{N}}\\
\Theta^{\vec{i}_4, \, \mathbb{N}}\\
\Theta^{\vec{i}_5, \, \mathbb{N}}
\end{pmatrix},
\end{gather}
though this sextet is decomposed into two singlets
\begin{gather}
\bm{1} : \quad \Theta^{\vec{i}_0, \, \mathbb{N}}, \,\, \Theta^{\vec{i}_1, \, \mathbb{N}},
\end{gather}
and two doublets
\begin{gather}
\bm{2} : \quad
\begin{pmatrix}
\Theta^{\vec{i}_2, \, \mathbb{N}}\\
\Theta^{\vec{i}_4, \, \mathbb{N}}
\end{pmatrix}, \,\,
\begin{pmatrix}
\Theta^{\vec{i}_3, \, \mathbb{N}}\\
\Theta^{\vec{i}_5, \, \mathbb{N}}
\end{pmatrix}.
\end{gather}
For $|\det\mathbb{N}|>6$, we can not obtain the other irreducible representation, i.e., the remaining singlet $\bm{1}'$.
The above representations $\bm{1}$ and $\bm{2}$ appear repeatedly.
We summarize the representations for $|\det\mathbb{N}|>6$ in Table \ref{tab:representation2}.

\begin{table}[h]
\centering
\begin{tabular}{cc} \hline
$|\det\mathbb{N}|$ & Representation under $S_3$\\ \hline
$3$ & $\bm{1}$, $\bm{2}$\\
$6$ & $2 \times \{\bm{1}, \bm{2}\}$\\
$9$ & $3 \times \{\bm{1}, \bm{2}\}$\\
$12$ & $4 \times \{\bm{1}, \bm{2}\}$\\ \hline
\end{tabular}
\caption{Examples of irreducible representations under $S_3 \cong Z_3 \rtimes Z_2$.}
\label{tab:representation2}
\end{table}

\subsection{Magnetized orbifold model with non-factorizable fluxes}
In this section, we study the flavor symmetry realized from the magnetized orbifold model with non-factorizable fluxes.
This model is obtained after the $Z_2$ projection by which the symmetry $(Z_g \times Z_g') \rtimes Z_g^{(C)}$ before orbifolding is broken into its subgroup.

First, we study an illustrating model, the model with $g=4$, where we set $\det\mathbb{N}_{ab}= \det\mathbb{N}_{ca}=-4$ and $\det\mathbb{N}_{bc}=8$.
In such a model, the flavor symmetry is $(Z_4 \times Z_4') \rtimes Z_4^{(C)}$ before the $Z_2$ projection.
We consider the zero-modes for $\det\mathbb{N}_{ab}=-4$ where three $Z_2$-even zero-modes survive, while the $Z_2$-odd zero-mode is projected out.
The basis of these even modes is written as
\begin{gather}
| \Theta^{4}_\textrm{even} \rangle =
\begin{pmatrix}
\Theta^{\vec{i}_0, \, \mathbb{N}}\\
\Theta^{\vec{i}_1, \, \mathbb{N}}+\Theta^{\vec{i}_3, \, \mathbb{N}}\\
\Theta^{\vec{i}_2, \, \mathbb{N}}
\end{pmatrix}
.
\end{gather}
The same holds for the zero-modes with $\det\mathbb{N}_{ca}=-4$.
For such a basis, we can define the generator of a $Z_4$ transformation as
\begin{gather}
Z=
\begin{pmatrix}
i & 0 & 0\\
0 & -1 & 0\\
0 & 0 & -i
\end{pmatrix}
,
\end{gather}
which is equivalent to the generator (\ref{Zg_nonfact}).
In addition to this operator, we can also define the generator of a cyclic permutation as
\begin{gather}
C=
\begin{pmatrix}
0 & 0 & 1\\
0 & 1 & 0\\
1 & 0 & 0
\end{pmatrix}
.
\end{gather}
The closed algebra for these generators is $D_4$.
This implies that $(Z_4 \times Z_4') \rtimes Z_4^{(C)}$ breaks into its subgroup $D_4$.
Notice that $D_4$ is the symmetry for interchanging $\Theta^{\vec{i}_0, \, \mathbb{N}}$ with $\Theta^{\vec{i}_2, \, \mathbb{N}}$.
Thus, for $\textrm{gcd}(\det\mathbb{N}_{ab}, \det\mathbb{N}_{ca}, \det\mathbb{N}_{bc})=2k \,\, (k : \mbox{integer})$, the flavor symmetry $D_4$ is invariably realized in 4D effective theory.
Such a result is quite similar to the one stated in Ref.~\cite{Marchesano:2013ega} in type IIA intersecting brane models on $T^6/Z_2$ or $T^6/Z_2 \times Z'_2$ orbifolds.

As stated in the previous section, there exist the exceptions on the magnetized orbifold model with non-factorizable fluxes.
We consider the case with $\textrm{gcd}(\det\mathbb{N}_{ab}, \det\mathbb{N}_{ca}, \det\mathbb{N}_{bc})=4$ and the same generation-types (\ref{exception}) at least for the two of three matters.
Then we have $(Z_4 \times Z_4') \rtimes Z_4^{(C)}$ as the flavor symmetry before the $Z_2$ projection.
After the projection, we obtain $(Z_4 \times Z_4') \rtimes Z_4^{(C)}$ in 4D effective theory.
Namely, the orbifold projection does not affect the flavor structure for such a setup.

\section{Non-Abelian discrete flavor symmetry from gauge symmetry breaking} \label{sec:gaugesymmetry}
In Ref.~\cite{Abe:2009uz}, it was mentioned that the non-Abelian discrete flavor symmetry is originated from the remnant of gauge symmetries, and recently the method to study the discrete flavor symmetry has been developed in Refs.~\cite{BerasaluceGonzalez:2011wy, Marchesano:2013ega}.
In this section, we study the appearance of the non-Abelian discrete flavor symmetry in the magnetized torus model with non-factorizable fluxes, by using the method proposed in Ref.~\cite{BerasaluceGonzalez:2011wy}.
They restrict their analysis to the magnetized torus model with factorizable fluxes in Ref.~\cite{BerasaluceGonzalez:2011wy} and to the magnetized orbifold model in Ref.~\cite{Marchesano:2013ega}.
We apply their method to the magnetized torus model with non-factorizable fluxes in this section.

In the following, we assume the complex structures as $\tau_1=\tau_2=i$, without loss of generality.
It is due to the fact that the values of the complex structure parameters do not affect the flavor symmetry in 4D effective field theory.
In addition, we also assume vanishing Wilson-lines, i.e., $\bar{\zeta}_i = 0 \,\, (i=1,2)$ for simplicity.
We consider $T^2 \times T^2$ toroidal compactifications with two $U(1)$ gauge field backgrounds,
\begin{align}
A_1 &= \pi M^{(1)} \, \textrm{Im} \, (\bar{z}_1 dz_1) + \pi M^{(12)} \, \textrm{Im} \, (\bar{z}_2 dz_2),\\
A_2 &= \pi M^{(21)} \, \textrm{Im} \, (\bar{z}_1 dz_1) + \pi M^{(2)} \, \textrm{Im} \, (\bar{z}_2 dz_2),
\end{align}
in differential forms, so that
\begin{align}
F_1 &= 2\pi M^{(1)} dx_1 \wedge dy_1 + 2\pi M^{(12)} dx_2 \wedge dy_2,\\
F_2 &= 2\pi M^{(21)} dx_1 \wedge dy_1 + 2\pi M^{(2)} dx_2 \wedge dy_2.
\end{align}
The above expressions are the straightforward extensions of those appearing in Refs.~\cite{Cremades:2004wa, BerasaluceGonzalez:2011wy}.
For vanishing $F_i \,\, (i=1,2)$, the model possesses the translational invariances generated by $\partial_{x_1}$ and $\partial_{y_1}$ on the first torus and by $\partial_{x_2}$ and $\partial_{y_2}$ on the second torus.
For non-vanishing $F_i \,\, (i=1,2)$, the model no longer has such invariances, because gauge fields $A_i \,\, (i=1,2)$ depend explicitly on the coordinates $x_i$, $y_i \,\, (i=1,2)$,
\begin{align}
A_1(x_1+\lambda, y_1, x_2, y_2) &= A_1(x_1, y_1, x_2, y_2) + \lambda \chi^{(1)}_{x_1}, \qquad \chi^{(1)}_{x_1} = \pi M^{(1)}y_1,\\
A_1(x_1, y_1+\lambda, x_2, y_2) &= A_1(x_1, y_1, x_2, y_2) + \lambda \chi^{(1)}_{y_1}, \qquad \chi^{(1)}_{y_1} = -\pi M^{(1)}x_1,\\
A_1(x_1, y_1, x_2+\lambda, y_2) &= A_1(x_1, y_1, x_2, y_2) + \lambda \chi^{(1)}_{x_2}, \qquad \chi^{(1)}_{x_2} = \pi M^{(12)}y_2,\\
A_1(x_1, y_1, x_2, y_2+\lambda) &= A_1(x_1, y_1, x_2, y_2) + \lambda \chi^{(1)}_{y_2}, \qquad \chi^{(1)}_{y_2} = -\pi M^{(12)}x_2,
\end{align}
and similarly
\begin{align}
A_2(x_1+\lambda, y_1, x_2, y_2) &= A_2(x_1, y_1, x_2, y_2) + \lambda \chi^{(2)}_{x_1}, \qquad \chi^{(2)}_{x_1} = \pi M^{(21)}y_1,\\
A_2(x_1, y_1+\lambda, x_2, y_2) &= A_2(x_1, y_1, x_2, y_2) + \lambda \chi^{(2)}_{y_1}, \qquad \chi^{(2)}_{y_1} = -\pi M^{(21)}x_1,\\
A_2(x_1, y_1, x_2+\lambda, y_2) &= A_2(x_1, y_1, x_2, y_2) + \lambda \chi^{(2)}_{x_2}, \qquad \chi^{(2)}_{x_2} = \pi M^{(2)}y_2,\\
A_2(x_1, y_1, x_2, y_2+\lambda) &= A_2(x_1, y_1, x_2, y_2) + \lambda \chi^{(2)}_{y_2}, \qquad \chi^{(2)}_{y_2} = -\pi M^{(2)}x_2.
\end{align}
In order to preserve the action unchanged, we need to perform gauge transformations that compensate the changes in $A_i \,\, (i=1,2)$.
That is, we perform the following operations for a wavefunction of charge $q$, 
\begin{align}
\psi(x_1, y_1, x_2, y_2) \rightarrow e^{-iq \lambda \chi^{(1)}_{x_1}} e^{-iq \lambda \chi^{(2)}_{x_1}} \psi(x_1+\lambda, y_1, x_2, y_2) = e^{q \lambda X^{(1)}_{x}} \psi(x_1, y_1, x_2, y_2),\\
\psi(x_1, y_1, x_2, y_2) \rightarrow e^{-iq \lambda \chi^{(1)}_{y_1}} e^{-iq \lambda \chi^{(2)}_{y_1}} \psi(x_1, y_1+\lambda, x_2, y_2) = e^{q \lambda X^{(1)}_{y}} \psi(x_1, y_1, x_2, y_2),\\
\psi(x_1, y_1, x_2, y_2) \rightarrow e^{-iq \lambda \chi^{(1)}_{x_2}} e^{-iq \lambda \chi^{(2)}_{x_2}} \psi(x_1, y_1, x_2+\lambda, y_2) = e^{q \lambda X^{(2)}_{x}} \psi(x_1, y_1, x_2, y_2),\\
\psi(x_1, y_1, x_2, y_2) \rightarrow e^{-iq \lambda \chi^{(1)}_{y_2}} e^{-iq \lambda \chi^{(2)}_{y_2}} \psi(x_1, y_1, x_2, y_2+\lambda) = e^{q \lambda X^{(2)}_{y}} \psi(x_1, y_1, x_2, y_2).
\end{align}
The above compensations are generated by the operators $X^{(i)}_x$ and $X^{(i)}_y \,\, (i=1,2)$, which are defined as
\begin{gather}
X^{(1)}_x = \partial_{x_1} -i\pi M^{(1)}y_1 -i\pi M^{(21)} y_2, \qquad X^{(1)}_y = \partial_{y_1} +i\pi M^{(1)}x_1 +i\pi M^{(21)} x_2,\\
X^{(2)}_x = \partial_{x_2} -i\pi M^{(12)}y_1 -i\pi M^{(2)} y_2, \qquad X^{(2)}_y = \partial_{y_2} +i\pi M^{(12)}x_1 +i\pi M^{(2)} x_2.
\end{gather}
These generators satisfy the Heisenberg algebras,
\begin{gather}
[X^{(i)}_x, X^{(i)}_y] = M^{(i)} X^{(i)}_Q,\\
[X^{(i)}_x, X^{(j)}_x]=[X^{(i)}_y, X^{(j)}_y]=0,
\end{gather}
where $i,j=1,2$ and we define $X^{(i)}_Q \equiv 2\pi i$.
The Heisenberg algebras exponentiate to the group element,
\begin{align}
&g(\epsilon^{(1)}_x, \epsilon^{(2)}_x, \epsilon^{(1)}_y, \epsilon^{(2)}_y, \epsilon^{(1)}_Q, \epsilon^{(2)}_Q) \notag\\
&\hspace{40pt}= \exp\left( \frac{\epsilon^{(1)}_x}{N} X^{(1)}_x + \frac{\epsilon^{(2)}_x}{N} X^{(2)}_x + \frac{\epsilon^{(1)}_y}{N} X^{(1)}_y + \frac{\epsilon^{(2)}_y}{N} X^{(2)}_y + \frac{\epsilon^{(1)}_Q}{N} X^{(1)}_Q + \frac{\epsilon^{(2)}_Q}{N} X^{(2)}_Q \right),\label{group}
\end{align}
where $N \equiv \det\mathbb{N}$ and 
\begin{gather}
\mathbb{N}=
\begin{pmatrix}
M^{(1)} & M^{(21)}\\
M^{(12)} & M^{(2)}
\end{pmatrix}
.
\end{gather}
Accordingly, the following relation is satisfied :
\begin{align}
&g(\epsilon^{(1)\prime}_x, \epsilon^{(2)\prime}_x, \epsilon^{(1)\prime}_y, \epsilon^{(2)\prime}_y, \epsilon^{(1)\prime}_Q, \epsilon^{(2)\prime}_Q) \, g(\epsilon^{(1)}_x, \epsilon^{(2)}_x, \epsilon^{(1)}_y, \epsilon^{(2)}_y, \epsilon^{(1)}_Q, \epsilon^{(2)}_Q) \notag\\
&\hspace{20pt}= g \biggl( \epsilon^{(1)\prime}_x+\epsilon^{(1)}_x, \epsilon^{(2)\prime}_x+\epsilon^{(2)}_x, \epsilon^{(1)\prime}_y+\epsilon^{(1)}_y, \epsilon^{(2)\prime}_y+\epsilon^{(2)}_y, \notag\\
&\hspace{55pt}\epsilon^{(1)\prime}_Q+\epsilon^{(1)}_Q+\frac{1}{2N^2} (\epsilon^{(1)\prime}_x \epsilon^{(1)}_y - \epsilon^{(1)\prime}_y \epsilon^{(1)}_x) M^{(1)}, \epsilon^{(2)\prime}_Q+\epsilon^{(2)}_Q+\frac{1}{2N^2} (\epsilon^{(2)\prime}_x \epsilon^{(2)}_y - \epsilon^{(2)\prime}_y \epsilon^{(2)}_x) M^{(2)} \biggr).
\end{align}
This implies that there exist discrete symmetries with respect to parameters $\epsilon^{(1)}_x$, $\epsilon^{(1)}_y$, $\epsilon^{(2)}_x$, $\epsilon^{(2)}_y$, $\epsilon^{(1)}_Q$ and $\epsilon^{(2)}_Q$.
Since two tori are compactified, we must impose periodic boundary conditions, namely
\begin{gather}
\psi(x_1+1,y_1,x_2,y_2) = e^{iq \chi^{(i)}_{x_1}} \psi(x_1,y_1,x_2,y_2), \quad 
\psi(x_1,y_1+1,x_2,y_2) = e^{iq \chi^{(i)}_{y_1}} \psi(x_1,y_1,x_2,y_2),\\
\psi(x_1,y_1,x_2+1,y_2) = e^{iq \chi^{(i)}_{x_2}} \psi(x_1,y_1,x_2,y_2), \quad 
\psi(x_1,y_1,x_2,y_2+1) = e^{iq \chi^{(i)}_{y_2}} \psi(x_1,y_1,x_2,y_2),
\end{gather}
for $i=1,2$.
The generators $X^{(i)}_x$, $X^{(i)}_y$ and $X^{(i)}_Q \,\, (i=1,2)$ must be compatible with the above conditions.
The generator $X^{(i)}_Q \,\, (i=1,2)$ satisfies automatically the above requirement, while the others not so.
Since the following condition :
\begin{gather}
e^{iq X^{(1)}_x} e^{iq \chi^{(1)}_y} \psi(x_1,y_1,x_2,y_2) = e^{iq \chi^{(1)}_y} e^{iq X^{(1)}_x} \psi(x_1,y_1,x_2,y_2),
\end{gather}
must be satisfied, the magnetic flux is quantized as $qM^{(1)} \in \mathbb{Z}$.
In particular, we have $M^{(1)} \in \mathbb{Z}$ for a wavefunction with $q=1$.
The same holds for the other magnetic fluxes, i.e., $M^{(12)}$, $M^{(21)}$, $M^{(2)} \in \mathbb{Z}$.
After all, we obtain $N \in \mathbb{Z}$.
This is exactly the same as the Dirac's quantization condition.
For particles with charge $q=1$, the discrete symmetry corresponds to the following set characterized by discrete parameters :
\begin{align}
\bold{P} &= \bigl\{ g(n^{(1)}_x, n^{(2)}_x, n^{(1)}_y, n^{(2)}_y, \epsilon^{(1)}_Q, \epsilon^{(2)}_Q) \,| \notag\\
&\hspace{70pt} n^{(i)}_X=0,1, \ldots, N-1 \,\, (i=1,2, \, X=x,y) \,;\, \epsilon^{(i)}_Q \,\, (i=1,2) \in \mathbb{R} \bigr\}.\label{setofgroup}
\end{align}

In fact, we have the zero-mode wavefunction on magnetized tori, which is written as
\begin{gather}
\psi^{\vec{j}, \, \mathbb{N}} (\vec{z}, \Omega) = e^{\pi i (\mathbb{N} \cdot \vec{z}) \cdot (\textrm{Im} \, \Omega)^{-1} \cdot (\textrm{Im} \, \vec{z})} \cdot \vartheta
\begin{bmatrix}
\vec{j}\\[5pt] 0
\end{bmatrix}
(\mathbb{N} \cdot \vec{z}, \mathbb{N} \cdot \Omega),
\end{gather}
up to a normalization factor.
For simplicity we set $N=3$.
We can straightforwardly check that the action of the group element (\ref{group}) is calculated as
\begin{align}
&g(n^{(1)}_x, n^{(2)}_x, n^{(1)}_y, n^{(2)}_y, \epsilon^{(1)}_Q, \epsilon^{(2)}_Q) \psi^{\vec{j}, \, \mathbb{N}} (\vec{z}, \Omega) \notag\\
&\hspace{20pt} = \exp\left[ 2\pi i \cdot \frac{j_1}{N} (M^{(1)}n^{(1)}_x + M^{(21)}n^{(2)}_x) \right] \exp\left[ 2\pi i \cdot \frac{j_2}{N} (M^{(12)}n^{(1)}_x + M^{(2)}n^{(2)}_x) \right] \notag\\
&\hspace{80pt} \times \exp\left[2\pi i \left(\frac{\epsilon^{(1)}_Q+\epsilon^{(2)}_Q} {N}+ \frac{n^{(1)}_x n^{(1)}_y}{2N^2} M^{(1)} + \frac{n^{(2)}_xn^{(2)}_y}{2N^2} M^{(2)} \right)\right] \psi^{\vec{j}+\vec{n}, \, \mathbb{N}} (\vec{z}, \Omega),\label{gpsi}
\end{align}
where $\vec{j} \equiv (j_1, j_2)$ and $\vec{n} \equiv (n^{(1)}_y/N, n^{(2)}_y/N)$.
The above relation holds only if the discrete parameters $n^{(i)}_X \,\, (i=1,2, \, X=x,y)$ satisfy the constraint, which is summarized in Table \ref{tab:gpsi_constraints}.

\begin{table}[h]
\centering
\begin{tabular}{ccc} \hline
Generation-type of $\vec{j}$ & $(n^{(1)}_x, n^{(2)}_x)$& $(n^{(1)}_y, n^{(2)}_y)$\\ \hline
Type 1 & $(1, 0)$ or $(2, 0)$ & $(1, 0)$ or $(2,0)$\\
Type 2 & $(1,1)$ or $(2,2)$ & $(1, 1)$ or $(2,2)$\\
Type 3 & $(1,2)$ or $(2,1)$& $(1, 2)$ or $(2,1)$\\
Type 4 & $(0,1)$ or $(0,2)$ & $(0,1)$ or $(0,2)$\\ \hline
\end{tabular}
\caption{The constraints that is indispensable for the equality in Eq.~(\ref{gpsi}).}
\label{tab:gpsi_constraints}
\end{table}

\noindent We can interpret the group element (\ref{group}) as the generator of the non-Abelian discrete flavor symmetry.
Thus, the discrete parameters are mapped into the representations of the generators appearing in the flavor symmetry.
Let us study an example.
We consider the following matrix of magnetic fluxes,
\begin{gather}
\mathbb{N} =
\begin{pmatrix}
2 & 1\\
1 & 2
\end{pmatrix}
.
\end{gather}
For the labels of Type 2, the group element
\begin{gather}
g(2, 2, 0, 0, 0, 0) =
\begin{pmatrix}
1 & 0 & 0\\
0 & \omega & 0\\
0 & 0 & \omega^2
\end{pmatrix}
,
\end{gather}
corresponds to the $Z_3$ generator $Z$, with $\omega \equiv e^{2\pi i /3}$.
Similarly,
\begin{gather}
g(0, 0, 1, 1, 0, 0) =
\begin{pmatrix}
0 & 1 & 0\\
0 & 0 & 1\\
1 & 0 & 0
\end{pmatrix}
,
\end{gather}
corresponds to the $Z_3^{(C)}$ generator $C$.
Then, the last group element
\begin{gather}
g(0,0,0,0,1, 0) =
\begin{pmatrix}
\omega & 0 & 0\\
0 & \omega & 0\\
0 & 0 & \omega
\end{pmatrix}
,
\end{gather}
is necessary for the closed algebra generated by the above generators.
In the end, we obtain the non-Abelian discrete flavor symmetry,
\begin{gather}
\bold{P} = \left\{Z=
\begin{pmatrix}
1 & 0 & 0\\
0 & \omega & 0\\
0 & 0 & \omega^2
\end{pmatrix}
, \, Z' =
\begin{pmatrix}
\omega & 0 & 0\\
0 & \omega & 0\\
0 & 0 & \omega
\end{pmatrix}
, \, C=
\begin{pmatrix}
0 & 1 & 0\\
0 & 0 & 1\\
1 & 0 & 0
\end{pmatrix}
\right\} = \Delta(27),
\end{gather}
in the 4D effective theory.
The same holds for the other generation-types, with replacing the arguments of the group elements (\ref{group}).
One can apply the above method to the other magnetized models with non-factorizable fluxes and obtain the generators of the other non-Abelian discrete flavor symmetries.

\section{Conclusions and discussions} \label{sec:conclusion}
We have studied the non-Abelian discrete flavor symmetries from magnetized brane models.
We have found that $Z_g \times Z_g$, $(Z_g \times Z_g) \rtimes Z_2$, $(Z_g \times Z_g) \rtimes Z_g$ and $(Z_g \times Z_g) \rtimes D_g$ symmetries appear from the magnetized torus model with non-factorizable fluxes, if generation-types in three sectors forming Yukawa couplings are aligned.
In three-generation models of quarks and leptons, $Z_3 \times Z_3$, $S_3 \times Z_3$, $\Delta(27)$ and $\Delta(54)$ symmetries can appear.
On the other hand, if the generation-types are not aligned, $Z_3 \times Z_3$ and $S_3 \times Z_3$ symmetries can appear.
The flavor symmetries obtained from non-factorizable fluxes are phenomenologically attractive.
Such results can become a clue when we reveal the property of the magnetized brane models.
In studying the flavor symmetry, we investigated the label $\vec{i}$, the generation-types of $\vec{i}$ and the number of zero-modes ($\det\mathbb{N}$).
In addition, we focused on the selection rule and the character of Riemann $\vartheta$-function.
We have studied the number of the generation-types and the classification for $|\det\mathbb{N}|=3$.

We have studied the non-Abelian discrete flavor symmetry from the magnetized orbifold model with non-factorizable fluxes.
We have found that $D_4$ and $(Z_g \times Z_g) \rtimes Z_g \,\, (g=4k)$ symmetries can appear from such a model.
Unlike the magnetized torus model only with factorizable fluxes, $(Z_g \times Z_g) \rtimes Z_g \,\, (g=4k)$ can survive after the orbifold projection.

We have also analyzed the non-Abelian discrete flavor symmetry from the perspective of gauge symmetry breaking.
Especially, we applied the method developed in Ref.~\cite{BerasaluceGonzalez:2011wy} to the model with non-factorizable fluxes, and confirmed the reappearance of $\Delta(27)$ flavor symmetry.

Here, we discuss phenomenological implications of our results.
The analyses in this paper show that one can derive several flavor structures from the torus compactification with non-factorizable magnetic fluxes as well as the orbifold compactification.
The Yukawa couplings among left-handed and right-handed fermions and Higgs fields in each of the up-type quark sector, down-type quark sector, charged lepton sector and neutrino sector can have non-Abelian discrete flavor symmetries such as $\Delta (54)$, $\Delta (27)$ and $S_3 \times Z_3$ as well as Abelian flavor symmetries such as $Z_3 \times Z_3$.

As shown in Ref.~\cite{Abe:2013bba}, non-factorizable magnetic fluxes make it possible to construct the models, where the charged lepton (up-type quark) sector and the neutrino (down-type quark) sector have flavor symmetries different from each other.
Then, such symmetries are broken down into their subgroup, which is common in all of the sectors.
This is quite interesting.
For example, one tries to understand the lepton mixing angles by using non-Abelian discrete flavor symmetries in field-theoretical model building as follows \cite{Altarelli:2010gt, Ishimori:2010au, King:2013eh}.
First, one assumes that there is a larger flavor symmetry in the full Lagrangian.
Then, one breaks it by vacuum expectation values of scalar fields such that the charged lepton sector and the neutrino sector (the up-type sector and the down-type sector) have different unbroken symmetries.
For instance, one can derive the tri-bimaximal mixing matrix, when the charged lepton mass terms and the neutrino mass terms have certain $Z_3$ and $Z_2$ symmetries, respectively.
Following such process, one can obtain other mixing angles.\footnote{See for a bottom-up type of systematic studies, e.g., Ref.~\cite{Holthausen:2012wt} including a study on the quark sector.}

{}Form such a viewpoint of model building, our results are fascinating.
As mentioned above, non-factorizble magnetic fluxes can lead to different flavor symmetries between the charged lepton sector and the neutrino sector, and also the flavor symmetries between the up-type quarks and down-type quarks can be different from each other.
That is, the gauge backgroup in extra dimensions breaks a larger symmetry and leads to different flavor symmetries between the charged leptons and neutrinos, up-type quarks and down-types quarks.\footnote{The orbifold embedding in a flavor symmetry may also breaks it and leads to different symmetries between the charged leptons and neutrinos,up-type quarks and down-type quarks \cite{Kobayashi:2008ih}.}
In the above sense, even the Abelian symmetries in some of the charged lepton, neutrino, up-type quark, and down-type quark sectors are interesting.
When non-Abelian discrete flavor symmetries remain in one sector of the up-type quarks, down-type quarks, charged leptons and neutrinos, the Higgs scalar fields are also multiplets under these symmetries.
A certain pattern of the VEVs of Higgs multiplet would break non-Abelian flavor symmetries into $Z_3$, $Z_2$ or the other Abelian discrete symmetry.
Then, we would find realistic mixing angles.
We would study such analysis systematically including the right-handed Majorana neutrino masses\footnote{See, e.g., Ref.~\cite{Hamada:2014hpa} for patterns of Majorana neutrino masses, which can be induced by stringy non-perturbative effects.} elsewhere.

\section*{Acknowledgement}
H.A. was supported in part by the Grant-in-Aid for Scientific Research No.~25800158 from the Ministry of Education, Culture, Sports, Science and Technology (MEXT) in Japan. 
T.K. was supported in part by the Grant-in-Aid for Scientific Research No.~25400252 from the MEXT in Japan. 
H.O. was supported in part by the JSPS Grant-in-Aid for Scientific Research (S) No.~22224003, for young Scientists (B) No.~25800139 from the MEXT in Japan. 
K.S. was supported in part by a Grant-in-Aid for JSPS Fellows No.~25$\cdot$4968 and a Grant for Excellent Graduate Schools from the MEXT in Japan. 
Y.T. was supported in part by a Grant for Excellent Graduate Schools from the MEXT in Japan. 
Y.T. would like to thank Yusuke Shimizu for fruitful discussions in Summer Institute 2013.

\appendix 
\section{The generation-types for $\det\mathbb{N}=n$} \label{appendix}
We refer to the generation-types for $\det\mathbb{N}=n$.
The number of generation-types is given as the sum of divisors of $\det\mathbb{N}=n$, as stated.
Although the strict proof is beyond the scope of this paper, in this appendix, we provide a certain aspect of this fact based on the label vector $\vec{i}$. 

First, we consider $\det\mathbb{N}=p$, where $p$ is a prime number.
For this case, the generation-type is classified by the direction of the label $\vec{i} \equiv (i_1, i_2)$.
For example, for $p=3$, four generation-types are schematized as follows.

\vspace{20pt}
\begin{minipage}{\textwidth}
\hspace{-15pt}
\begin{minipage}{0.245\textwidth}
\centering\setlength{\unitlength}{0.6mm}
\begin{picture}(35,37)(0,0)
\put(0,0){\vector(1,0){32}}
\put(0,0){\vector(0,1){32}}
\put(34,-1){$i_1$}
\put(-1,34){$i_2$}
\put(0,0){\circle*{3}}
\put(10,0){\circle*{3}}
\put(20,0){\circle*{3}} 
\end{picture}
\end{minipage}
\begin{minipage}{0.245\textwidth}
\centering\setlength{\unitlength}{0.6mm}
\begin{picture}(35,37)(0,0)
\put(0,0){\vector(1,0){32}}
\put(0,0){\vector(0,1){32}}
\put(34,-1){$i_1$}
\put(-1,34){$i_2$}
\put(0,0){\circle*{3}}
\put(10,20){\circle*{3}}
\put(20,10){\circle*{3}} 
\end{picture}
\end{minipage}
\begin{minipage}{0.245\textwidth}
\centering\setlength{\unitlength}{0.6mm}
\begin{picture}(35,37)(0,0)
\put(0,0){\vector(1,0){32}}
\put(0,0){\vector(0,1){32}}
\put(34,-1){$i_1$}
\put(-1,34){$i_2$}
\put(0,0){\circle*{3}}
\put(10,10){\circle*{3}}
\put(20,20){\circle*{3}} 
\end{picture}
\end{minipage}
\begin{minipage}{0.245\textwidth}
\centering\setlength{\unitlength}{0.6mm}
\begin{picture}(35,37)(0,0)
\put(0,0){\vector(1,0){32}}
\put(0,0){\vector(0,1){32}}
\put(34,-1){$i_1$}
\put(-1,34){$i_2$}
\put(0,0){\circle*{3}}
\put(0,10){\circle*{3}}
\put(0,20){\circle*{3}} 
\end{picture}
\end{minipage}\vspace{20pt}
\end{minipage}

\noindent The gradients on $(i_1, i_2)$-plane of these generation-types are $0$, $1/2$, $1$ and $\infty$, respectively from the left.
The above gradients can be written as
\begin{gather}
0, \qquad 1/n \quad (n=0,1,2).
\end{gather}
Then the number of generation-types is $1+3=4$.
For $\det\mathbb{N}=5$, generation-types are shown as follows.

\vspace{20pt}
\begin{minipage}{\textwidth}
\hspace{-20pt}
\begin{minipage}{0.165\textwidth}
\centering\setlength{\unitlength}{0.6mm}
\begin{picture}(35,37)(0,0)
\put(0,0){\vector(1,0){32}}
\put(0,0){\vector(0,1){32}}
\put(34,-1){$i_1$}
\put(-1,34){$i_2$}
\put(0,0){\circle*{3}}
\put(6,0){\circle*{3}}
\put(12,0){\circle*{3}} 
\put(18,0){\circle*{3}}
\put(24,0){\circle*{3}} 
\end{picture}
\end{minipage}
\begin{minipage}{0.165\textwidth}
\centering\setlength{\unitlength}{0.6mm}
\begin{picture}(35,37)(0,0)
\put(0,0){\vector(1,0){32}}
\put(0,0){\vector(0,1){32}}
\put(34,-1){$i_1$}
\put(-1,34){$i_2$}
\put(0,0){\circle*{3}}
\put(24,6){\circle*{3}}
\put(18,12){\circle*{3}} 
\put(12,18){\circle*{3}}
\put(6,24){\circle*{3}} 
\end{picture}
\end{minipage}
\begin{minipage}{0.165\textwidth}
\centering\setlength{\unitlength}{0.6mm}
\begin{picture}(35,37)(0,0)
\put(0,0){\vector(1,0){32}}
\put(0,0){\vector(0,1){32}}
\put(34,-1){$i_1$}
\put(-1,34){$i_2$}
\put(0,0){\circle*{3}}
\put(18,6){\circle*{3}}
\put(6,12){\circle*{3}} 
\put(24,18){\circle*{3}}
\put(12,24){\circle*{3}} 
\end{picture}
\end{minipage}
\begin{minipage}{0.167\textwidth}
\centering\setlength{\unitlength}{0.6mm}
\begin{picture}(35,37)(0,0)
\put(0,0){\vector(1,0){32}}
\put(0,0){\vector(0,1){32}}
\put(34,-1){$i_1$}
\put(-1,34){$i_2$}
\put(0,0){\circle*{3}}
\put(6,18){\circle*{3}}
\put(12,6){\circle*{3}} 
\put(18,24){\circle*{3}}
\put(24,12){\circle*{3}} 
\end{picture}
\end{minipage}
\begin{minipage}{0.1657\textwidth}
\centering\setlength{\unitlength}{0.6mm}
\begin{picture}(35,37)(0,0)
\put(0,0){\vector(1,0){32}}
\put(0,0){\vector(0,1){32}}
\put(34,-1){$i_1$}
\put(-1,34){$i_2$}
\put(0,0){\circle*{3}}
\put(6,6){\circle*{3}}
\put(12,12){\circle*{3}} 
\put(18,18){\circle*{3}}
\put(24,24){\circle*{3}} 
\end{picture}
\end{minipage}
\begin{minipage}{0.165\textwidth}
\centering\setlength{\unitlength}{0.6mm}
\begin{picture}(35,37)(0,0)
\put(0,0){\vector(1,0){32}}
\put(0,0){\vector(0,1){32}}
\put(34,-1){$i_1$}
\put(-1,34){$i_2$}
\put(0,0){\circle*{3}}
\put(0,6){\circle*{3}}
\put(0,12){\circle*{3}} 
\put(0,18){\circle*{3}}
\put(0,24){\circle*{3}} 
\end{picture}
\end{minipage}\vspace{20pt}
\end{minipage}

\noindent Also in this case, the gradients can be written as
\begin{gather}
0, \qquad 1/n \quad (0,1, \ldots, 4).
\end{gather}
For arbitrary $p$, the gradients of the label are written as,
\begin{gather}
0, \qquad 1/n \quad (0,1, \ldots, |p|-1).
\end{gather}
Thus, for $\det\mathbb{N}=p$, we can classify the generation-types by their gradients and there are $(1+|p|)$ generation-types.

Next, we consider the case with $\det\mathbb{N}=pq$, where $p$ and $q$ are prime numbers.
For $|p|=|q|$, let us show you an example, $\det\mathbb{N}=4$.
Then, generation-types are found as follows. 

\vspace{20pt}
\begin{minipage}{\textwidth}
\hspace{-15pt}
\begin{minipage}{0.2\textwidth}
\centering\setlength{\unitlength}{0.6mm}
\begin{picture}(35,37)(0,0)
\put(0,0){\vector(1,0){32}}
\put(0,0){\vector(0,1){32}}
\put(34,-1){$i_1$}
\put(-1,34){$i_2$}
\put(0,0){\circle*{3}}
\put(8,0){\circle*{3}}
\put(16,0){\circle*{3}} 
\put(24,0){\circle*{3}}
\end{picture}
\end{minipage}
\begin{minipage}{0.2\textwidth}
\centering\setlength{\unitlength}{0.6mm}
\begin{picture}(35,37)(0,0)
\put(0,0){\vector(1,0){32}}
\put(0,0){\vector(0,1){32}}
\put(34,-1){$i_1$}
\put(-1,34){$i_2$}
\put(0,0){\circle*{3}}
\put(8,24){\circle*{3}}
\put(16,16){\circle*{3}} 
\put(24,8){\circle*{3}} 
\end{picture}
\end{minipage}
\begin{minipage}{0.2\textwidth}
\centering\setlength{\unitlength}{0.6mm}
\begin{picture}(35,37)(0,0)
\put(0,0){\vector(1,0){32}}
\put(0,0){\vector(0,1){32}}
\put(34,-1){$i_1$}
\put(-1,34){$i_2$}
\put(0,0){\circle*{3}}
\put(0,16){\circle*{3}}
\put(16,8){\circle*{3}} 
\put(16,24){\circle*{3}} 
\end{picture}
\end{minipage}
\begin{minipage}{0.2\textwidth}
\centering\setlength{\unitlength}{0.6mm}
\begin{picture}(35,37)(0,0)
\put(0,0){\vector(1,0){32}}
\put(0,0){\vector(0,1){32}}
\put(34,-1){$i_1$}
\put(-1,34){$i_2$}
\put(0,0){\circle*{3}}
\put(8,8){\circle*{3}}
\put(16,16){\circle*{3}} 
\put(24,24){\circle*{3}} 
\end{picture}
\end{minipage}
\begin{minipage}{0.2\textwidth}
\centering\setlength{\unitlength}{0.6mm}
\begin{picture}(35,37)(0,0)
\put(0,0){\vector(1,0){32}}
\put(0,0){\vector(0,1){32}}
\put(34,-1){$i_1$}
\put(-1,34){$i_2$}
\put(0,0){\circle*{3}}
\put(0,8){\circle*{3}}
\put(0,16){\circle*{3}} 
\put(0,24){\circle*{3}} 
\end{picture}
\end{minipage}
\end{minipage}

\vspace{20pt}
\begin{minipage}{\textwidth}\centering
\begin{minipage}{0.25\textwidth}
\centering\setlength{\unitlength}{0.6mm}
\begin{picture}(35,37)(0,0)
\put(0,0){\vector(1,0){32}}
\put(0,0){\vector(0,1){32}}
\put(34,-1){$i_1$}
\put(-1,34){$i_2$}
\put(0,0){\circle*{3}}
\put(0,16){\circle*{3}}
\put(16,0){\circle*{3}} 
\put(16,16){\circle*{3}} 
\end{picture}
\end{minipage}
\begin{minipage}{0.25\textwidth}
\centering\setlength{\unitlength}{0.6mm}
\begin{picture}(35,37)(0,0)
\put(0,0){\vector(1,0){32}}
\put(0,0){\vector(0,1){32}}
\put(34,-1){$i_1$}
\put(-1,34){$i_2$}
\put(0,0){\circle*{3}}
\put(8,16){\circle*{3}}
\put(16,0){\circle*{3}} 
\put(24,16){\circle*{3}} 
\end{picture}
\end{minipage}\vspace{20pt}
\end{minipage}
The first five generation-types can be classified by their gradients, as above.
There are two more generation-types, where two of four points reside in the $i_1$-axis.
Note that when $\det\mathbb{N}$ equals to composite number, we can not classify the generation-types only by their gradients.

For $|p| \neq |q|$, we show an example, $\det\mathbb{N}=6$.
Then, generation-types are depicted as follows. 

\vspace{20pt}
\begin{minipage}{\textwidth}
\hspace{-15pt}
\begin{minipage}{0.245\textwidth}
\centering\setlength{\unitlength}{0.6mm}
\begin{picture}(35,37)(0,0)
\put(0,0){\vector(1,0){32}}
\put(0,0){\vector(0,1){32}}
\put(34,-1){$i_1$}
\put(-1,34){$i_2$}
\put(0,0){\circle*{3}}
\put(5,0){\circle*{3}}
\put(10,0){\circle*{3}} 
\put(15,0){\circle*{3}}
\put(20,0){\circle*{3}} 
\put(25,0){\circle*{3}}
\end{picture}
\end{minipage}
\begin{minipage}{0.245\textwidth}
\centering\setlength{\unitlength}{0.6mm}
\begin{picture}(35,37)(0,0)
\put(0,0){\vector(1,0){32}}
\put(0,0){\vector(0,1){32}}
\put(34,-1){$i_1$}
\put(-1,34){$i_2$}
\put(0,0){\circle*{3}}
\put(5,25){\circle*{3}}
\put(10,20){\circle*{3}} 
\put(15,15){\circle*{3}}
\put(20,10){\circle*{3}} 
\put(25,5){\circle*{3}}
\end{picture}
\end{minipage}
\begin{minipage}{0.245\textwidth}
\centering\setlength{\unitlength}{0.6mm}
\begin{picture}(35,37)(0,0)
\put(0,0){\vector(1,0){32}}
\put(0,0){\vector(0,1){32}}
\put(34,-1){$i_1$}
\put(-1,34){$i_2$}
\put(0,0){\circle*{3}}
\put(0,15){\circle*{3}}
\put(10,10){\circle*{3}} 
\put(10,25){\circle*{3}}
\put(20,5){\circle*{3}} 
\put(20,20){\circle*{3}}
\end{picture}
\end{minipage}
\begin{minipage}{0.245\textwidth}
\centering\setlength{\unitlength}{0.6mm}
\begin{picture}(35,37)(0,0)
\put(0,0){\vector(1,0){32}}
\put(0,0){\vector(0,1){32}}
\put(34,-1){$i_1$}
\put(-1,34){$i_2$}
\put(0,0){\circle*{3}}
\put(0,10){\circle*{3}}
\put(0,20){\circle*{3}} 
\put(15,5){\circle*{3}}
\put(15,15){\circle*{3}} 
\put(15,25){\circle*{3}} 
\end{picture}
\end{minipage}
\end{minipage}

\vspace{20pt}
\begin{minipage}{\textwidth}\centering
\begin{minipage}{0.25\textwidth}
\centering\setlength{\unitlength}{0.6mm}
\begin{picture}(35,37)(0,0)
\put(0,0){\vector(1,0){32}}
\put(0,0){\vector(0,1){32}}
\put(34,-1){$i_1$}
\put(-1,34){$i_2$}
\put(0,0){\circle*{3}}
\put(0,15){\circle*{3}}
\put(10,5){\circle*{3}} 
\put(10,20){\circle*{3}}
\put(20,10){\circle*{3}} 
\put(20,25){\circle*{3}} 
\end{picture}
\end{minipage}
\begin{minipage}{0.25\textwidth}
\centering\setlength{\unitlength}{0.6mm}
\begin{picture}(35,37)(0,0)
\put(0,0){\vector(1,0){32}}
\put(0,0){\vector(0,1){32}}
\put(34,-1){$i_1$}
\put(-1,34){$i_2$}
\put(0,0){\circle*{3}}
\put(5,5){\circle*{3}}
\put(10,10){\circle*{3}} 
\put(15,15){\circle*{3}}
\put(20,20){\circle*{3}} 
\put(25,25){\circle*{3}} 
\end{picture}
\end{minipage}
\begin{minipage}{0.25\textwidth}
\centering\setlength{\unitlength}{0.6mm}
\begin{picture}(35,37)(0,0)
\put(0,0){\vector(1,0){32}}
\put(0,0){\vector(0,1){32}}
\put(34,-1){$i_1$}
\put(-1,34){$i_2$}
\put(0,0){\circle*{3}}
\put(0,5){\circle*{3}}
\put(0,10){\circle*{3}} 
\put(0,15){\circle*{3}}
\put(0,20){\circle*{3}} 
\put(0,25){\circle*{3}} 
\end{picture}
\end{minipage}\vspace{20pt}
\end{minipage}
These generation-types can be classified by their gradients.
The number of generation-types with two points existing on the $i_1$-axis are three, as shown in the graphs below.

\vspace{20pt}
\begin{minipage}{\textwidth}\centering
\begin{minipage}{0.25\textwidth}
\centering\setlength{\unitlength}{0.6mm}
\begin{picture}(35,37)(0,0)
\put(0,0){\vector(1,0){32}}
\put(0,0){\vector(0,1){32}}
\put(34,-1){$i_1$}
\put(-1,34){$i_2$}
\put(0,0){\circle*{3}}
\put(0,10){\circle*{3}}
\put(0,20){\circle*{3}} 
\put(15,0){\circle*{3}}
\put(15,10){\circle*{3}} 
\put(15,20){\circle*{3}} 
\end{picture}
\end{minipage}
\begin{minipage}{0.25\textwidth}
\centering\setlength{\unitlength}{0.6mm}
\begin{picture}(35,37)(0,0)
\put(0,0){\vector(1,0){32}}
\put(0,0){\vector(0,1){32}}
\put(34,-1){$i_1$}
\put(-1,34){$i_2$}
\put(0,0){\circle*{3}}
\put(5,10){\circle*{3}}
\put(10,20){\circle*{3}} 
\put(15,0){\circle*{3}}
\put(20,10){\circle*{3}} 
\put(25,20){\circle*{3}} 
\end{picture}
\end{minipage}
\begin{minipage}{0.25\textwidth}
\centering\setlength{\unitlength}{0.6mm}
\begin{picture}(35,37)(0,0)
\put(0,0){\vector(1,0){32}}
\put(0,0){\vector(0,1){32}}
\put(34,-1){$i_1$}
\put(-1,34){$i_2$}
\put(0,0){\circle*{3}}
\put(5,20){\circle*{3}}
\put(10,10){\circle*{3}} 
\put(15,0){\circle*{3}}
\put(20,20){\circle*{3}} 
\put(25,10){\circle*{3}} 
\end{picture}
\end{minipage}\vspace{20pt}
\end{minipage}
While the number of generation-types with three points existing on the $i_1$-axis are two, as shown in the graphs below.

\vspace{20pt}
\begin{minipage}{\textwidth}\centering
\begin{minipage}{0.25\textwidth}
\centering\setlength{\unitlength}{0.6mm}
\begin{picture}(35,37)(0,0)
\put(0,0){\vector(1,0){32}}
\put(0,0){\vector(0,1){32}}
\put(34,-1){$i_1$}
\put(-1,34){$i_2$}
\put(0,0){\circle*{3}}
\put(0,15){\circle*{3}}
\put(10,0){\circle*{3}} 
\put(10,15){\circle*{3}}
\put(20,0){\circle*{3}} 
\put(20,15){\circle*{3}} 
\end{picture}
\end{minipage}
\begin{minipage}{0.25\textwidth}
\centering\setlength{\unitlength}{0.6mm}
\begin{picture}(35,37)(0,0)
\put(0,0){\vector(1,0){32}}
\put(0,0){\vector(0,1){32}}
\put(34,-1){$i_1$}
\put(-1,34){$i_2$}
\put(0,0){\circle*{3}}
\put(5,15){\circle*{3}}
\put(10,0){\circle*{3}} 
\put(15,15){\circle*{3}}
\put(20,0){\circle*{3}} 
\put(25,15){\circle*{3}} 
\end{picture}
\end{minipage}\vspace{20pt}
\end{minipage}

\noindent Notice that the number of generation-types with $|p|$ points existing on the $i_1$-axis are $|q|$ types, and vice versa.
Thus, we can classify all the generation-types by their gradients and the number of the points existing on the $i_1$-axis.

We can straightforwardly apply the above argument to the case with $\det\mathbb{N} = pqr, pqrs, \ldots$.
Accordingly, we would demonstrate that the number of generation-types is given as the sum of divisors of $\det\mathbb{N}=n$.

\section{More about flavor symmetries in three-generation models : aligned generation-types}\label{sec:egs}
It is shown that there exists $\Delta(27)$ or $Z_3^{(C)} \times Z_3'$ ($\Delta(54)$ or $S_3 \times Z_3'$) as the flavor symmetries in the case with aligned generation-types.
In this appendix, some other configurations of magnetic fluxes for $g=3$ and the resultant flavor symmetries are analyzed.
These results are enumerated in Table \ref{tab:enumeration1} and \ref{tab:enumeration2}.

\begin{table}[H]
\centering\small
\begin{tabular}{cccccc} \hline
 \# of zero-modes & $\mathbb{N}_{ab}$ & $\mathbb{N}_{ca}$ & $\mathbb{N}_{bc}$ & generation-types & flavor symmetry \\ \hline
\multirow{5}{*}{3-3-3} & $\begin{pmatrix}3&3\\ 0&-1\end{pmatrix}$ & $\begin{pmatrix}6&-5\\ -3&2\end{pmatrix}$ & $\begin{pmatrix}-9&2\\ 3&-1\end{pmatrix}$ & 1 & $Z_3^{(C)} \times Z_3'$ \\
 & $\begin{pmatrix}3&3\\ 2&1\end{pmatrix}$ & $\begin{pmatrix}-1&-2\\ -1&1\end{pmatrix}$ & $\begin{pmatrix}-2&-1\\ -1&-2\end{pmatrix}$ & 2 & $Z_3^{(C)} \times Z_3'$ \\
 & $\begin{pmatrix}-1&0\\ -1&3\end{pmatrix}$ & $\begin{pmatrix}5&-3\\ -6&3\end{pmatrix}$ & $\begin{pmatrix}-4&3\\ 7&-6\end{pmatrix}$ & 4 & $Z_3^{(C)} \times Z_3'$ \\ \hline
\multirow{11}{*}{3-3-6} & $\begin{pmatrix}0&-1\\ -3&-1\end{pmatrix}$ & $\begin{pmatrix}3&3\\ 6&5\end{pmatrix}$ & $\begin{pmatrix}-3&-2\\ -3&-4\end{pmatrix}$ & 1 & $Z_3^{(C)} \times Z_3'$ \\
 & $\begin{pmatrix}3&-4\\ 0&-1\end{pmatrix}$ & $\begin{pmatrix}6&5\\ 3&2\end{pmatrix}$ & $\begin{pmatrix}-9&-1\\ -3&-1\end{pmatrix}$ & 1 & $\Delta(27)$ \\
 & $\begin{pmatrix}5&2\\ 4&1\end{pmatrix}$ & $\begin{pmatrix}2&-1\\ -3&0\end{pmatrix}$ & $\begin{pmatrix}-7&-1\\ -1&-1\end{pmatrix}$ & 3 & $Z_3^{(C)} \times Z_3'$ \\
 & $\begin{pmatrix}-1&-1\\ -2&1\end{pmatrix}$ & $\begin{pmatrix}5&2\\ 4&1\end{pmatrix}$ & $\begin{pmatrix}-4&-1\\ -2&-2\end{pmatrix}$ & 3 & $\Delta(27)$ \\
 & $\begin{pmatrix}3&6\\ 2&3\end{pmatrix}$ & $\begin{pmatrix}0&-3\\ -1&0\end{pmatrix}$ & $\begin{pmatrix}-3&-3\\ -1&-3\end{pmatrix}$ & 4 & $Z_3^{(C)} \times Z_3'$ \\
 & $\begin{pmatrix}5&-3\\ -1&0\end{pmatrix}$ & $\begin{pmatrix}-1&0\\ -1&3\end{pmatrix}$ & $\begin{pmatrix}-4&3\\ 2&-3\end{pmatrix}$ & 4 & $\Delta(27)$ \\ \hline
\multirow{7}{*}{3-3-9} & $\begin{pmatrix}0&1\\ 3&5\end{pmatrix}$ & $\begin{pmatrix}3&0\\ 0&-1\end{pmatrix}$ & $\begin{pmatrix}-3&-1\\ -3&-4\end{pmatrix}$ & 1 & $Z_3^{(C)} \times Z_3'$ \\
 & $\begin{pmatrix}3&3\\ 5&4\end{pmatrix}$ & $\begin{pmatrix}4&-1\\ 1&-1\end{pmatrix}$ & $\begin{pmatrix}-7&-2\\ -6&-3\end{pmatrix}$ & 2 & $Z_3^{(C)} \times Z_3'$ \\
 & $\begin{pmatrix}0&-3\\ -1&5\end{pmatrix}$ & $\begin{pmatrix}5&-4\\ -2&1\end{pmatrix}$ & $\begin{pmatrix}-5&7\\ 3&-6\end{pmatrix}$ & 3 & $Z_3^{(C)} \times Z_3'$ \\
 & $\begin{pmatrix}3&3\\ 4&3\end{pmatrix}$ & $\begin{pmatrix}-1&0\\ -3&3\end{pmatrix}$ & $\begin{pmatrix}-2&-3\\ -1&-6\end{pmatrix}$ & 4 & $Z_3^{(C)} \times Z_3'$ \\ \hline
\multirow{7}{*}{3-3-12} & $\begin{pmatrix}3&5\\ 3&4\end{pmatrix}$ & $\begin{pmatrix}3&-3\\ 0&-1\end{pmatrix}$ & $\begin{pmatrix}-6&-2\\ -3&-3\end{pmatrix}$ & 1 & $Z_3^{(C)} \times Z_3'$ \\
 & $\begin{pmatrix}3&3\\ 5&4\end{pmatrix}$ & $\begin{pmatrix}1&-1\\ -3&0\end{pmatrix}$ & $\begin{pmatrix}-4&-2\\ -2&-4\end{pmatrix}$ & 2 & $Z_3^{(C)} \times Z_3'$ \\
 & $\begin{pmatrix}3&-3\\ -5&4\end{pmatrix}$ & $\begin{pmatrix}1&1\\ 3&0\end{pmatrix}$ & $\begin{pmatrix}-4&2\\ 2&-4\end{pmatrix}$ & 3 & $Z_3^{(C)} \times Z_3'$ \\
 & $\begin{pmatrix}0&3\\ 1&3\end{pmatrix}$ & $\begin{pmatrix}4&-3\\ -1&0\end{pmatrix}$ & $\begin{pmatrix}-4&0\\ 0&-3\end{pmatrix}$ & 4 & $Z_3^{(C)} \times Z_3'$ \\ \hline
\end{tabular}
\caption{The configurations of magnetic fluxes and flavor symmetries. Flavor symmetries are written in the case with non-vanishing Wilson-lines.}
\label{tab:enumeration1}
\end{table}
\begin{table}[H]
\centering\small
\begin{tabular}{cccccc} \hline
 \# of zero-modes & $\mathbb{N}_{ab}$ & $\mathbb{N}_{ca}$ & $\mathbb{N}_{bc}$ & generation-types & flavor symmetry \\ \hline
\multirow{7}{*}{3-6-6} & $\begin{pmatrix}3&-5\\ -3&4\end{pmatrix}$ & $\begin{pmatrix}3&3\\ 0&-2\end{pmatrix}$ & $\begin{pmatrix}-6&2\\ 3&-2\end{pmatrix}$ & 1 & $Z_3^{(C)} \times Z_3'$ \\
 & $\begin{pmatrix}3&3\\ 5&4\end{pmatrix}$ & $\begin{pmatrix}-1&-2\\ -3&0\end{pmatrix}$ & $\begin{pmatrix}-2&-1\\ -2&-4\end{pmatrix}$ & 2 & $Z_3^{(C)} \times Z_3'$ \\
 & $\begin{pmatrix}3&-3\\ -5&4\end{pmatrix}$ & $\begin{pmatrix}-1&2\\ 3&0\end{pmatrix}$ & $\begin{pmatrix}-2&1\\ 2&-4\end{pmatrix}$ & 3 & $Z_3^{(C)} \times Z_3'$ \\
 & $\begin{pmatrix}3&-3\\ -1&0\end{pmatrix}$ & $\begin{pmatrix}-1&3\\ 1&3\end{pmatrix}$ & $\begin{pmatrix}-2&0\\ 0&-3\end{pmatrix}$ & 4 & $Z_3^{(C)} \times Z_3'$ \\ \hline
\multirow{15}{*}{3-6-9} & $\begin{pmatrix}3&4\\ 0&-1\end{pmatrix}$ & $\begin{pmatrix}3&-5\\ -3&3\end{pmatrix}$ & $\begin{pmatrix}-6&1\\ 3&-2\end{pmatrix}$ & 1 & $Z_3^{(C)} \times Z_3'$ \\
 & $\begin{pmatrix}0&1\\ 3&4\end{pmatrix}$ & $\begin{pmatrix}3&-1\\ -3&-1\end{pmatrix}$ & $\begin{pmatrix}-3&0\\ 0&-3\end{pmatrix}$ & 1 & $\Delta(27)$ \\
 & $\begin{pmatrix}1&-1\\ 0&3\end{pmatrix}$ & $\begin{pmatrix}5&-2\\ -3&0\end{pmatrix}$ & $\begin{pmatrix}-4&1\\ 3&-3\end{pmatrix}$ & 2 & $Z_3^{(C)} \times Z_3'$ \\
 & $\begin{pmatrix}7&2\\ 5&1\end{pmatrix}$ & $\begin{pmatrix}-2&-1\\ -4&1\end{pmatrix}$ & $\begin{pmatrix}-5&-1\\ -1&-2\end{pmatrix}$ & 2 & $\Delta(27)$ \\
 & $\begin{pmatrix}1&1\\ 3&0\end{pmatrix}$ & $\begin{pmatrix}5&-4\\ -4&2\end{pmatrix}$ & $\begin{pmatrix}-6&3\\ 1&-2\end{pmatrix}$ & 3 & $Z_3^{(C)} \times Z_3'$ \\
 & $\begin{pmatrix}-1&-1\\ 2&5\end{pmatrix}$ & $\begin{pmatrix}4&-2\\ -5&1\end{pmatrix}$ & $\begin{pmatrix}-3&3\\ 3&-6\end{pmatrix}$ & 3 & $\Delta(27)$ \\
 & $\begin{pmatrix}-1&0\\ -4&3\end{pmatrix}$ & $\begin{pmatrix}3&3\\ 5&3\end{pmatrix}$ & $\begin{pmatrix}-2&-3\\ -1&-6\end{pmatrix}$ & 4 & $Z_3^{(C)} \times Z_3'$ \\
 & $\begin{pmatrix}1&-3\\ -2&3\end{pmatrix}$ & $\begin{pmatrix}2&3\\ 2&0\end{pmatrix}$ & $\begin{pmatrix}-3&0\\ 0&-3\end{pmatrix}$ & 4 & $\Delta(27)$ \\ \hline
\end{tabular}
\caption{The configurations of magnetic fluxes and flavor symmetries. Flavor symmetries are written in the case with non-vanishing Wilson-lines.}
\label{tab:enumeration2}
\end{table}

\section{Flavor symmetries in four-generation models}\label{sec:4gene}
We study the flavor symmetries for $g=4$, i.e., $Z_4^{(C)}$ and $(Z_4 \times Z_4') \rtimes Z_4^{(C)}$ with non-vanishing Wilson-lines or $D_4$ and $(Z_4 \times Z_4') \rtimes D_4$ without Wilson-lines, depending on the zero-mode degeneracies, the combination of generation-types and the existence of non-vanishing Wilson-lines.
In the following, we assume the vanishing Wilson-lines in the expressions of Yukawa couplings.
First, we consider the configuration of magnetic fluxes, which is given as
\begin{gather}
\mathbb{N}_{ab}=
\begin{pmatrix}
-2&-2\\
0&2
\end{pmatrix}, \qquad
\mathbb{N}_{ca}=
\begin{pmatrix}
6&4\\
4&2
\end{pmatrix}, \qquad
\mathbb{N}_{bc}=
\begin{pmatrix}
-4&-2\\
-4&-4
\end{pmatrix}.
\end{gather}
Then, the labels of zero-modes are given by
\begin{align}
\vec{i}_{ab,0}=
\begin{pmatrix}
0\\ 0
\end{pmatrix}, \qquad
\vec{i}_{ab,1}&=
\begin{pmatrix}
0\\ 1/2
\end{pmatrix}, \qquad
\vec{i}_{ab,2}=
\begin{pmatrix}
1/2\\ 0
\end{pmatrix}, \qquad
\vec{i}_{ab,3}=
\begin{pmatrix}
1/2\\ 1/2
\end{pmatrix},\\
\vec{i}_{ca,0}=
\begin{pmatrix}
0\\ 0
\end{pmatrix}, \qquad
\vec{i}_{ca,1}&=
\begin{pmatrix}
0\\ 1/2
\end{pmatrix}, \qquad
\vec{i}_{ca,2}=
\begin{pmatrix}
1/2\\ 0
\end{pmatrix}, \qquad
\vec{i}_{ca,3}=
\begin{pmatrix}
1/2\\ 1/2
\end{pmatrix}.
\end{align}
and
\begin{gather}
\vec{i}_{bc,0}=
\begin{pmatrix}
0\\ 0
\end{pmatrix}, \qquad
\vec{i}_{bc,1}=
\begin{pmatrix}
1/2\\ 0
\end{pmatrix}, \qquad
\vec{i}_{bc,2}=
\begin{pmatrix}
1/4\\ 0
\end{pmatrix}, \qquad
\vec{i}_{bc,3}=
\begin{pmatrix}
3/4\\ 0
\end{pmatrix},\\
\vec{i}_{bc,4}=
\begin{pmatrix}
0\\ 1/2
\end{pmatrix}, \qquad
\vec{i}_{bc,5}=
\begin{pmatrix}
1/2\\ 1/2
\end{pmatrix}, \qquad
\vec{i}_{bc,6}=
\begin{pmatrix}
1/4\\ 1/2
\end{pmatrix}, \qquad
\vec{i}_{bc,7}=
\begin{pmatrix}
3/4\\ 1/2
\end{pmatrix}.
\end{gather}
This is exactly the case with aligned generation-types.
Then, Yukawa couplings are written as
\begin{align}
\lambda_{\vec{i}_{ab}\vec{i}_{ca}\vec{i}_{bc,0}}=\lambda_{\vec{i}_{ab}\vec{i}_{ca}\vec{i}_{bc,1}}=
\begin{pmatrix}
\lambda_0 & 0 & 0 & 0\\
0 & 0 & 0 & \lambda_1\\
0 & 0 & \lambda_0 & 0\\
0 & \lambda_1 & 0 & 0
\end{pmatrix}, \quad
\lambda_{\vec{i}_{ab}\vec{i}_{ca}\vec{i}_{bc,2}}=\lambda_{\vec{i}_{ab}\vec{i}_{ca}\vec{i}_{bc,3}}=
\begin{pmatrix}
0 & 0 & \lambda_1 & 0\\
0 & \lambda_0 & 0 & 0\\
\lambda_1 & 0 & 0 & 0\\
0 & 0 & 0 & \lambda_0
\end{pmatrix},\\
\lambda_{\vec{i}_{ab}\vec{i}_{ca}\vec{i}_{bc,4}}=\lambda_{\vec{i}_{ab}\vec{i}_{ca}\vec{i}_{bc,5}}=
\begin{pmatrix}
0 & 0 & 0 & \lambda_2\\
\lambda_3 & 0 & 0 & 0\\
0 & \lambda_2 & 0 & 0\\
0 & 0 & \lambda_3 & 0
\end{pmatrix}, \quad
\lambda_{\vec{i}_{ab}\vec{i}_{ca}\vec{i}_{bc,6}}=\lambda_{\vec{i}_{ab}\vec{i}_{ca}\vec{i}_{bc,7}}=
\begin{pmatrix}
0 & \lambda_3 & 0 & 0\\
0 & 0 & \lambda_2 & 0\\
0 & 0 & 0 & \lambda_3\\
\lambda_2 & 0 & 0 & 0
\end{pmatrix}.
\end{align}
Thus, there do exist the symmetries under the three $Z_4$ generators, i.e., $Z$, $Z'$ and $C$ in these Yukawa couplings, and therefore we obtain $(Z_4 \times Z_4') \rtimes Z_4^{(C)}$ with non-vanishing Wilson-lines or $(Z_4 \times Z_4') \rtimes (Z_4^{(C)} \rtimes Z_2)$ with vanishing Wilson-lines.

Next, the magnetic fluxes
\begin{gather}
\mathbb{N}_{ab}=
\begin{pmatrix}
4&4\\
5&4
\end{pmatrix}, \qquad
\mathbb{N}_{ca}=
\begin{pmatrix}
-1&0\\
-1&4
\end{pmatrix}, \qquad
\mathbb{N}_{bc}=
\begin{pmatrix}
-3&-4\\
-4&-8
\end{pmatrix},
\end{gather}
lead to the following aligned generation-types :
\begin{align}
\vec{i}_{ab,0}=
\begin{pmatrix}
0\\ 0
\end{pmatrix}, \qquad
\vec{i}_{ab,1}&=
\begin{pmatrix}
0\\ 1/4
\end{pmatrix}, \qquad
\vec{i}_{ab,2}=
\begin{pmatrix}
0\\ 1/2
\end{pmatrix}, \qquad
\vec{i}_{ab,3}=
\begin{pmatrix}
0\\ 3/4
\end{pmatrix},\\
\vec{i}_{ca,0}=
\begin{pmatrix}
0\\ 0
\end{pmatrix}, \qquad
\vec{i}_{ca,1}&=
\begin{pmatrix}
0\\ 1/4
\end{pmatrix}, \qquad
\vec{i}_{ca,2}=
\begin{pmatrix}
0\\ 1/2
\end{pmatrix}, \qquad
\vec{i}_{ca,3}=
\begin{pmatrix}
0\\ 3/4
\end{pmatrix},
\end{align}
and
\begin{gather}
\vec{i}_{bc,0}=
\begin{pmatrix}
0\\ 0
\end{pmatrix}, \qquad
\vec{i}_{bc,1}=
\begin{pmatrix}
0\\ 1/2
\end{pmatrix}, \qquad
\vec{i}_{bc,2}=
\begin{pmatrix}
1/2\\ 0
\end{pmatrix}, \qquad
\vec{i}_{bc,3}=
\begin{pmatrix}
1/2\\ 1/2
\end{pmatrix},\\
\vec{i}_{bc,4}=
\begin{pmatrix}
0\\ 1/4
\end{pmatrix}, \qquad
\vec{i}_{bc,5}=
\begin{pmatrix}
0\\ 3/4
\end{pmatrix}, \qquad
\vec{i}_{bc,6}=
\begin{pmatrix}
1/2\\ 1/4
\end{pmatrix}, \qquad
\vec{i}_{bc,7}=
\begin{pmatrix}
1/2\\ 3/4
\end{pmatrix}.
\end{gather}
Then, the selection rule does not rule out any coupling, namely, Yukawa couplings have all non-vanishing elements, which are written as
\begin{align}
\lambda_{\vec{i}_{ab}\vec{i}_{ca}\vec{i}_{ab,0}}&=
\begin{pmatrix}
\lambda_0 & \lambda_1 & \lambda_4 & \lambda_1\\
\lambda_3 & \lambda_2 & \lambda_3 & \lambda_7\\
\lambda_8 & \lambda_6 & \lambda_5 & \lambda_6\\
\lambda_3 & \lambda_7 & \lambda_3 & \lambda_2
\end{pmatrix}, \qquad
\lambda_{\vec{i}_{ab}\vec{i}_{ca}\vec{i}_{ab,1}}=
\begin{pmatrix}
\lambda_5 & \lambda_6 & \lambda_8 & \lambda_6\\
\lambda_3 & \lambda_2 & \lambda_3 & \lambda_7\\
\lambda_4 & \lambda_1 & \lambda_0 & \lambda_1\\
\lambda_3 & \lambda_7 & \lambda_3 & \lambda_2
\end{pmatrix}, \\
\lambda_{\vec{i}_{ab}\vec{i}_{ca}\vec{i}_{ab,2}}&=
\begin{pmatrix}
\lambda_9 & \lambda_{10} & \lambda_{13} & \lambda_{10}\\
\lambda_{12} & \lambda_{11} & \lambda_{12} & \lambda_{16}\\
\lambda_{17} & \lambda_{15} & \lambda_{14} & \lambda_{15}\\
\lambda_{12} & \lambda_{16} & \lambda_{12} & \lambda_{11}
\end{pmatrix}, \qquad
\lambda_{\vec{i}_{ab}\vec{i}_{ca}\vec{i}_{ab,3}}=
\begin{pmatrix}
\lambda_{14} & \lambda_{15} & \lambda_{17} & \lambda_{15}\\
\lambda_{12} & \lambda_{11} & \lambda_{12} & \lambda_{16}\\
\lambda_{13} & \lambda_{10} & \lambda_9 & \lambda_{10}\\
\lambda_{12} & \lambda_{16} & \lambda_{12} & \lambda_{11}
\end{pmatrix},\\
\lambda_{\vec{i}_{ab}\vec{i}_{ca}\vec{i}_{ab,4}}&=
\begin{pmatrix}
\lambda_2 & \lambda_3 & \lambda_7 & \lambda_3\\
\lambda_6 & \lambda_5 & \lambda_6 & \lambda_8\\
\lambda_7 & \lambda_3 & \lambda_2 & \lambda_3\\
\lambda_1 & \lambda_4 & \lambda_1 & \lambda_0
\end{pmatrix}, \qquad
\lambda_{\vec{i}_{ab}\vec{i}_{ca}\vec{i}_{ab,5}}=
\begin{pmatrix}
\lambda_2 & \lambda_3 & \lambda_7 & \lambda_3\\
\lambda_1 & \lambda_0 & \lambda_1 & \lambda_4\\
\lambda_7 & \lambda_3 & \lambda_2 & \lambda_3\\
\lambda_6 & \lambda_8 & \lambda_6 & \lambda_5
\end{pmatrix}, \\
\lambda_{\vec{i}_{ab}\vec{i}_{ca}\vec{i}_{ab,6}}&=
\begin{pmatrix}
\lambda_{11} & \lambda_{12} & \lambda_{16} & \lambda_{12}\\
\lambda_{15} & \lambda_{14} & \lambda_{15} & \lambda_{17}\\
\lambda_{16} & \lambda_{12} & \lambda_{11} & \lambda_{12}\\
\lambda_{10} & \lambda_{13} & \lambda_{10} & \lambda_9
\end{pmatrix}, \qquad
\lambda_{\vec{i}_{ab}\vec{i}_{ca}\vec{i}_{ab,7}}=
\begin{pmatrix}
\lambda_{11} & \lambda_{12} & \lambda_{16} & \lambda_{12}\\
\lambda_{10} & \lambda_{9} & \lambda_{10} & \lambda_{13}\\
\lambda_{16} & \lambda_{12} & \lambda_{11} & \lambda_{12}\\
\lambda_{15} & \lambda_{17} & \lambda_{15} & \lambda_{14}
\end{pmatrix},
\end{align}
where values of $\lambda_n \,\, (n=0,1, \ldots, 17)$ are different from each other.
These Yukawa couplings do not allow the invariance under the $Z_4$ transformation $Z$, and therefore the flavor symmetry is $Z_4^{(C)}$, or $D_4$ with the existence of non-vanishing Wilson-lines.

Finally, we consider the configuration of fluxes
\begin{gather}
\mathbb{N}_{ab}=
\begin{pmatrix}
5&-1\\
1&-1
\end{pmatrix}, \qquad
\mathbb{N}_{ca}=
\begin{pmatrix}
0&-1\\
-4&3
\end{pmatrix}, \qquad
\mathbb{N}_{bc}=
\begin{pmatrix}
-5&2\\
3&-2
\end{pmatrix},
\end{gather}
which lead to the not-aligned generation-types, i.e.,
\begin{align}
\vec{i}_{ab,0}=
\begin{pmatrix}
0\\ 0
\end{pmatrix}, \qquad
\vec{i}_{ab,1}&=
\begin{pmatrix}
1/4\\ 1/4
\end{pmatrix}, \qquad
\vec{i}_{ab,2}=
\begin{pmatrix}
1/2\\ 1/3
\end{pmatrix}, \qquad
\vec{i}_{ab,3}=
\begin{pmatrix}
3/4\\ 3/4
\end{pmatrix},\\
\vec{i}_{ca,0}=
\begin{pmatrix}
0\\ 0
\end{pmatrix}, \qquad
\vec{i}_{ca,1}&=
\begin{pmatrix}
1/4\\ 0
\end{pmatrix}, \qquad
\vec{i}_{ca,2}=
\begin{pmatrix}
1/2\\ 0
\end{pmatrix}, \qquad
\vec{i}_{ca,3}=
\begin{pmatrix}
3/4\\ 0
\end{pmatrix},\\
\vec{i}_{bc,0}=
\begin{pmatrix}
0\\ 0
\end{pmatrix}, \qquad
\vec{i}_{bc,1}&=
\begin{pmatrix}
0\\ 1/2
\end{pmatrix}, \qquad
\vec{i}_{bc,2}=
\begin{pmatrix}
1/2\\ 1/4
\end{pmatrix}, \qquad
\vec{i}_{bc,3}=
\begin{pmatrix}
1/2\\ 3/4
\end{pmatrix},
\end{align}
and Yukawa couplings are given by
\begin{gather}
\lambda_{\vec{i}_{ab}\vec{i}_{ca}\vec{i}_{ab,0}}=
\begin{pmatrix}
\lambda_0 & 0 & 0 & 0\\
0 & 0 & 0 & \lambda_1\\
0 & 0 & \lambda_2 & 0\\
0 & \lambda_1 & 0 & 0
\end{pmatrix}, \qquad
\lambda_{\vec{i}_{ab}\vec{i}_{ca}\vec{i}_{ab,1}}=
\begin{pmatrix}
0 & 0 & \lambda_2 & 0\\
0 & \lambda_1 & 0 & 0\\
\lambda_0 & 0 & 0 & 0\\
0 & 0 & 0 & \lambda_1
\end{pmatrix},\\
\lambda_{\vec{i}_{ab}\vec{i}_{ca}\vec{i}_{ab,2}}=
\begin{pmatrix}
0 & \lambda_1 & 0 & 0\\
\lambda_4 & 0 & 0 & 0\\
0 & 0 & 0 & \lambda_1\\
0 & 0 & \lambda_3 & 0
\end{pmatrix}, \qquad
\lambda_{\vec{i}_{ab}\vec{i}_{ca}\vec{i}_{ab,3}}=
\begin{pmatrix}
0 & 0 & 0 & \lambda_1\\
0 & 0 & \lambda_3 & 0\\
0 & \lambda_1 & 0 & 0\\
\lambda_4 & 0 & 0 & 0
\end{pmatrix}.
\end{gather}
Therefore, we obtain $(Z_4 \rtimes Z_2) \times Z_4' \cong D_4 \times Z'_4$, or $Z_4 \times Z_4'$ with the existence of non-vanishing Wilson-lines.


\end{document}